\documentclass[final,12pt]{elsarticle}%

\usepackage[font=small,skip=0pt]{caption}
\usepackage{hyperref}
\usepackage{lineno}
\usepackage{graphicx}%
\usepackage{multirow}%
\usepackage{amsmath,amssymb,amsfonts}%
\usepackage{amsthm}%
\usepackage{mathrsfs}%
\usepackage[title]{appendix}%
\usepackage[dvipsnames]{xcolor}%
\usepackage{textcomp}%
\usepackage{manyfoot}%
\usepackage{booktabs}%
\usepackage{algorithm}%
\usepackage{algorithmicx}%
\usepackage{algpseudocode}%
\usepackage{listings}%
\usepackage{tikz}
\usetikzlibrary {perspective, patterns.meta, patterns}
\usepackage{pgfplots}
\pgfplotsset{compat=newest} 
\usepgfplotslibrary{units} 
\usepackage{siunitx}
\usepackage{xurl}
\usepackage{longtable}
\usepackage{soul}

\newcommand{\ddroit}{\mathrm{d}}
\newcommand{\us}[0]{{\color{black} \mathbf{u}^s}}
\newcommand{\vs}{{\color{black} \mathbf{v}^s}}
\newcommand{\plc}[0]{{\color{black} p^{lc}}}

\newcommand{\pc}[0]{{\color{black} p^c}}
\newcommand{\pl}[0]{{\color{black} p^l}}
\newcommand{\pb}[0]{{\color{black} p^b}}

\newcommand{\omegaodeux}[0]{{\color{black} \omega^{O2,l}}}

\usepackage{slashbox}
\usepackage{bbding}
\usepackage{pifont}
\usepackage{tcolorbox}
\tcbuselibrary{skins,breakable}

\DeclareSIUnit\mmHg{mmHg}

    {\endtcolorbox}

    {\endtcolorbox}

    {\endtcolorbox}

\begin{document}

\begin{frontmatter}
\title{Hierarchical poromechanical approach to investigate the impact of mechanical loading on human skin micro-circulation.}


\author[1,2,3]{Thomas Lavigne} 

\author[1]{Stéphane Urcun}

\author[5,6]{Bérengère Fromy}

\author[5,6]{Audrey Josset-Lamaugarny}

\author[2,3]{Alexandre Lagache}

\author[1]{Camilo A. Suarez-Afanador}

\author[1]{Stéphane P. A. Bordas}

\author[2]{Pierre-Yves Rohan}

\author[3,4]{Giuseppe Sciumè*}

\ead{giuseppe.sciume@u-bordeaux.fr}
\cortext[*]{Corresponding author}

\affiliation[1]{organization={Institute of Computational Engineering, Department of Engineering, University of Luxembourg},
addressline={2 place de l'université},
city={Esch-sur-Alzette},
postcode={L-4365},
country={Luxembourg}}

\affiliation[2]{organization={Institut de Biomécanique Humaine Georges Charpak, Arts et Métiers Institute of Technology},
addressline={151 boulevard de l'hôpital},
city={Paris},
postcode={F-75013},
country={France}}

\affiliation[3]{organization={CNRS, Bordeaux INP, I2M, UMR 5295, I2M Bordeaux, Arts et Metiers Institute of Technology, University of Bordeaux},
city={Talence},
postcode={F-33400},
country={France}}

\affiliation[4]{organization={Institut Universitaire de France (IUF)}}

\affiliation[5]{organization={CNRS UMR 5305, Tissue Biology and Therapeutic Engineering Laboratory (LBTI)},
city={Lyon},
postcode={F-69007},
country={France}}

\affiliation[6]{organization={Claude Bernard University Lyon},
city={Villeurbanne},
postcode={F-69100},
country={France}}


\begin{abstract}
Extensive research on human skin anatomy has revealed that the skin functions as a complex multi-scale and multi-phase system, containing up to $70\si{\percent}$ of bounded and free circulating water. The presence of moving fluids significantly influences the mechanical and biological responses of the skin, affecting its time-dependent behaviour and the transport of essential nutrients and oxygen to cells.

Poroelastic modelling emerges as a promising approach to investigate biologically relevant phenomena at finer scales while embedding crucial mechanisms at larger scales as it facilitates the integration of multi-scale and multi-physics processes. Despite extensive use of poromechanics in other tissues, no hierarchical multi-compartment porous model that incorporates blood supply has yet been experimentally evaluated to simulate the in vivo mechanical and micro-circulatory response of human skin.

This paper introduces a hierarchical two-compartment model that accounts for fluid distribution within the interstitium and the micro-circulation of blood. A general theoretical framework, which includes a biphasic interstitium (comprising interstitial fluid and non-structural cells), is formulated and studied through a one-dimensional consolidation test of a $100~\si{\micro\meter}$ column. The inclusion of a biphasic interstitium allows the model to account separately for the motion of cells and interstitial fluid, recognising their differing characteristic times. An extension of the model to include biological exchanges such as oxygen transport is discussed in the appendix. The preliminary evaluation demonstrated that cell viscosity introduces a second characteristic time beyond that of interstitial fluid movement. However, at high cell viscosity values and short time scales, cells exhibit behaviour akin to that of solid materials.

Based on these observations, a simplified version of the model was used to replicate an experimental campaign carried out on short time scales. Local pressure (up to $31$ kPa) was applied to the skin of the dorsal face of the middle finger through a laser Doppler probe PF801 (Perimed Sweden) attached to an apparatus as previously described (Fromy Brain Res 1998).

The model demonstrated its qualitative ability to represent both ischaemia and post-occlusive reactive hyperaemia, aligning with experimental observations.

All numerical simulations were performed using the open source software FEniCSx v0.9.0. To promote transparency and reproducibility, the anonymised experimental data and the corresponding finite element codes are publicly available on GitHub.
\end{abstract}



\begin{highlights}
\item 2-compartment porous media mechanics including both biphasic and mono-phasic interstitium,
\item \textit{In vivo} experimental LDF response of indented phalanx skin to indentation (loading),
\item 1D consolidation problem and 3D comparison to the experiment,
\item Porous media mechanics to account for the vascular micro-circulation of the skin.
\end{highlights}

\end{frontmatter}

\section{Introduction}\label{sec:introduction}


The role of the skin as the primary protective barrier to the external environment makes the skin highly susceptible to pathologies such as pressure ulcers (PU). 
The National Pressure Injury Advisory Panel redefined the definition of a pressure injury in 2016 as \textit{"A pressure injury [which] is localised damage to the skin and underlying soft tissue, usually over a bony prominence or related to a medical or other device. [...] The injury occurs as a result of intense and/or prolonged pressure or pressure in combination with shear. The tolerance of soft tissue for pressure and shear may also be affected by microclimate, nutrition, perfusion, co-morbidities, and condition of the soft tissue."}.
The prevalence of pressure ulcers varies widely among at-risk populations, such as the elderly, diabetic patients, and amputees. For example, 
between $9$ \si{\percent} and $20$ \si{\percent} of hospitalised patients in Europe are reported to have pressure ulcers (\citet{Vanderwee2007}). In particular, superficial pressure ulcers account for 70 \% of all cases (\citet{Kim2020, Moore2019}). Despite extensive research, current qualitative risk assessment methods for PUs lack reliability and remain insufficiently accurate (\citet{Anthony2008}). Preventing and managing skin injuries requires a deeper understanding of the skin structural composition and mechanical behaviour.

Anatomically, the skin comprises three primary layers: the epidermis, dermis, and hypodermis (\citet{Hsu2021}). The epidermis, the outermost layer, usually measures 50 to 150 \si{\micro\meter} in thickness and consists predominantly of keratinocytes (\citet{Burns2010, Wong2015, Joodaki2018, MostafaviYazdi2022, Abdo2020}). Beneath the epidermis lies the dermis, connected via a basement membrane. Together, these layers are often referred to as the cutis. The dermis, significantly thicker than the epidermis (0.5 to 5 \si{\milli\meter}), is composed of the papillary and reticular layers and serves as a structural matrix. This matrix retains water (about  $60$ \% of the total weight of the dermis) and is predominantly made of collagen ($80- 85$ \% dry weight), elastin ($2- 4$ \%) and reticulin (\citet{Burns2010, Dwivedi2022}). In contrast to the avascular epidermis, the dermis is highly vascularised (via blood and lymphatic vessels), containing sweat glands, hair follicles (approximately 11 per \si{\square\centi\meter}), nerves (\citet{Liao2013, Braverman1989}). The innermost layer, the hypodermis (subcutaneous tissue), primarily comprises adipose tissue and loose areolar tissue, which stores approximately $80$ \% of body fat in non-obese individuals (\citet{Querleux2002}). The composition of this layer significantly influences the biomechanical properties of the skin, as detailed by \citet{Querleux2002}, who characterised it as ${7.4}$ \si{\percent} water, ${7.3}$ \si{\percent} unsaturated lipids, and ${85.3}$ \si{\percent} saturated lipids.

Understanding the skin's behaviour and the risk to develop a PU during sustained loads thus necessitates multiscale and multiphysics approaches. Injury risk factors are classified into intrinsic parameters (e.g., skin morphology, oxygen levels) and extrinsic factors (e.g., external loading). At the macroscopic level, both excessive and sustained moderate loads can initiate PU formation (\citet{Ceelen2008, Loerakker2010, Loerakker2011, Stekelenburg2006, vanNierop2010, Traa2019, Tsuji2005}). At the microscopic level, ischaemia-reperfusion injury - characterised by oxygen deprivation followed by sudden re-oxygenation - causes cell damage (\citet{Stekelenburg2006, Loerakker2011, Tsuji2005}). Temporal factors, including duration of sustained compression, are critical determinants of tissue damage, highlighting the need for a holistic understanding of risk factors for PU formation (\citet{Coleman2013}). Consequently, investigating the time-dependent mechanical behaviour of the soft tissues accounting for their time-dependent response is essential to improve the prevention of PU.

Another crucial aspect is the incorporation of micro-circulation dynamics (\citet{Liao2013, Sree2019b, NguyenTu2013, Tsuji2005}). This supports various applications beyond PU prevention, including oncology (\citet{FUKUMURA200772, Russell2015, Tesselaar2017}), dermatology (\citet{Kelly1995, Baran2015, Braverman1989}), and hypertension risk assessment (\citet{Farkas2004, Rossi2006b}). Blood flow, a key determinant of tissue health, is evaluated in clinical routine using advanced methodologies such as laser Doppler imaging (\citet{Svedman1998, Kubli2000, Millet2011}) and optical coherence tomography (OCT) (\citet{An2010}). Laser Doppler flowmetry is a well-established, non-invasive, real-time technique for measuring micro-circulatory blood flow of the skin (\citet{Poffo2014, Varghese2009317, FolgosiCorrea2012}). The laser Doppler flowmetry method further presents the advantage of a precise temporal evaluation, but includes an averaged spatial response. In this context, computational models, including porous media approaches, offer a robust framework for coupling mechanical and biochemical processes in soft tissues. 

3D-1D coupled models were introduced by \citet{Bauer2005} to gain a deeper understanding of the underlying physiological mechanisms of the initial micro-circulatory response to mechanical skin irritation. \citet{Mithraratne20121071} proposed a coupled 3D-1D computational model of the foot, in which hydrostatic pressure acts on the external surface of blood vessels, leading to a reduction in the flow cross-sectional area and, consequently, to a decrease in blood supply. Similarly, the study by \citet{Sree2019b} investigated oxygen diffusion in human skin using a multilayered model that includes the stratum corneum, the viable epidermis, and the dermis. The model considers a volume $1~\si{\cubic\milli\meter}$ that required $1,208,437$ elements, which is considerably heavy and limits the scale of the sample studied. Computational homogenisation schemes (in particular, the FE$^2$ method) have been proposed to bridge the scales between cells and tissue using the strain-time cell death threshold (\citet{Breuls2003,Breuls2002,Lustig2021}).

Alternatively, mechanical approaches based on porous media mechanics allow the coupling of the behaviour of the solid phase with that of the fluid in the pores, where biochemical exchanges occur. Such models, originally developed for soil mechanics (\citet{Terzaghi_1943}) are now quite extensively applied in biomechanics in different contexts. Given the multiphasic and multiphysics nature of soft tissues, such models can be transferred to study their behaviour with two important differences: a large strain regime can be reached and there is a strong coupling during the loading between the solid and the fluid phases. These models have been applied to various tissues, including the brain (\citet{Budday2019, HosseiniFarid2020, Greiner2021, Urcun2022a, urcun_2023, HervasRaluy2023, CarrascoMantis2023,deLucio2024}), the liver (\citet{ricken2019computational}), the meniscus (\citet{Kazemi2013,Uzuner2020, bulle:tel-03652547, Bulle2021, Uzuner2022}), and muscle tissue (\citet{Lavigne2022}). Porous models also present an opportunity to current practice in including biochemistry and diffusion processes through a strong coupling as developed in \citet{Scium2021,Urcun2021} for tumour growth and as proposed by \citet{deLucio2023,deLucio2024} for drug delivery. Although poromechanics theory has been extensively applied in other soft biological tissues, only a few studies have been published for the skin, and these were mainly limited to \textit{in silico} and \textit{ex vivo} studies. Furthermore, to the best of the authors' knowledge, no hierarchical multi-compartment porous model including blood supply has been evaluated with regard to experiments yet.

This paper investigates the feasibility of using a two-compartment porous model to simulate the \textit{in vivo} mechanical and micro-circulatory response of human skin. The proposed theory incorporates a biphasic interstitium and a vascular network. All simulations have been performed using open-source software FEniCSx, which supports parallel computation. Sensitivity analysis of material parameters and qualitative comparisons with experimental data, including laser Doppler flowmetry, are presented to evaluate the model's potential for advancing our understanding of skin mechanics and pathology.

\section{Materials and methods}\label{matmeth}


\subsection{Data Acquisition}
\label{sec:experimental}


The cutaneous perfusion state of the middle finger's mid-phalanx under indentation has been assessed using a Laser Doppler Flowmeter (PERIMED PERIFLUX 5000). Smoking volunteers were excluded from this study. The LDF technique is based on the Doppler effect, observing a frequency shift when there is relative movement between the target and the source (\citet{Fang2024,Pedanekar2018}). Given the size and speed of red blood cells, the LDF probe uses a fixed wavelength of $780$ \si{\nano\meter}. The shift in wavelength after hitting red blood cells is proportional to their number and speed, regardless of the direction of movement (\citet{Fang2024,Pedanekar2018}).

A total of \textcolor{black}{11} healthy volunteers (6 men and 5 women), aged $25 \pm 2$ years, were recruited for the study after informed consent was provided. This study received ethical approval (national registration number RCB: 2023-A00418-37). Volunteers who smoked were excluded from this study. They were seated comfortably and the room temperature was meticulously regulated at $20.9 \pm 0.6$°C. The probe was calibrated prior to measurements using the PERIMED PF1001 calibration tool. In addition, the skin temperature on the dorsum of the hand was continuously monitored to ensure stable thermal conditions. A 10-minute acclimatisation period was observed to allow stabilisation of both skin temperature and LDF readings prior to data collection. The temperature and LDF signals were recorded at 2 Hz intervals.

\begin{figure}[ht!]
    \centering
    \includegraphics[width=\linewidth]{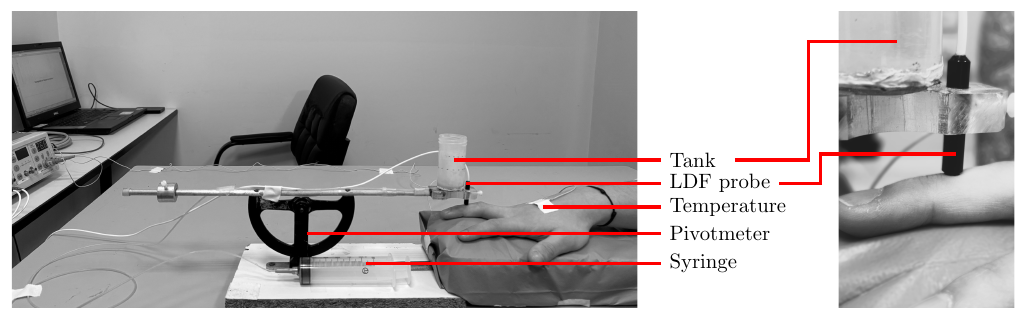}
    \caption{Experimental set-up. The load is directly applied to the skin using the LDF probe. The probe is placed at the extremity of the pivotmeter which is initially set at its equilibrium point. A controlled volume of water is then added/removed in the tank aside the probe to increase/decrease the load magnitude.}
    \label{fig:pivotmeter}
\end{figure}

The load was applied to the midpoint of the phalanx directly via the LDF probe using a pivotmeter. The pivotmeter, depicted in Figure \ref{fig:pivotmeter}, is an in-house loading device developed by \citet{fromy1998,fromy2000}. Initially, the device was set to its equilibrium state, with the probe gently contacting the skin under the influence of an equilibrium mass. The loading application was achieved by filling the tank adjacent to the probe with water, allowing for precise and controlled loading conditions.

\begin{figure}[ht!]
    \centering
    \includegraphics[width=0.85
    \linewidth]{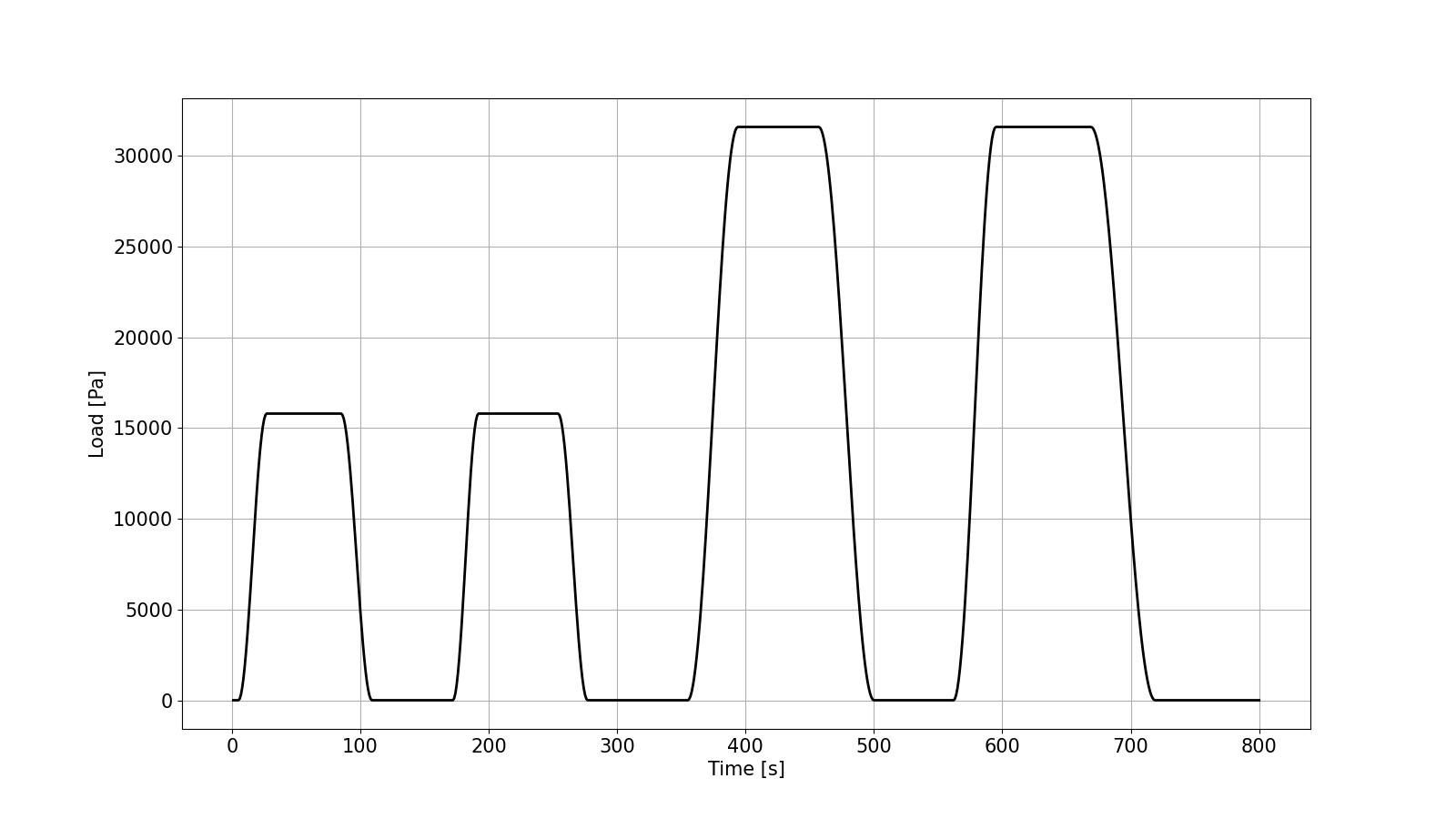}
    \caption{Four cycles of loading-sustaining-unloading-sustaining were performed. The first two cycles respectively corresponded to a $15~\si{\kilo\pascal}$ load and the two last ones to a $31~\si{\kilo\pascal}$ load.}
    \label{fig:load}
\end{figure}

Measurements were carried out over two consecutive days, with two sessions per day, each separated by at least a 15-minute rest period. Each session consisted of a sequence of four loading-sustaining-unloading-sustaining cycles. The initial two cycles involved a water volume of $20~\si{\milli\litre}$ corresponding to a load of $15~\si{\kilo\pascal}$, while the subsequent two cycles used a volume of $40~\si{\milli\litre}$ corresponding to a load of $31~\si{\kilo\pascal}$ (Figure \ref{fig:load}).

The water was added and removed from the tank using a syringe and capillary, ensuring precise control of the process at a consistent rate of $1~\si{\milli\litre\per\second}$. This controlled loading and unloading protocol facilitated an accurate and reproducible assessment of perfusion changes during each cycle and computational modelling.

\subsection{The mathematical model}

\begin{figure}[ht!]
    \centering
    \includegraphics[width=\textwidth,trim={0 1cm 0cm 0cm},clip]{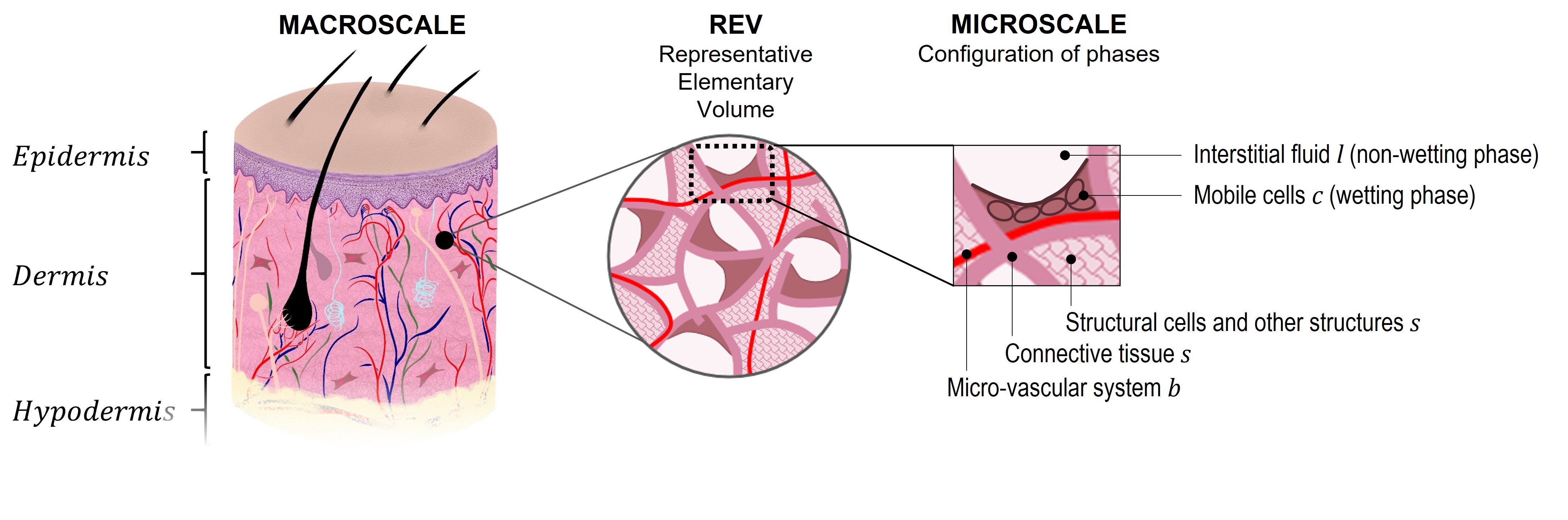}
    \caption{Description of the Representative Elementary Volume: the solid scaffold comprising the connective tissue, appendices and stroma cells (pink), the circulating cells (brown), the interstitial fluid (white), and the micro-vascular system (red). Nutrient and oxygen exchanges occur between the fluid phases.}
    \label{fig:REV}
\end{figure}

A 2-compartment model has been developed to incorporate a biphasic interstitium. The following developments are based on the previous work of \citet{Scium2021,Lavigne2023}. As detailed in the anatomy of the skin aforementioned in the introduction, the model considers several distinct phases, which are detailed in Table \ref{tab:species_two_comp}. The Representative Elementary Volume (REV) is provided in Figure \ref{fig:REV}.

\begin{table}[ht!]
\centering
\resizebox{\columnwidth}{!}{%
\begin{tabular}{|l|c|c|c|c|c|c|}
\hline
\backslashbox{Phase}{Species} & Connective Tissue       & \begin{tabular}[c]{@{}c@{}}Other structures \\ (glands, hair, etc.)\end{tabular} & \begin{tabular}[c]{@{}c@{}}Cells in the\\ interstitium \end{tabular} & Water        & Oxygen    & Other Species \\ \hline
Solid (s)                     & \ding{51} & \ding{51}                                                   &                                                                &     \ding{51}    &           &        \ding{51}       \\
Liquid (l)                    &           &                                                             &                                                                & \ding{51} & \ding{51} & \ding{51}     \\
Cell (c)                      &           &                                                             & \ding{51}                                                      & \ding{51}          & \ding{51} & \ding{51}     \\ 
Vascular (blood) (b)                      &           &                                                             &                                                       & \ding{51}           & \ding{51} & \ding{51}     \\ \hline
\end{tabular}%
}
\caption{Description of the model phases: the 2-comparment model is assumed to be composed of a solid phase, an interstitium composed of an interstitial fluid phase and a cell phase, and a vascular phase.}
\label{tab:species_two_comp}
\end{table}

The water absorbed by the connective tissue constituents is assumed to be locally in equilibrium with the pressure of the interstitial fluid. This equilibrium ensures that vessels within connective tissue are subjected to compression (or dilation) according to the average pressure, $p^s$, of the fluids in the interstitium of the solid scaffold. The solid scaffold of tissue can be conceptualised as comprising four key components: connective tissue itself, resident cells, skin appendices, and absorbed fluid, which together define the biomechanical and physiological properties of the structure. The solid scaffold has two other independent compartments: the interstitium and the vascular compartment.

The liquid phase of the interstitium is modelled as a water-like medium that is responsible for transporting oxygen and other nutrients. The cellular phase encompasses non-structural cells within the skin, as well as embedded water, oxygen, and nutrient elements. Lastly, blood in the vascular compartment is characterised as Newtonian and contains red blood cells, which facilitate the delivery of oxygen and nutrients to the interstitial compartment, ensuring sustained metabolic function. 

These elements collectively form the foundation of a hierarchical 3D poroelastic model that accounts for haemodynamic processes.

Similar to the classical theory of multiphasic porous media, the proposed model is developed based on the concept of a wetting phase, a non-wetting phase, and their associated capillary pressure within the biphasic interstitium (\citet{Scium2021}).
Within a pore, the interstitial fluid and cells coexist, but typically one phase adheres more closely to the solid scaffold, whereas the other occupies the central region of the pore. These are termed the wetting phase and the non-wetting phase, respectively. In the present case, we consider the interstitial fluid as the non-wetting phase and the cells as the wetting phase. The capillary pressure, $p^{lc}$, is defined in Equation \ref{eq:capillarypressure} as the difference between the pressures of the non-wetting ($p^l$) and wetting phases ($p^c$). This pressure difference is intrinsically related to the saturation levels of each phase, which provides a framework for describing the fluid dynamics within the biphasic interstitium.
\begin{align}
\plc=\pl-\pc\label{eq:capillarypressure}
\end{align}

\noindent where $\pc=\pl - \plc$ is the wetting phase pressure, \textit{e.g.} the cell pressure in the present case.

The primary variables of the mathematical model are the displacement field of the solid scaffold $\us$, the capillary pressure $\plc$, the pressure of the interstitial fluid (non-wetting phase) $\pl$ and the blood pressure $\pb$.

\subsubsection{Governing Equations}
Each phase occupies part of the volume of a Representative Elementary Volume. Within the REV, the volume fractions are defined according to Equation \ref{eq:volume_frac} and respect the constraint Equation \ref{eq:2comp:volfrac}.
\begin{align}
	&\varepsilon^\alpha = \frac{\mathrm{Volume}^\alpha}{\sum_{\mathrm{phases}}\mathrm{Volume}^{\mathrm{phases}}} \label{eq:volume_frac} \\
	&\varepsilon^s + \underbrace{\varepsilon^c + \varepsilon^{l}}_{\mathrm{extra-vascular\,porosity}}+\underbrace{\varepsilon^{b}}_{\mathrm{vascular porosity}} = 1
\label{eq:2comp:volfrac}
\end{align}

Specifically, the vascular porosity $\varepsilon^b$ is introduced along with the extra-vascular porosity $\varepsilon$:
\begin{align}
	\varepsilon &=  \varepsilon^c + \varepsilon^{l} \label{eq:twocomp:extravascporo}
\end{align}

We further introduce the saturation $S^\alpha=\frac{\varepsilon^\alpha}{\varepsilon}$ of the wetting and non-wetting non-miscible phases in the interstitium and their constraint Equation \ref{eq:saturation_constrain}.
\begin{align} 
	S^{c} + S^{l} = 1 \label{eq:saturation_constrain}
\end{align}

The porous medium reads the classical conservation laws of mechanics. The \ul{momentum conservation} of the multiphase continuum gives:
\begin{align}
	\mathbf{\nabla}\cdot\mathbf{t}^{\text{tot}} + \mathbf{b} = \mathbf{0}\label{eq:twocomp:momentum}
\end{align}

\noindent where $\mathbf{t}^{\text{tot}}$ is the total Cauchy stress tensor. 

For the fluid phases, the momentum conservation reads:
\begin{align}
    -\frac{\mathbf{K}^f}{\mu^f}\,\mathbf{\nabla}p^f=\varepsilon^f\left(\mathbf{v}^f-\vs \right),\;(f=l,c,b)
    \label{eq:momentum_fluid}
\end{align}
\noindent where $\frac{\mathbf{K}^f}{\mu^f}$ is referred to as hydraulic conductivity, which is influenced by the characteristics of both the moving fluid and the porous solid material. Specifically, it is determined by the dynamic viscosity of the fluid $\mu^f$ and the intrinsic permeability $\mathbf{K}^f$ of the solid matrix. In the case of blood, the momentum Equation \ref{eq:momentum_fluid} holds true under the assumptions of slow laminar flow with negligible inertial effects.

The \ul{mass conservation} Equations are introduced for each phase of the medium using the material derivative operator $\frac{\mathrm{D}^s \cdot}{\mathrm{D}t}$ to describe the movement of the fluids with respect to the solid scaffold (\citet{Scium2021}):
\begin{align}
    \frac{\mathrm{D}^s}{\mathrm{D}t}(\rho^s \varepsilon^s) + \rho^s \varepsilon^s \nabla \cdot\vs = 0 \label{eq:masssolid10}\; & \text{solid phase},\\
    \underbrace{\frac{\mathrm{D}^s}{\mathrm{D}t}( \varepsilon^f)}_{\text{Accumulation rate}} + \underbrace{\nabla \cdot( \varepsilon^f(\mathbf{v}^f-\vs))}_{\text{Infiltration}} + \underbrace{\varepsilon^f \nabla \cdot \vs}_{\text{ECM deformation}} = 0 \label{eq:massfluid10}\; & (f=l,c,b),
\end{align}

\subsubsection{Constitutive Equations}

To simplify the mathematical model, the following assumptions were established:
\begin{enumerate}
    \item Blood vessels mainly interact with fluid constituents, and so they have no relevant mechanical interaction with the "structural" connective tissue fibbers.
    \item The solid pressure, $p^s$, is assumed to be related to the pressure of the fluids in the extravascular space only and respects Equation \ref{eq:total_fluid_pressure}.
    \item Consistently with hypothesis 1, we suppose that $\varepsilon^b$ depends on $( \pb - p^{s} )$.
\end{enumerate}

As a result, the solid scaffold interacts with the blood vessels through extravascular fluids. This is a reasonable assumption for the system considered and significantly simplifies the mathematical formulation.

Based on Hypothesis 3, we can introduce a state Equation for the vascular volume fraction (which will drive ischaemia and reperfusion): 
\begin{align}
	\varepsilon^b = \varepsilon^b_0 \cdot \left( 1 - \frac{2}{\pi}\arctan \left(\frac{p^s-\pb}{K}\right)\right)
\label{eq:twocomp:porobstatelaw}
\end{align}
\noindent where $\varepsilon^b_0$ stands for the initial blood volume fraction and $K$ is the compressibility of the vessels. Increasing the compressibility value of the vessel requires a larger pressure gap to reduce the porosity value. The arc-tangent formulation allows us to prevent negative values. 

To account for the non-newtonian behaviour of the blood, the following constitutive Equation based on the porosity evolution for the vascular permeability is proposed to introduce non-linearity:
\begin{align}
	k^b = k^b_0\,\left(\frac{\varepsilon^b}{\varepsilon^b_0}\right)^{\alpha_b},~\alpha_b\geq 2
\label{eq:twocomp:vascpermea}
\end{align}
\noindent where $k^b_0$ and $\varepsilon^b_0$, respectively, denote the initial vascular permeability and the initial vascular porosity. The vascular permeability can further be anisotropic and a ratio has been introduced such that the permeability matrix reads:
\begin{align}
\mathbf{K^b_0} = \begin{bmatrix}
\frac{k^b_0}{ratio_x} & 0 & 0\\
0 & \frac{k^b_0}{ratio_y}  & 0 \\
0 & 0 & \frac{k^b_0}{ratio_z} 
\end{bmatrix}
\label{eq:permeabilitymatrix}
\end{align}

The state law for the saturation-capillary pressure relationship (according to the wetting / non-wetting description) is defined by Equations \ref{eq:capillary_saturation_cell}.
\begin{align}
S^{c}=1 - \left[ \frac{2}{\pi} \arctan\left(\frac{\plc}{a} \right)\right]  \label{eq:capillary_saturation_cell}
\end{align}

\noindent where $a$ is a constant parameter depending on the microstructure of the connective tissue. The initial saturation in cells is obtained by appropriately selecting of the initial value of $\plc$.

The total Cauchy stress tensor can be defined as:
\begin{align}
	\mathbf{t}^{\text{tot}} &= \sum_{\alpha=s,c,l,b}\varepsilon^\alpha \mathbf{t}^\alpha = \underbrace{\varepsilon^s\mathbf{\tau}^s}_{\mathbf{t}^{\text{eff}}(\us)}-\varepsilon^s p^s\mathbf{1} - \varepsilon (S^c \pc +S^l \pl) \mathbf{1} - \varepsilon^b \pb \mathbf{1}\\
 & = \underbrace{\varepsilon^s\mathbf{\tau}^s}_{\mathbf{t}^{\text{eff}}(\us)}-\varepsilon^s p^s\mathbf{1} - \varepsilon (\pl - S^c \plc) \mathbf{1} - \varepsilon^b \pb \mathbf{1} \label{eq:twocomp:tot_stress_tensor}
\end{align}

\noindent where $\varepsilon^s\mathbf{\tau}^s$ is the effective stress tensor, $p^s$ is the empirically defined solid pressure Equation \ref{eq:total_fluid_pressure}.
\begin{align}
	p^s=S^c p^c + S^{l} p^{l}\label{eq:total_fluid_pressure0}\\
  \underset{(\ref{eq:capillarypressure})+(\ref{eq:saturation_constrain})}{\implies} p^s = \pl - S^c \plc \label{eq:total_fluid_pressure}
\end{align}

As a result, introducing Equations \ref{eq:2comp:volfrac} and \ref{eq:total_fluid_pressure} in Equation \ref{eq:twocomp:tot_stress_tensor}:
\begin{align}
	\mathbf{t}^{\text{tot}} = \mathbf{t}^\text{eff} - \underbrace{(1-\varepsilon^b)}_{=\varepsilon^s+\varepsilon} \, (\pl - S^c \plc) \mathbf{1} - \varepsilon^b \pb\mathbf{1}\label{eq:twocomp:momentum2}
\end{align}

In the experiment conducted in this study, the solid scaffold remains in the small strain regime. For the sake of simplicity of the analysis, a linear elastic behaviour is therefore considered such that:
\begin{align}
    \mathbf{t}^\text{eff} = 2 \mu \left(\frac{1}{2}(\mathbf{\nabla u}+\mathbf{\nabla u}^T)\right) + \lambda \mathrm{tr}   \left(\frac{1}{2}(\mathbf{\nabla u}+\mathbf{\nabla u}^T)\right) \mathbf{1}
\end{align}
\noindent where $\lambda,\,\mu$ are the Lamé coefficients depending on the Young modulus and the Poisson ratio.

\subsubsection{Derivation of the final formulation of the system of partial differential Equations}

This section outlines the systematic process of integrating the governing and constitutive Equations to derive the final system of Equations for solution.

Starting with the solid phase mass Equation, we have:
\begin{align}
\frac{\mathrm{D}^s}{\mathrm{D}t}(\rho^s \varepsilon^s) + \rho^s \varepsilon^s \nabla \cdot\vs = 0 \label{eq:masssolid1}\\
	\underset{(\ref{eq:2comp:volfrac}) + (\ref{eq:masssolid1})*\frac{1}{\rho^s}}{\implies} 
-\frac{\mathrm{D}^s}{\mathrm{D}t}(\varepsilon+\varepsilon^{b}) + (1-\varepsilon-\varepsilon^{b}) \nabla \cdot\vs = 0
\label{eq:twocomp:solidmass0}\\
\implies \frac{\mathrm{D}^s}{\mathrm{D}t}(\varepsilon) = -\frac{\mathrm{D}^s}{\mathrm{D}t}(\varepsilon^{b}) + (1-\varepsilon-\varepsilon^{b}) \nabla \cdot\vs
\label{eq:twocomp:solidmass1}
\end{align}

This Equation is used to update the extravascular porosity, which is an internal variable of the problem. A first order approximation in time allows to assess the porosity at the new time-step such that (with $\ddroit \us(t) =(\us(t)-\us(t-dt))$):
 \begin{equation}
 \varepsilon(t) = \frac{\varepsilon(t - \ddroit t)+\varepsilon^b(t - \ddroit t)-\varepsilon^b(t) (1+\nabla \cdot \ddroit \us(t))+ \nabla \cdot \ddroit \us(t) }{1+\nabla \cdot \ddroit \us(t)} 
 \end{equation}



Introducing Equation \ref{eq:momentum_fluid} into Equation \ref{eq:massfluid10}, we can eliminate the fluid velocity and introduce the fluid pressures. As a result, the blood mass conservation gives:
\begin{align} 
	\frac{\mathrm{D}^s}{\mathrm{D}t}( \varepsilon^b) - \nabla \cdot\left(\frac{k^{b}}{\mu^{b}}\mathbf{\nabla}\pb\right)+ \varepsilon^b \nabla \cdot \vs = 0 \label{eq:massblood1}
\end{align}

According to the state law defined defined for the vascular porosity Equation \ref{eq:twocomp:porobstatelaw}, its derivative can also be expressed as:
\begin{align}
	\frac{\mathrm{D}^s\varepsilon^{b}}{\mathrm{D}t}=\frac{\mathrm{D}^s\varepsilon^{b}}{\mathrm{D}p^s}\frac{\mathrm{D}^s p^s}{\mathrm{D}t} + \frac{\mathrm{D}^s\varepsilon^{b}}{\mathrm{D}\pb}\frac{\mathrm{D}^s\pb}{\mathrm{D}t}
\label{eq:twocomp:derivativeporobstatelaw}
\end{align}

Furthermore, using Equations \ref{eq:total_fluid_pressure}, we get:
\begin{align}
\frac{\mathrm{D}^s p^s}{\mathrm{D}t} &= \frac{\mathrm{D}^s \pl}{\mathrm{D}t}-\frac{\mathrm{D}^s S^c \plc}{\mathrm{D}t}
\label{eq:twocomp:derivsolidpress0}\\
& =  \frac{\mathrm{D}^s \pl}{\mathrm{D}t}-S^c \frac{\mathrm{D}^s \plc}{\mathrm{D}t}- \plc \frac{\mathrm{D}^s S^c}{\mathrm{D}\plc}\frac{\mathrm{D}^s \plc}{\mathrm{D}t}   \label{eq:twocomp:derivsolidpress1}\\
& =  \frac{\mathrm{D}^s \pl}{\mathrm{D}t} - \underbrace{\left( S^c + \plc\frac{\mathrm{D}^s S^c}{\mathrm{D}\plc}\right)}_{=C_{state}} \frac{\mathrm{D}^s \plc}{\mathrm{D}t} \label{eq:twocomp:derivsolidpress3} 
\end{align}
\noindent where $C_{state}$ is a function defined for ease of reading depending on $\frac{\mathrm{D}^s S^{c}}{\mathrm{D}\plc} = -\frac{2}{a \pi} \frac{1}{1+\left( \frac{\plc}{a} \right)^2}$.

Furthermore, we can compute $C_{e,p}=\frac{\mathrm{D}^s\varepsilon^{b}}{\mathrm{D}p^s}$ such that:
\begin{align}
    C_{e,p}=\frac{\mathrm{D}^s\varepsilon^{b}}{\mathrm{D}p^s}= -\frac{2 \varepsilon^b_0}{\pi K}\frac{1}{1+\left(\frac{\pl - S^c \plc -\pb}{K} \right)^2}
    \label{eq:partialderivepsbps}
\end{align}

Immediately, we also have:
\begin{align}
    \frac{\mathrm{D}^s\varepsilon^{b}}{\mathrm{D}\pb}= -C_{e,p}
    \label{eq:partialderivepsbpb}
\end{align}

Finally, the derivative of the state function reads:
\begin{align}
	\frac{\mathrm{D}^s\varepsilon^{b}}{\mathrm{D}t}=C_{e,p} \left(  \frac{\mathrm{D}^s \pl}{\mathrm{D}t} - C_{state}  \frac{\mathrm{D}^s \plc}{\mathrm{D}t}- \frac{\mathrm{D}^s\pb}{\mathrm{D}t} \right)
\label{eq:twocomp:dporo2}
\end{align}

Introducing Equation \ref{eq:twocomp:dporo2}, Equation \ref{eq:massblood1} becomes: 
\begin{align}
\begin{split}
    	C_{e,p} \left(  \frac{\mathrm{D}^s \pl}{\mathrm{D}t} - C_{state}  \frac{\mathrm{D}^s \plc}{\mathrm{D}t}- \frac{\mathrm{D}^s\pb}{\mathrm{D}t} \right)- \nabla \cdot\left(\frac{k^{b}}{\mu^{b}}\mathbf{\nabla}\pb\right)+ \varepsilon^b \nabla \cdot \vs = 0
\end{split}
    \label{eq:massblood2}
\end{align}

Performing similar operations for the cells' mass conservation Equation, with $k^c$ and $\mu^c$ respectively denoting the cell phase permeability and viscosity and introducing the function $C_{m,c} = \varepsilon \frac{\mathrm{D}^s S^{c}}{\mathrm{D}\plc}$:
\begin{align}
    &	\frac{\mathrm{D}^s}{\mathrm{D}t}(S^{c}\varepsilon) - \nabla \cdot\left(\frac{k^{c}}{\mu^{c}}\mathbf{\nabla}\pc\right)+S^{c}\varepsilon \nabla \cdot \vs  = 0
\label{eq:twocomp:cellmass0}\\
\implies & \left(S^{c}\frac{\mathrm{D}^s}{\mathrm{D}t}(\varepsilon)+ \varepsilon\frac{\mathrm{D}^s}{\mathrm{D}t}(S^{c})\right) - \nabla \cdot\left(\frac{k^{c}}{\mu^{c}}\mathbf{\nabla}(\pl-\plc)\right)+S^{c}\varepsilon \nabla \cdot \vs = 0 \label{eq:twocomp:cellmass1}\\
\begin{split}
\underset{(\ref{eq:twocomp:solidmass1})}{\implies} \left(S^{c}\left[ -\frac{\mathrm{D}^s}{\mathrm{D}t}(\varepsilon^{b}) + (1-\varepsilon-\varepsilon^{b}) \nabla \cdot\vs \right]+ \varepsilon\frac{\mathrm{D}^s}{\mathrm{D}t}(S^{c})\right) \\ - \nabla \cdot\left(\frac{k^{c}}{\mu^{c}}\mathbf{\nabla}\pl\right)+ \nabla \cdot\left(\frac{k^{c}}{\mu^{c}}\mathbf{\nabla}\plc\right)+S^{c}\varepsilon \nabla \cdot \vs = 0 \label{eq:twocomp:cellmass2}
\end{split}\\
\begin{split}
\implies
 -S^{c} \frac{\mathrm{D}^s \varepsilon^{b}}{\mathrm{D}t} + C_{m,c} \frac{\mathrm{D}^s \plc}{\mathrm{D}t} - \nabla \cdot\left(\frac{k^{c}}{\mu^{c}}\mathbf{\nabla}\pl\right)+ \nabla \cdot\left(\frac{k^{c}}{\mu^{c}}\mathbf{\nabla}\plc\right)\\+S^{c}(1-\varepsilon^{b}) \nabla \cdot \vs =0 \label{eq:twocomp:cellmass4}
\end{split}
\end{align}

Introducing Equation \ref{eq:massblood1} in Equation \ref{eq:twocomp:cellmass4}, we get:
\begin{align}
    \begin{split}
       & S^c \left( - \nabla \cdot\left(\frac{k^{b}}{\mu^{b}}\mathbf{\nabla}\pb\right)+ \varepsilon^b \nabla \cdot \vs \right)
       + C_{m,c} \frac{\mathrm{D}^s \plc}{\mathrm{D}t}\\
       &- \nabla \cdot\left(\frac{k^{c}}{\mu^{c}}\mathbf{\nabla}\pl\right)+ \nabla \cdot\left(\frac{k^{c}}{\mu^{c}}\mathbf{\nabla}\plc\right)+S^{c}(1-\varepsilon^{b}) \nabla \cdot \vs =0 \label{eq:twocomp:cellmass4b}
    \end{split}
\end{align}

Finally the mass conservation of the cell phase reads:
\begin{align}
    \begin{split} 
       C_{m,c} \frac{\mathrm{D}^s \plc}{\mathrm{D}t} - S^c \nabla \cdot\left(\frac{k^{b}}{\mu^{b}}\mathbf{\nabla}\pb\right) - \nabla \cdot\left(\frac{k^{c}}{\mu^{c}}\mathbf{\nabla}\pl\right)\\
       + \nabla \cdot\left(\frac{k^{c}}{\mu^{c}}\mathbf{\nabla}\plc\right)
       +S^{c}\nabla \cdot \vs =0 \label{eq:twocomp:cellmass4c}
    \end{split}
\end{align}

Finally, the fluid phase in the interstitium mass conservation becomes, with $k^l$ and $ \mu^l$ respectively denoting the interstitial fluid phase permeability and viscosity:
\begin{align}
	& \frac{\mathrm{D}^s}{\mathrm{D}t}(S^{l}\varepsilon) - \nabla \cdot\left(\frac{k^{l}}{\mu^{l}}\mathbf{\nabla}\pl\right)+S^{l}\varepsilon \nabla \cdot \vs = 0
\label{eq:twocomp:IFmass0}\\
\begin{split}
 \underset{(\ref{eq:capillarypressure})+(\ref{eq:saturation_constrain})}{\implies} \frac{\mathrm{D}^s}{\mathrm{D}t}(\varepsilon)-\frac{\mathrm{D}^s}{\mathrm{D}t}(S^{c}\varepsilon) - \nabla \cdot\left(\frac{k^{l}}{\mu^{l}}\mathbf{\nabla}\pl\right)+(1-S^{c})\varepsilon \nabla \cdot \vs = 0
\end{split}
\label{eq:twocomp:IFmass0b}\\
\begin{split}
\underset{(\ref{eq:twocomp:solidmass1}) + (\ref{eq:massblood1}) + (\ref{eq:twocomp:cellmass0})}{\implies} - \nabla \cdot\left(\frac{k^{b}}{\mu^{b}}\mathbf{\nabla}\pb\right) + (1-\varepsilon) \nabla \cdot\vs 
- \nabla \cdot\left(\frac{k^{c}}{\mu^{c}}\mathbf{\nabla}{\underbrace{(\pl-\plc)}_{=\pc}}\right) \\ 
+S^{c}\varepsilon \nabla \cdot \vs  - \nabla \cdot\left(\frac{k^{l}}{\mu^{l}}\mathbf{\nabla}\pl\right)
+(1-S^{c})\varepsilon \nabla \cdot \vs = 0
\end{split}
\label{eq:twocomp:IFmass1}
\end{align}

Finally, we obtain for the mass conservation of the IF phase:
\begin{align}
    \begin{split}
- \nabla \cdot\left(\frac{k^{b}}{\mu^{b}}\mathbf{\nabla}\pb\right) - \nabla \cdot\left(\left[\frac{k^{c}}{\mu^{c}} + \frac{k^{l}}{\mu^{l}} \right]\mathbf{\nabla}\pl\right) + \nabla \cdot\left(\frac{k^{c}}{\mu^{c}}\mathbf{\nabla}\plc\right)
+ \nabla \cdot \vs = 0 \label{eq:twocomp:IFmass2}
\end{split}
\end{align}


\textbf{Remark:} The extension of this model to include biological exchanges is straightforward. An example of mathematical developments for oxygen biochemistry is provided \ref{appendix:02}.


\subsubsection{Weak form of the final system of partial differential Equations}
Mathematical developments have been introduced in the FEniCSx v0.9.0 (\citet{barratta,FEniCS}) environment to numerically solve a consolidation test in a 1D column. The finite element framework consists of solving a variational form defined as follows.
Consider ($q^{c}$,$q^{l}$, $q^b$,$\mathbf{w}$) the test functions defined in the mixed space $L_0^2(\Omega)\times L_0^2(\Omega)\times[H^1(\Omega)]^3$. The solutions of the problem are the capillary pressure, the cell pressure, the blood pressure, and the displacement of the solid: ($\pl,\,\plc,\,\pb,\,\us$).

Using Equations \ref{eq:twocomp:momentum}, \ref{eq:massblood2}, \ref{eq:twocomp:cellmass4c}, \ref{eq:twocomp:IFmass2}, the problem to be solved reads:
\begin{align}
\us&=\mathbf{u}_{\text{imposed}}\,\text{on}\, \partial\Omega_u \label{eq:twocomp:dirichlet}\\
{\color{black} p^\alpha} &= p_{\text{imposed}}\,\text{on}\, \partial\Omega_p^\alpha \label{eq:twocomp:bc_pressure}
\end{align}
\begin{align}
\begin{split}
\int_\Omega C_{m,c} \frac{\mathrm{D}^s \plc}{\mathrm{D}t} q^c \ddroit \Omega 
+ \int_\Omega  S^{c} \frac{k^{b}}{\mu^{b}} \nabla \pb \nabla q^c \ddroit \Omega 
 + \int_\Omega \frac{k^{c}}{\mu^{c}}\mathbf{\nabla}\pl \nabla q^c \ddroit \Omega \\
 - \int_\Omega \frac{k^{c}}{\mu^{c}}\mathbf{\nabla}\plc \nabla q^c \ddroit \Omega 
+ \int_\Omega S^c \nabla \cdot \left(\frac{\mathrm{D}^s \us}{\mathrm{D}t} \right) q^c \ddroit \Omega = 0,\,\forall~q^c\in {L}_0^2(\Omega)
\end{split}  \label{eq:twocomp:variationalmasscell}
\end{align}
\begin{align}
\begin{split}
\int_\Omega \frac{k^{b}}{\mu^{b}} \nabla \pb \nabla q^l \ddroit \Omega 
+  \int_\Omega \left[\frac{k^{c}}{\mu^{c}} + \frac{k^{l}}{\mu^{l}} \right] \nabla \pl \nabla q^l \ddroit \Omega \\
- \int_\Omega \frac{k^{c}}{\mu^{c}} \nabla \plc \nabla q^l \ddroit \Omega 
 + \int_\Omega \nabla \cdot \left(\frac{\mathrm{D}^s \us}{\mathrm{D}t} \right) q^l \ddroit \Omega = 0,\,\forall~q^l\in {L}_0^2(\Omega)
\end{split}  \label{eq:twocomp:variationalmassIF}
\end{align}
\begin{align}
\begin{split}
\int_\Omega C_{e,p} \frac{\mathrm{D}^s \pl}{\mathrm{D}t} q^b \ddroit \Omega 
- \int_\Omega C_{e,p}C_{state} \frac{\mathrm{D}^s \plc}{\mathrm{D}t} q^b \ddroit \Omega 
- \int_\Omega C_{e,p} \frac{\mathrm{D}^s \pb}{\mathrm{D}t} q^b \ddroit \Omega \\
+ \int_\Omega \frac{k^{b}}{\mu^{b}}\mathbf{\nabla}\pb \nabla q^b \ddroit \Omega 
+ \int_\Omega \varepsilon^b \nabla \cdot \left(\frac{\mathrm{D}^s \us}{\mathrm{D}t} \right) q^b \ddroit \Omega = 0,\,\forall~q^b\in {L}_0^2(\Omega)
\end{split}  \label{eq:twocomp:variationalmassblood}
\end{align}
\begin{align}
\begin{split}
\int_\Omega \mathbf{t}^{\text{eff}}(\us) : \nabla \mathbf{w} \ddroit \Omega 
- \int_\Omega \left(1-\varepsilon^b\right) \left(\pl - S^c \plc \right) \nabla \cdot \mathbf{w} \ddroit \Omega \\
- \int_\Omega \varepsilon^b \pb \nabla \cdot \mathbf{w} \ddroit \Omega 
- \int_{\Gamma_s} \mathbf{t}^{\text{imposed}} \cdot \mathbf{w} \ddroit \Gamma_s =0, \forall~\mathbf{w}\in[H^1(\Omega)]^3 \label{eq:twocomp:variationalmomentum}
\end{split}
\end{align}


\subsection{Evaluation of the model and application to the LDF experiment}

\subsubsection{1D column}

A one-dimensional consolidation in a column is computed in a manner similar to the approach presented by \citet{Lavigne2023}, which examined a 1D consolidation test for a two-compartment model where the extra-vascular porosity is occupied by a single fluid. The present example aims to show the 2-compartment behaviour, including biphasic interstitium.

\begin{figure}[ht!]
    \centering
    \includegraphics[width=\linewidth]{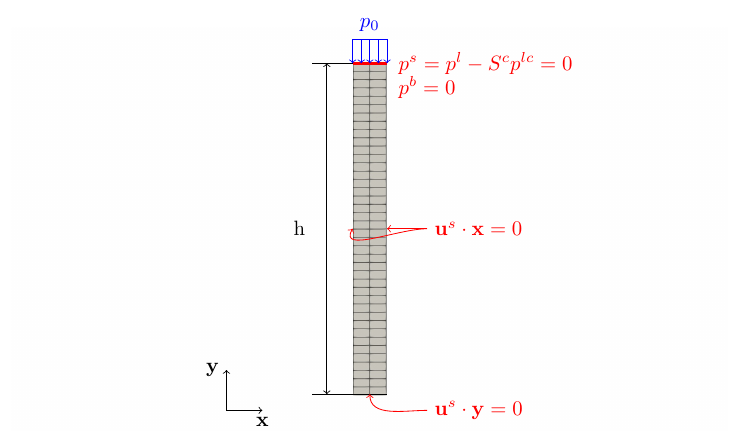}
    \caption{Load (blue), Boundary conditions (red) and mesh (gray) of the uni-axial confined compression of a porous 2D column of height h. For the mono-phasic interstitium model, $p^s=p^l=0$ Pa on the boundary (\citet{Lavigne2023}). A null pressure allows the leakage.}
    \label{fig:1D:column}
\end{figure}

The column was $100~\si{\micro\meter}$ in height and $20~\si{\micro\meter}$ in width.
The boundary conditions corresponding to a consolidation test are provided in Figure \ref{fig:1D:column}.

The tissue is supposed to have an extra-vascular porosity of $60~\%$, $50~\%$ interstitial fluid and $10~\%$ non-structural cells, and a vascular porosity of $4~\%$. The material parameters are provided in Table \ref{tab:column1D}. To appreciate the influence of the cell phase, a tissue of mono-phasic extravascular porosity of $50~\%$ of interstitial fluid (the remaining $10~\%$ cells being transferred to the solid phase) and a vascular porosity of $4~\%$ were also considered and compared. Given the significant difference in viscosities between the interstitial fluid and the cells, the objective is to analyse how the movement of the cells influences the characteristic times of the observed mechanical response.

An initial time of 10~s is initially set for the system to stabilise. In fact, a capillary pressure $p^{lc}_0$ is initially set to obtain the desired saturation in cell $S^c_0$ that provides the expected porosity $10~\%$. The initial pressure of the interstitial fluid $p^l_0$ is then calculated to ensure an initial state where $p^s_0=p^l_0-S^c_0p^{lc}_0=0$ Pa. For the mono-phasic interstitium model, this step is not required, as we directly consider $p^s_0=p^l_0=0$ Pa on the boundary.
\begin{align}
    S^c_0=1-\frac{\varepsilon^l}{\varepsilon}\approx0.17\\
    p^{lc}_0 = a \cdot \tan\left(\frac{\pi}{2} \cdot (1 - S^c_0)\right)\approx2240~\si{\pascal}\\
    p^l_0=S^c_0\cdot p^{lc}_0\approx 380~\si{\pascal}
\end{align}

Following initialisation, a 5-second ramp is applied at the top of the sample to reach a $p_0=200~\si{\pascal}$ load. The load is then sustained for 120 seconds.

\begin{table}[ht!]
\centering
\resizebox{\textwidth}{!}{%
\begin{tabular}{lllll}
\hline
Parameter & Symbol & Value & Value (\cite{Lavigne2023}) & Unit \\ \hline
Young modulus &  E      &   $5000$ &   $5000$  &  \si{\pascal}    \\
Poisson ratio  &  $\nu$      &  $0.2$ &  $0.2$   &   -   \\
IF viscosity &   $\mu^l$     &      $1$  &      $1$  &   \si{\pascal\second}   \\
IF intrinsic permeability &   $k^{l}$     &      $1.\times 10^{-14}$  &      $1.\times 10^{-14}$  &   \si{\square\meter}   \\
Cell viscosity &   $\mu^c$     &      $1,\,5,\,20$ &      - &   \si{\pascal\second}   \\
Cell intrinsic permeability &   $k^{c}$     &      $1.\times 10^{-14}$ &      - &   \si{\square\meter}   \\
Initial IF porosity  &   $\varepsilon^l$     &   0.5  &   0.5    &   -   \\
Initial Cell porosity  &   $\varepsilon^c$     &   0.1    &   -  &   -   \\
Coefficient pressure-saturation  &   $a$     &  600 & -   &   \si{\pascal}   \\
Vessel Bulk modulus  &   $K $     &   $1\times 10^{3}$   &   $1\times 10^{3}$  & \si{\pascal}  \\
Initial vessel Intrinsic permeability &   $k^\varepsilon_b$     &      $4\times 10^{-12}$ &      $4\times 10^{-12}$ &   \si{\square\meter}   \\
Blood viscosity &   $\mu^b$     &      $4.0\times 10^{-3}$ &      $4.0\times 10^{-3}$  &   \si{\pascal\second}   \\
Initial vascular porosity  &    $\varepsilon^b_0$    &  $4\%$  &  $4\%$   & -  \\ \hline
\end{tabular}%
}
\caption{Mechanical parameters for the bi-compartment model including a biphasic interstitium.}
\label{tab:column1D}
\end{table}

\newpage

\subsubsection{LDF Experiment: simplified interstitium}

Supported by the results of the 1D-Column consolidation test and the characteristic times of the experiment (about one minute), the cells are expected to stay close to their initial location because they can be assimilated to highly viscous fluids. As a result, the application to the LDF experiment considers a simplified case with a mono-phasic interstitium to assess the feasibility of using the 2-compartment model to include the haemodynamic response. 
The model therefore simplifies and $S^c=0$, $\varepsilon = \varepsilon^l$ and capillary pressure is no longer considered ($p^s=p^l$). Hence, the conservation Equations become:
$\bullet$ Solid phase:
\begin{align}
\frac{\mathrm{D}^s}{\mathrm{D}t}(\varepsilon) = -\frac{\mathrm{D}^s}{\mathrm{D}t}(\varepsilon^{b}) + (1-\varepsilon-\varepsilon^{b}) \nabla \cdot\vs
\label{eq:mono:twocomp:solidmass1}
\end{align}
$\bullet$ Vascular phase:
\begin{align}
\begin{split}
    	\tilde{C}_{e,p} \left(  \frac{\mathrm{D}^s \pl}{\mathrm{D}t}- \frac{\mathrm{D}^s\pb}{\mathrm{D}t} \right) - \nabla \cdot\left(\frac{k^{b}}{\mu^{b}}\mathbf{\nabla}\pb\right)+ \varepsilon^b \nabla \cdot \vs = 0
\end{split}
    \label{eq:mono:massblood2}
\end{align}
where $\frac{\mathrm{D}^s\varepsilon^{b}}{\mathrm{D}t}= \tilde{C}_{e,p} \left(  \frac{\mathrm{D}^s \pl}{\mathrm{D}t}- \frac{\mathrm{D}^s\pb}{\mathrm{D}t} \right)$ and $\tilde{C}_{e,p}=\frac{\mathrm{D}^s\varepsilon^{b}}{\mathrm{D}p^s}= -\frac{2 \varepsilon^b_0}{\pi K}\frac{1}{1+\left(\frac{\pl -\pb}{K} \right)^2}$.

$\bullet$ IF phase:
\begin{align}
    \begin{split}
- \nabla \cdot\left(\frac{k^{b}}{\mu^{b}}\mathbf{\nabla}\pb\right) - \nabla \cdot\left(\frac{k^{l}}{\mu^{l}}\mathbf{\nabla}\pl\right)
+ \nabla \cdot \vs = 0 \label{eq:mono:twocomp:IFmass2}
\end{split}
\end{align}
The corresponding variational forms are provided in \ref{appendix:mono:var}.

As depicted in Figure \ref{fig:BCs}, a generic model consisting of a quarter section of the second phalanx of the middle finger is considered. The simplified geometry representing the soft tissue of the finger was created using GMSH, consisting of a quarter cylinder minus bone (\citet{Geuzaine2009}). Measurements derived from echography —validated against literature values from \citet{Kallepalli2022}— determined that the diameter of the simplified phalanx was $2.8~\si{\centi\meter}$ and the thickness of the skin $4~\si{\milli\meter}$. An external force was applied over a cylindrical surface region at the top of the phalanx, corresponding to the area contacted by the LDF probe, which has a radius of $2~\si{\milli\meter}$. To avoid boundary effects, the mesh was sufficiently extruded along the z-direction.

LDF is sensitive to positioning and external factors and has a limited penetration depth of $0.5-1~\si{\milli\meter}$ (\citet{Heurtier2013, FREDRIKSSON20094,Poffo2014}). The relative LDF value has been computed as the product between the blood cells concentration (represented by the vascular porosity) and the blood velocity in a domain represented by a cylinder of $1~\si{\milli\meter}$ depth below the load application surface:
\begin{align}
    LDF\,[\%]\, = \frac{\int_{\Omega_{LDF}}  \sqrt{\varepsilon^b (v^b-v^s)\cdot \varepsilon^b (v^b-v^s)} d\Omega }{LDF_0}
\end{align}
\noindent where $LDF_0$ stands for the established value of the LDF, also known as basal skin blood flow.

\begin{figure}[ht!]
    \centering
    \includegraphics[width=\linewidth]{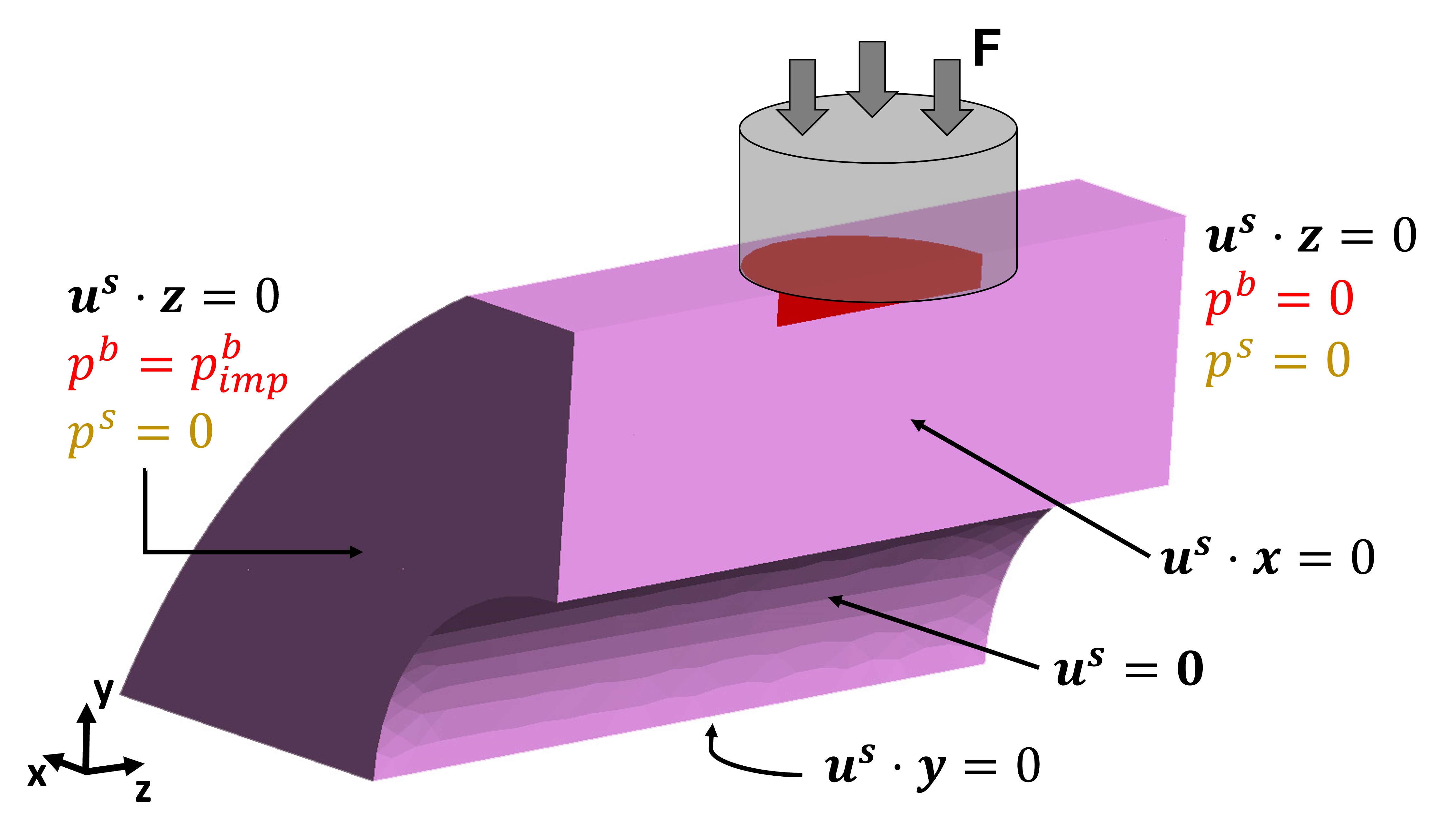}
    \caption{A quarter of a phalanx is modelled. Displacements are blocked on the symmetry planes and no flux are allowed. A blood pressure gradient is introduced on the z-axis to create the basal skin blood flow. The interstitial fluid is allowed to get in/out the sample by imposing $p^s=0$ Pa. The LDF signal is computed in the red domain directly below the load application surface.}
    \label{fig:BCs}
\end{figure}

The hydrostatic pressure in the skin capillaries varies significantly between segments. The mean hydrostatic pressure ranges from $32-38~\si{\mmHg}$ in the arterial part, decreases to $17-20~\si{\mmHg}$ in the transitional part, and reduces to $12-18~\si{\mmHg}$ in the venous part. Given that the physiological pressure of the whole blood remains approximately constant at $25~\si{\mmHg}$, the effective filtration and re-absorption pressure gradients in the arterial and venous parts of the capillaries are approximately $7-13~\si{\mmHg}$ ($933-1733~\si{\pascal}$), respectively (\citet{Fedorovich2018}).
As a result, the average Capillary Blood flow Velocity (CBV) in the capillaries of human skin under thermoneutral conditions ranges from 400 to 900~$\si{\micro\meter\per\second}$, showing an oscillatory nature (\citet{Fedorovich2018, Rossi2006}). \citet{Fagrell1977-ws} reported a mean resting CBV of $0.7\pm0.3~\si{\milli\meter\per\second}$ at an average skin temperature of $30.4\pm2.3$°C. Another study with 64 healthy subjects found a mean resting CBV of $0.6\pm0.5~\si{\milli\meter\per\second}$ at a skin temperature of $30.9\pm3.2$°c. 

As a result, boundary conditions ensured that normal displacements were constrained along the symmetric sides, while the interstitial fluid was allowed to flow freely along the z-direction. A blood pressure of $4000~\si{\pascal}$ was applied to the left side to reach its maximum value in a span of five seconds, achieving a basal skin blood flow flow rate of approximately $400~\si{\micro\meter\per\second}$.
All pressures were initialised to zero and a five-second stabilisation period preceded the application of loading to allow the model to reach a homeostatic state.


\subsection{Mesh Convergence}
\label{sec:MeshConvergence}
Before performing the sensitivity analysis, a comprehensive mesh convergence analysis was performed to ensure numerical accuracy and stability. The same set of parameters was evaluated for four different mesh configurations: 25k elements, 44k elements with local refinement, 125k elements, and 250k elements. The solution obtained with the 250k element mesh was considered the reference mesh for comparison. To quantify the deviation of other meshes from this reference, the root mean square error (RMSE) was calculated using the formula:
\begin{equation}
    \mathrm{RMSE} = \frac{1}{N} \sqrt{\sum(LDF-LDF_{250k})^2}
\end{equation}
where $N$ represents the number of time points.

\subsection{Sensitivity Study}
\label{sec:sensitivity}

The two-compartment model that includes a mono-phasic interstitium is driven by nine material parameters. Considering the interaction between some of these parameters, the sensitivity analysis is reduced to five parameters provided in Table \ref{tab:sobol:init}.

For the solid phase, the mechanical behaviour for the elastic law considered is controlled by two parameters: the Young's modulus $E$ and the Poisson ratio $\nu$. These parameters influence the computation of the effective solid stress and are related through the Lamé coefficients. The literature still lacks a clear value for the Poisson ratio, with models assuming values ranging from 0.3 to 0.5 (\citet{Mukhina2020,Kalra2016,PaillerMattei2008,raveh2004elastic,Lakhani2021,Sanders1973}). Therefore, the authors aimed to remain consistent with the values commonly used in previous studies and fixed its value at 0.42, as proposed by \citet{deLucio2024}. 

Furthermore, the reported elastic modulus values exhibit a wide range (5~\si{\kilo\pascal} to 196~\si{\mega\pascal}, according to \citet{Payan2017}), influenced by factors such as ageing, loading conditions, model choice, and geometry considerations. \citet{Kalra2016} reported Young's modulus values ranging from 4 to 15~\si{\mega\pascal} under quasi-static tensile conditions and from 14 to 100~\si{\mega\pascal} for dynamic tensile loading. Indentation, torsion, and suction tests yielded Young's modulus values of 10~\si{\kilo\pascal} to 2.4~\si{\mega\pascal}, 20~\si{\kilo\pascal} to 1.12~\si{\mega\pascal}, and 25~\si{\kilo\pascal} to 260~\si{\kilo\pascal}, respectively. Hence, this study considered values ranging from 50~\si{\kilo\pascal} to 1~\si{\mega\pascal}.

The mono-phasic interstitium is described as a single fluid whose behaviour is governed by Darcy's law. The corresponding material parameters are the intrinsic permeability $k^l$ and the viscosity $\mu^l$. These two parameters describe the ability of the fluid to flow within the domain, and their ratio can be referred to as the hydraulic permeability $L^l = \frac{k^l}{\mu^l}$. The sensitivity analysis focusses on using the ratio of these quantities directly.
\citet{Wahlsten2019} studied animal skin (Murine) and \textit{ex vivo} (abdominal and breast region) and \textit{in vivo} (forearm) human skin samples under tensile conditions. They reported that the mechanics of porous media allowed the inclusion of volume loss as a result of the expulsion of interstitial fluid. Using a single layer model, they identified elastic tensile moduli of $89 \pm 27$~\si{\kilo\pascal} for human skin and $288 \pm 178$~\si{\kilo\pascal} for murine skin, considering an \textit{ex vivo} hydraulic conductivity of $5\times10^{-13}$~\si{\square\meter\per\pascal\per\second} and a reference solid volume fraction of 0.30 (\citet{Nakagawa2010}). 
\citet{Oftadeh2018} investigated the role of ECM viscosity and fluid in the mechanical response of the mouse dermis. The hydraulic conductivity of the skin was set to $(1.47 \pm 0.23) \times 10^{-13}$~\si{\square\meter\per\pascal\per\second}. \citet{Oftadeh2023}, based on poroelastic considerations (poro-viscoelastic finite element modelling identification), identified the hydraulic permeability of water for three layers of skin (stratum corneum, epidermis, dermis). Analysing 14 \textit{ex vivo} human skin samples tested with macroscopic and nano-indentation methodologies, a hydraulic permeability of $5{-}15\times10^{-14}$~\si{\square\meter\per\pascal\per\second} was found on the nanoscopic scale and $0.5{-}2\times10^{-14}$~\si{\square\meter\per\pascal\per\second} on the mesoscopic scale. \citet{Leng2021,Han2023} also considered hydraulic permeabilities in the range $[0.5 \times 10^{-14}, 9.8 \times 10^{-10}]$~\si{\square\meter\per\pascal\per\second}. Therefore, this study considered values ranging from $0.5 \times 10^{-14}$~\si{\square\meter\per\pascal\per\second} to $1 \times 10^{-13}$~\si{\square\meter\per\pascal\per\second}.

Similarly, for the vascular phase, the intrinsic permeability $k^b_0$ and the blood viscosity $\mu^b$ are not treated separately, but considered through the hydraulic permeability of the blood $L^b_0$. To ensure the same initial blood flow, the blood pressure on the sample side was modified accordingly. The blood permeability was assumed to be higher than the permeability of the interstitial fluid (\citet{deLucio2024}). In the study, values were introduced that ranged from $1 \times 10^{-10}$~\si{\square\meter\per\pascal\per\second} to $1 \times 10^{-8}$~\si{\square\meter\per\pascal\per\second} were introduced.

The blood volume fraction was fixed to maintain the same geometry throughout the study. Regarding the volume fraction of the skin's vascular system, \citet{Kelly1995} used morphometry (based on fluorescein angiographic and native capillaroscopic fields) to evaluate the volume fraction of blood in 25 subjects (12 young, mean age 23.2 years, and 13 elderly, mean age 74.9 years) on the forehead and forearm. For young subjects, the volume fractions were $0.0791 \pm 0.0067$ and $0.0481 \pm 0.0036$ for the dermal papillary loops, and $0.0571 \pm 0.0123$ and $0.0223 \pm 0.0043$ for the horizontal vessels. In older subjects, these values were lower for the capillary loops but higher for horizontal vessels, respectively: $0.0494 \pm 0.0037$ and $0.0266 \pm 0.0029$ for the dermal papillary loops, and $0.1121 \pm 0.0124$ and $0.1013 \pm 0.0069$ for horizontal vessels. As the cohort age is 25$\pm$1 year, the authors fixed the initial vascular porosity at 8\%.

The two remaining parameters used in this sensitivity analysis are the exponent $\alpha$ and the compressibility of the vessel $K$. The exponent in the law of evolution of the permeability and compressibility of blood vessels arbitrarily varied between 2 to 6 and 500~\si{\pascal} to 5000~\si{\pascal}, respectively.

\begin{table}[ht!]
\centering
\resizebox{\textwidth}{!}{%
\begin{tabular}{cccc}
\hline
Parameter                          & Initial                                                   & Minimal                                                   & Maximal                                                   \\ \hline
Young Modulus (E)                  & $2.00\times10^{5} \si{\pascal}$                           & $5.00\times10^{4} \si{\pascal}$                           & $1.00\times10^{6} \si{\pascal}$                           \\
Hydraulic permeability ISF ($L^l$) & $1\times10^{-14}\si{\square\meter\per\pascal\per\second}$ & $1\times10^{-15}\si{\square\meter\per\pascal\per\second}$ & $1\times10^{-13}\si{\square\meter\per\pascal\per\second}$ \\
Hydraulic permeability Blood ($L^b$) & $1\times10^{-9}\si{\square\meter\per\pascal\per\second}$  & $1\times10^{-10}\si{\square\meter\per\pascal\per\second}$ & $1\times10^{-8}\si{\square\meter\per\pascal\per\second}$  \\
$\alpha$ (Eq. \ref{eq:twocomp:vascpermea})     & $3$                                                       & $2$                                                       & $6$                                                       \\
Vessel Compressibility ($K$)         & $1000$                                                    & $500$                                                     & $5000$                                                    \\ \hline
\end{tabular}%
}
\caption{Material parameters for the sensitivity analysis: initial, minimal and maximal values.}
\label{tab:sobol:init}
\end{table}

We propose using variance-based sensitivity analysis through the Sobol index 
to evaluate the influence of model parameters on the predicted micro-circulatory response under compression. An initial simulation was conducted using the parameter values specified in Table \ref{tab:sobol:init}, which yielded the reference skin blood flow response $R_0$ generated by the model. Subsequently, each of the five parameters was individually perturbed by a random value $\delta_i$ within the random range $[-95\%, 900\%]$. This perturbation produced a modified response $R_{i, \delta_i\%}$ for the parameter $i^{\text{th}}$. The relative variations in the model's response were then computed as follows:
\begin{equation}
  \text{Var}_{i \delta_i\%}=\frac{1}{N}\Sigma_{i=1}^n\frac{R_{i \delta_i\%}}{R_0} 
\end{equation}
In order to quantify the impact of each parameter, the following linear model was fit:
\begin{equation}
  \text{Var}_i= 1 + \sum_i \theta_i x_i
\end{equation}
where $x_i$ is the $i^{th}$ parameter variation $\in [\text{-95};\text{900}]\%$ and $\theta_i$ the slope of the variation. 
Then, the 15 parameter couples were perturbed similarly by random values $\delta_i, \delta_j$. The variances $\text{Var}_{i,j,\delta_i, \delta_j\%, i>j}$ were interpolated by a second-order polynomial model. 
\begin{equation}
  \text{Var}_{i,j,\delta_i, \delta_j\%, i>j}= 1 + \sum_i \theta_i \alpha_i + \sum_{ij,i>j} \theta_{ij} \alpha_i \alpha_j
\end{equation}
The first-order analysis give the Sobol indices:
\begin{equation}
  \text{S}_i=\frac{\theta_i^2}{\sum_i \theta_i^2}
\end{equation}
The second-order analysis give the Sobol indices:
\begin{equation}
  \text{S}_i=\frac{\theta_i^2}{\sum_i \theta_i^2 + \sum_{ij,i>j} \theta_{ij}^2} \quad \text{and} \quad \text{S}_{ij}=\frac{\theta_{ij}^2}{\sum_i \theta_i^2 + \sum_{ij,i>j} \theta_{ij}^2}
\end{equation}

\section{Results}\label{res__} 

\subsection{Experimental Laser Doppler Flowmetry: indentation test}

\begin{figure}[ht!]
    \centering
    \includegraphics[width=0.48\linewidth]{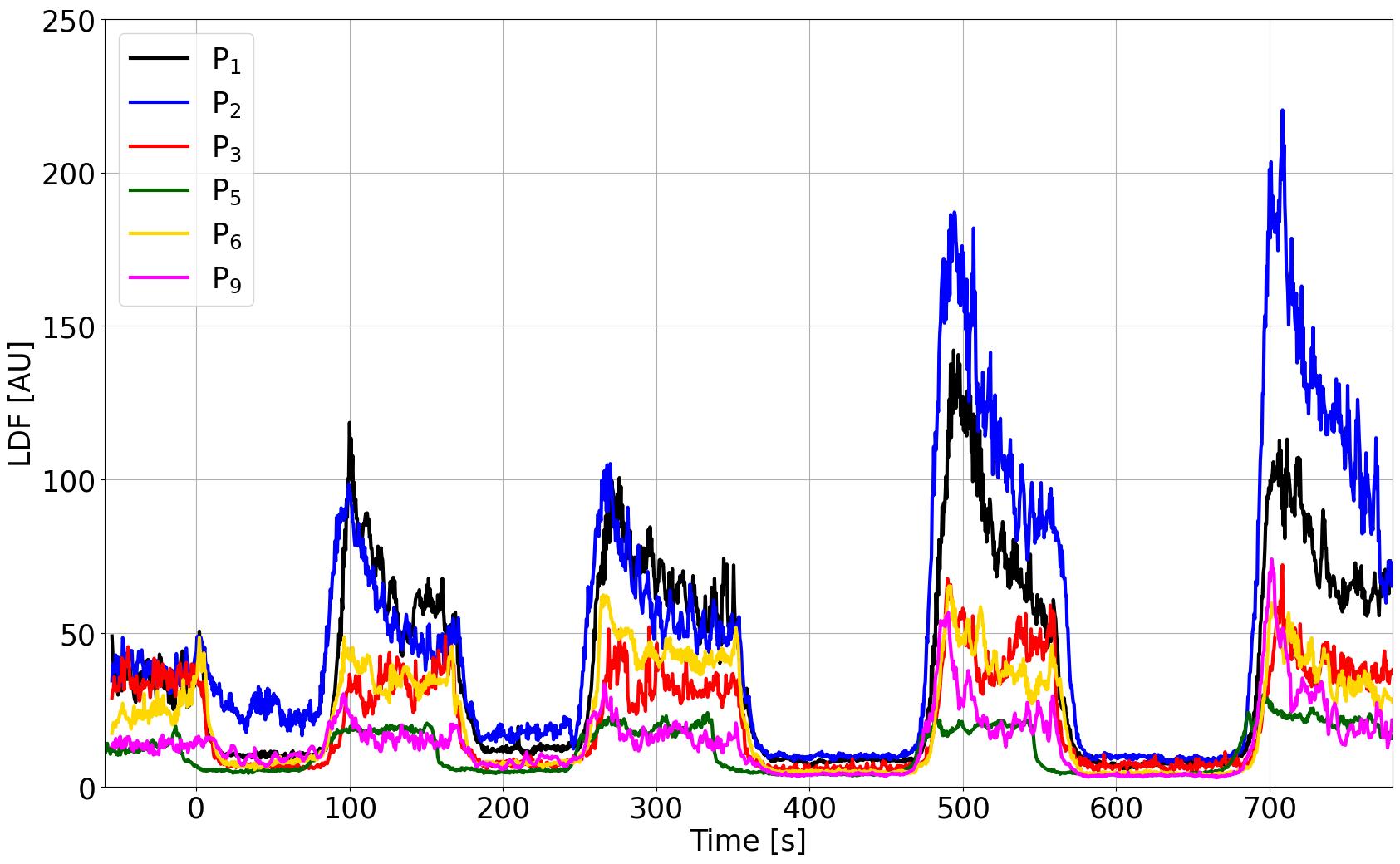}
    \includegraphics[width=0.48\textwidth]{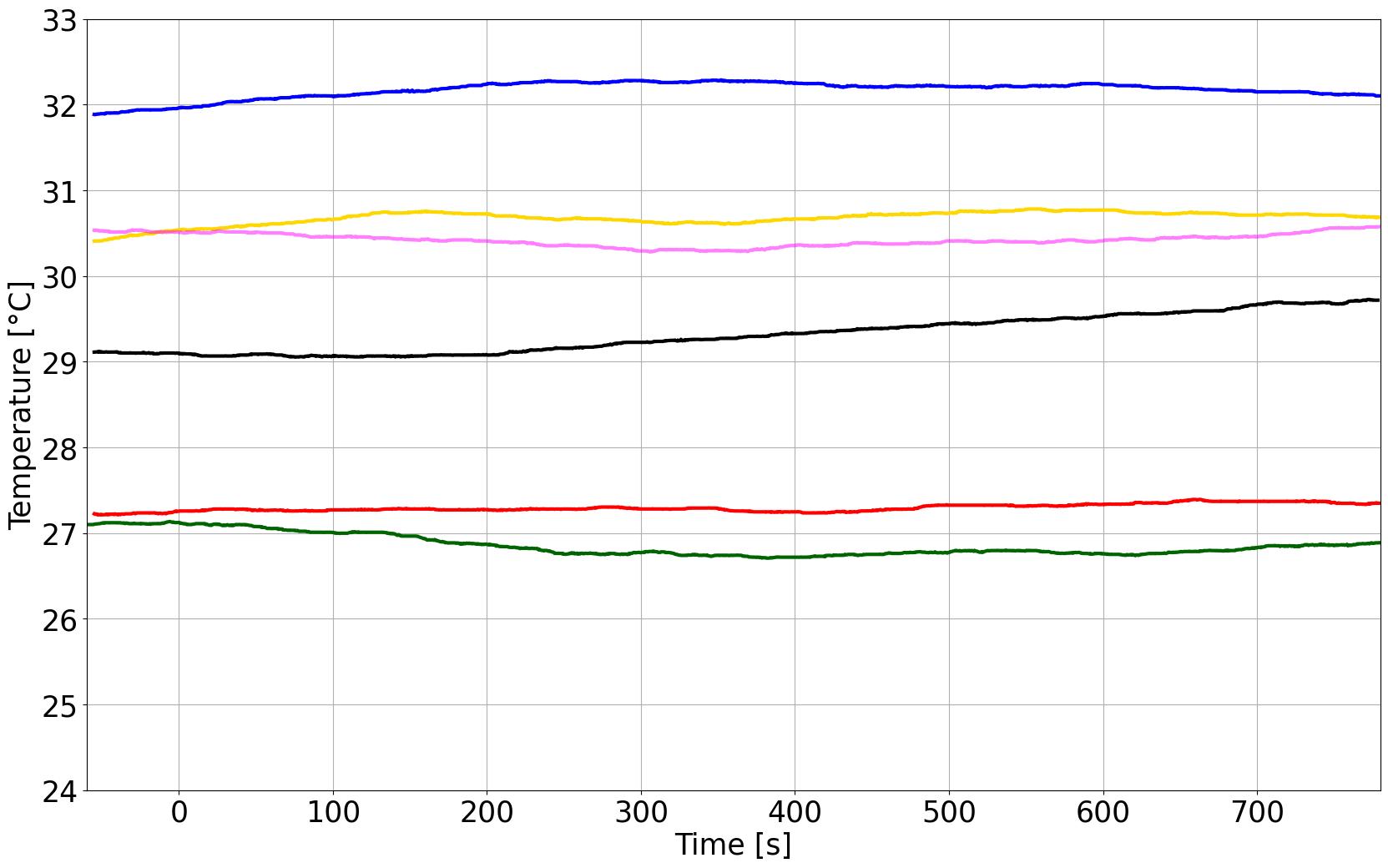}
    \includegraphics[width=0.48\linewidth]{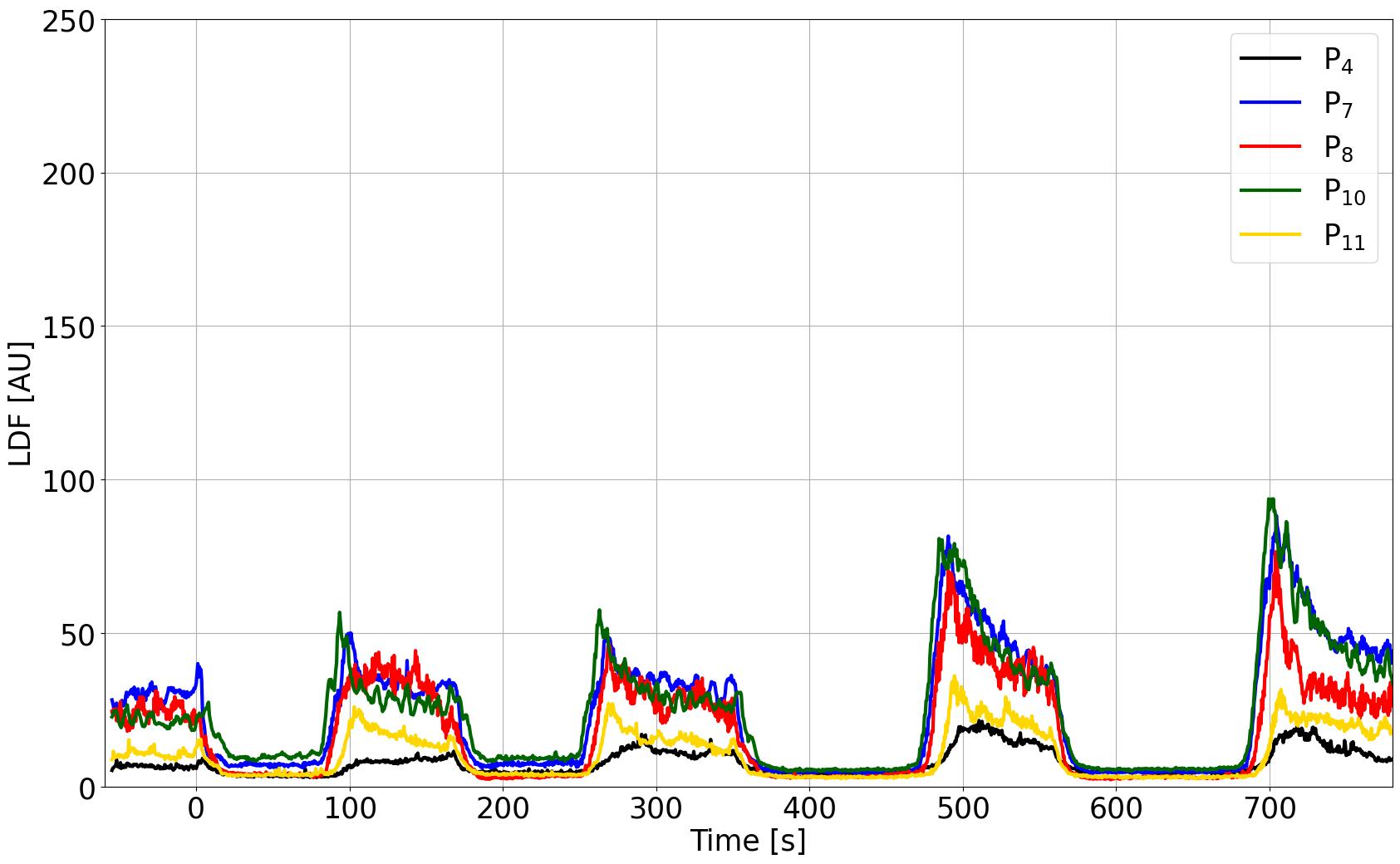}
    \includegraphics[width=0.48\textwidth]{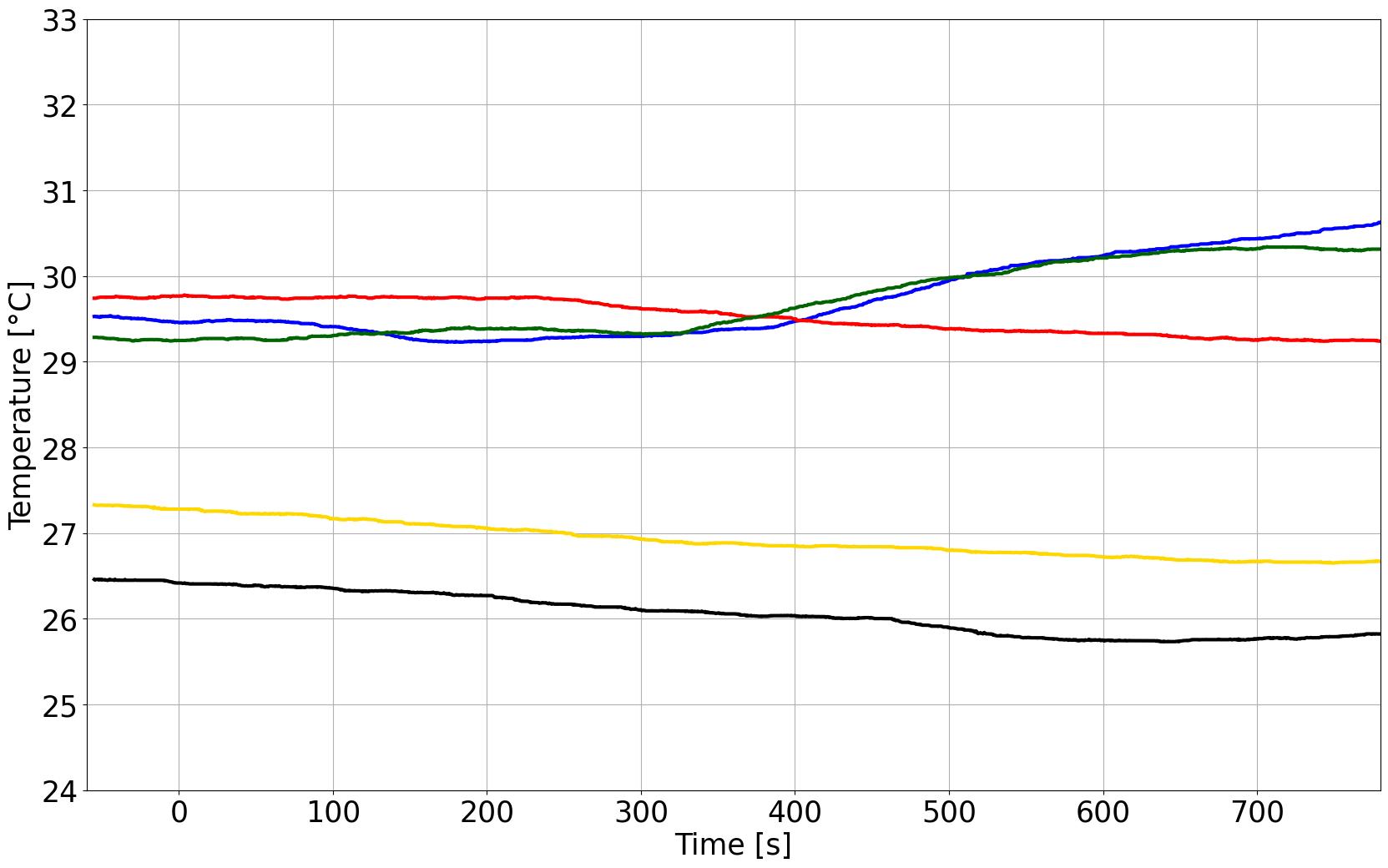}
    \caption{Median LDF signals (median$\pm$IQR are provided in \ref{appendix:LDF}) for each of the N=6 male (top) and N=5 female (bottom) patients (left). Dorsal hand temperature recorded during the experiment for each of the patient (right).}
    \label{fig:raw_LDF}
\end{figure}

Normality tests were conducted to assess the distribution of the data. As the variables were not normally distributed, nonparametric tests were used. A Wilcoxon rank-sum test was performed. Individual median $\pm$ interquartile range (IQR) experimental curves are provided in \ref{appendix:LDF}. The following aims to analyse the tendency in the ischaemic and hyperaemic responses of the median response ($\pm$ IQR). Figure \ref{fig:raw_LDF} presents the raw experimental LDF results together with the measured skin temperature.

\subsubsection{N=6 men}

For each cycle, measurements were repeated four times, resulting in a total of \(6 \times 4 = 24\) samples. Since each load was tested over two cycles, the \(48\) samples were analysed. Before the first load cycle, the skin temperature was \(29.5 \pm 2.8\)°C. The basal blood flow of the skin, measured during the 60~\si{\second} preceding the first application of local pressure, was \(24.7 \pm 19.9\) arbitrary units (AU).  

Upon application of the compressive load, the laser Doppler flowmetry (LDF) signal decreased, indicating a reduction in blood flow. This ischaemic response was dependent on the intensity of the load. During the first cycle, the LDF values dropped to \(8.0 \pm 5.5\)~AU (\(t\)-statistic = 36.3, \(p\)-value = \(1.3 \times 10^{-8}\)), representing a significant reduction to \(38.47 \pm 25.8~\%\) of the basal skin blood flow (\(t\)-statistic = 25.9, \(p\)-value = \(2.3 \times 10^{-6}\)). In the last two cycles, blood flow further declined to \(5.6 \pm 3.2\)~AU (\(t\)-statistic = 5.1, \(p\)-value = \(0.06\)), corresponding to \(22.8 \pm 9.4~\%\) of the basal blood flow (\(t\)-statistic = 57.3, \(p\)-value = \(3.5 \times 10^{-13}\)), indicating a progressive intensification of ischaemia in repeated cycles. The higher the load, the closer the signals became.  

Following the release of the compressive load, post-occlusive reactive hyperaemia (PORH) was observed (Figure~\ref{fig:raw_LDF}). Similarly to the ischaemic response, the hyperaemic peak depended on the applied load intensity. The first load led to an increase in LDF values up to \(66.7 \pm 76.1\)~AU (\(t\)-statistic = 26.0, \(p\)-value = \(2.2 \times 10^{-6}\)), corresponding to approximately \(750~\%\) of the basal blood flow (\(251.8 \pm 129.5~\%\); \(t\)-statistic = 23.5, \(p\)-value = \(7.9 \times 10^{-6}\)), demonstrating a strong hyperaemic response. With the second load intensity, LDF values peaked at \(94.1 \pm 89.3\)~AU (\(t\)-statistic = 5.8, \(p\)-value = \(0.06\)), reaching approximately \(1100~\%\) of the basal blood flow (\(353.0 \pm 270.5~\%\); \(t\)-statistic = 14.9, \(p\)-value = \(5.7 \times 10^{-4}\)). After reaching the PORH peak, the LDF signal gradually returned toward a value slightly above the initial basal level.

\subsubsection{N=5 women}

For each cycle, the measurements were repeated four times, resulting in a total of \(5 \times 4 = 20\) samples. Since each load was tested over two cycles, the \(40\) samples were analysed. Before the first load cycle, the skin temperature was \(27.8 \pm 2.6\)°C.  
The basal blood flow of the skin, measured during the 60~\si{\second} preceding the first application of local pressure, was \(16.6 \pm 17.3\) arbitrary units (AU).  

Upon application of the compressive load, the laser Doppler flowmetry (LDF) signal decreased, indicating a reduction in blood flow. This ischaemic response was dependent on the load intensity. During the first cycle, the LDF values dropped to \(4.9 \pm 3.8\)~AU (\(t\)-statistic = 10.9, \(p\)-value = \(0.004\)), representing a significant reduction to \(32.8 \pm 24.2~\%\) of the basal skin blood flow (\(t\)-statistic = 17.6, \(p\)-value = \(1.5 \times 10^{-4}\)), confirming a substantial decrease. In the last two cycles, blood flow further declined to \(4.3 \pm 1.9\)~AU (\(t\)-statistic = 0.9, \(p\)-value = \(0.6\)), corresponding to \(24.3 \pm 17.9~\%\) of the basal blood flow (\(t\)-statistic = 15.8, \(p\)-value = \(3.7 \times 10^{-4}\)), indicating a progressive intensification of ischaemia over repeated cycles.  

Following the release of the compressive load, post-occlusive reactive hyperaemia (PORH) was observed (Figure~\ref{fig:raw_LDF}). Similarly to the ischaemic response, the hyperaemic peak depended on the applied load intensity. The first load led to an increase in LDF values up to \(45.9 \pm 26.1\)~AU (\(t\)-statistic = 5.3, \(p\)-value = \(0.07\)), approximately corresponding to \(475\%\) of the basal blood flow (\(255.6 \pm 113.2~\%\); \(t\)-statistic = 2.4, \(p\)-value = 0.3), suggesting a moderate but statistically non-significant hyperaemic response. With the second load intensity, LDF values peaked at \(60.3 \pm 58.5\)~AU (\(t\)-statistic = 7.3, \(p\)-value = \(0.03\)), reaching approximately \(626\%\) of the basal blood flow (\(336.4 \pm 144.6~\%\); \(t\)-statistic = 2.6, \(p\)-value = 0.3). After reaching the PORH peak, the LDF signal gradually returned toward a value slightly above the initial basal level.  

\subsection{Differences Between Men and Women}

\begin{figure}[ht!]
    \centering
    \includegraphics[width=0.48\linewidth]{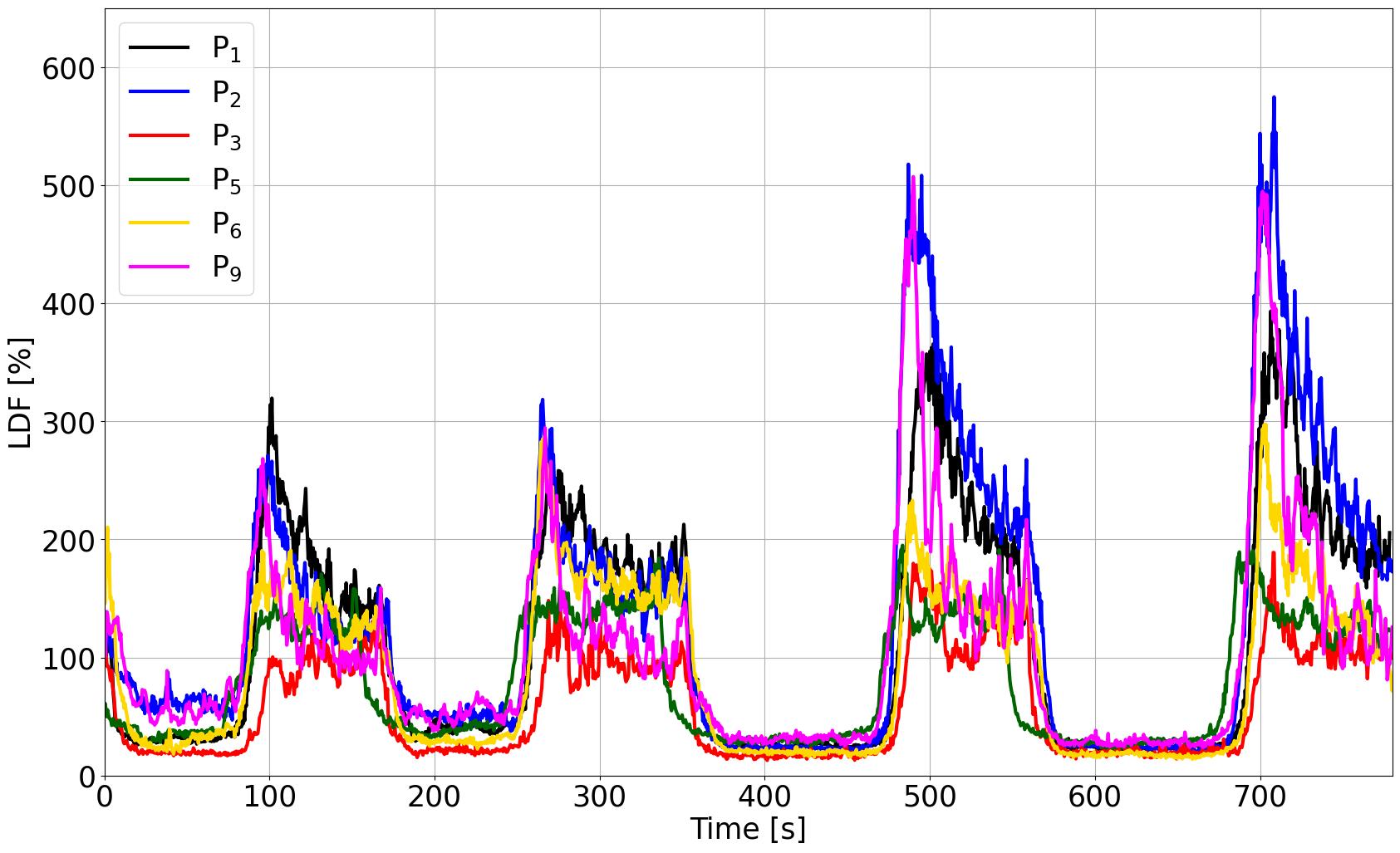}
    \includegraphics[width=0.48\linewidth]{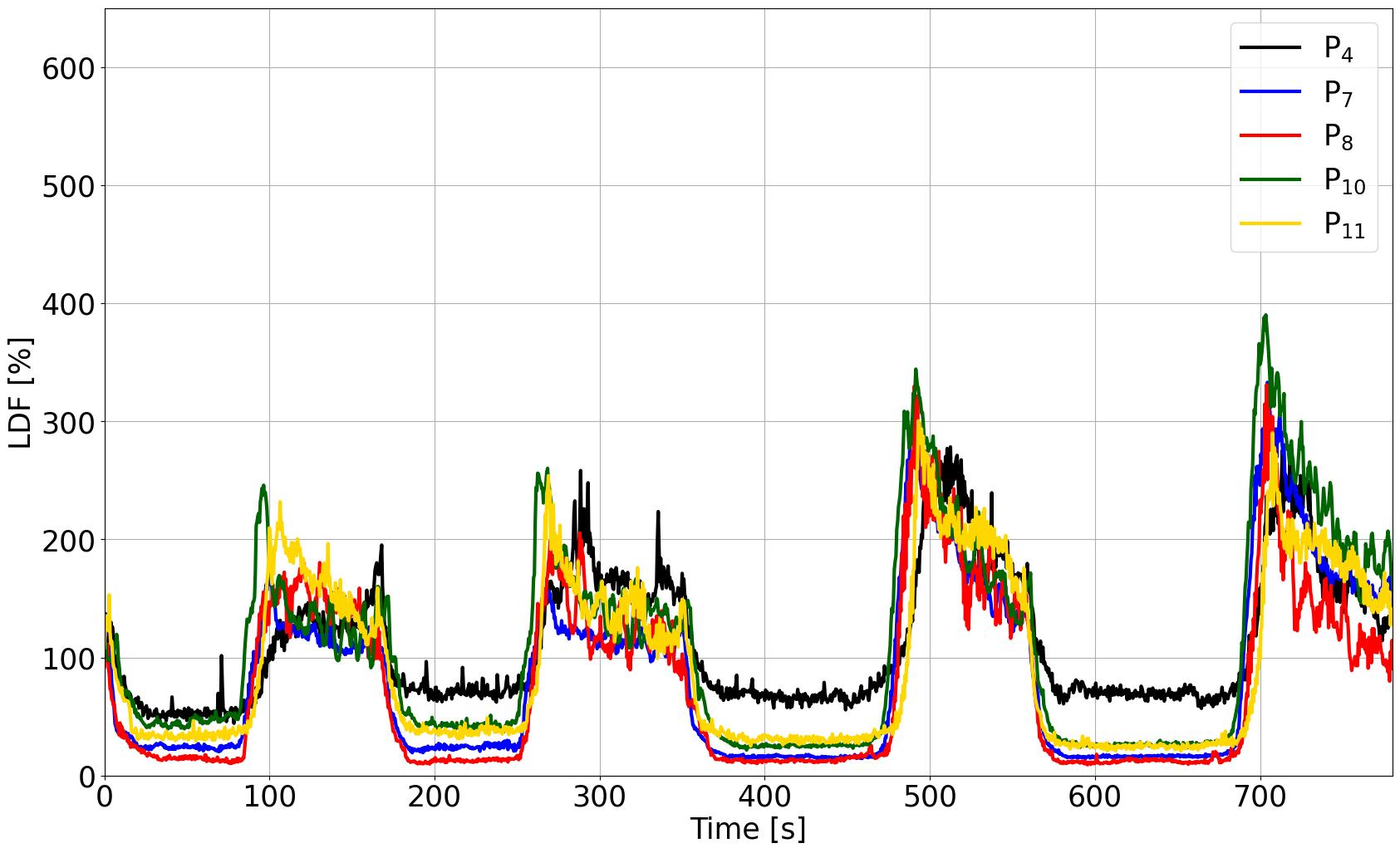}
    \caption{Median LDF signals normalised to the basal skin blood flow for each of the \(N=6\) male (left) and \(N=5\) female (right) participants. When expressed relative to baseline values, the difference in magnitude and behaviour between groups is reduced. Temperature primarily affects basal skin blood flow rather than the microvascular response to compressive loading.}
    \label{fig:pc_LDF}
\end{figure}

The lowest LDF values were observed in individuals with the lowest skin temperatures. However, as shown in Figure~\ref{fig:pc_LDF}, the difference in magnitude and behaviour is reduced when LDF values are expressed relative to basal blood flow of the skin. This normalisation accounts for inter-individual variability in absolute blood flow measurements.

Statistical analysis using independent Wilcoxon rank tests was performed to compare ischaemic and hyperaemic responses between male and female participants. The null hypothesis that two sets of measurements are drawn from the same distribution. The alternative hypothesis is that the values in one sample are more likely to be larger than the values in the other sample. The results showed significant differences in terms of raw AU signals but did not show significant differences between the groups for any of the conditions tested when looking to the relative evolution to the basal blood flow.

For the raw LDF signal (AU):  
\begin{itemize}
    \item 20 mL ischemia: \( t = 4.13 \), \( p = 3.6\times10^{-5} \)  
    \item 20 mL hyperemia: \( t = 3.41 \), \( p = 6.5\times10^{-4} \)  
    \item 40 mL ischemia: \( t = 3.76 \), \( p = 1.7\times10^{-4} \)  
    \item 40 mL hyperemia: \( t = 3.05 \), \( p = 2.3\times10^{-3}  \) 
\end{itemize}

For the relative LDF variation compared to the basal blood flow (in \%):  
\begin{itemize}
    \item 20 mL ischemia: \( t = 0.96 \), \( p = 0.34 \)  
    \item 20 mL hyperemia: \( t = 0.69 \), \( p = 0.49 \)  
    \item 40 mL ischemia: \( t = 0.41 \), \( p = 0.69 \)  
    \item 40 mL hyperemia: \( t = 0.46 \), \( p = 0.64 \) 
\end{itemize}

Since all \( p \)-values for the relative LDF variations are well above the standard significance threshold (\( \alpha = 0.05 \)), we do not observe statistically significant differences between men and women in their microvascular responses to compressive loading. These findings indicate that, while absolute LDF values may vary between individuals, relative responses to ischaemia and subsequent hyperaemia are comparable between sexes.

These results further suggest that temperature primarily affects basal skin blood flow rather than the microvascular response to load-induced ischaemia and subsequent hyperaemia. Consequently, expressing the LDF signal as a percentage of the individual’s basal blood flow provides a more consistent and reliable measure, reducing the variability associated with baseline differences in skin perfusion.

\subsection{Numerical 2-compartment biphasic 1D column}

\begin{figure}[ht!]
    \centering
    \includegraphics[width=0.4\linewidth]{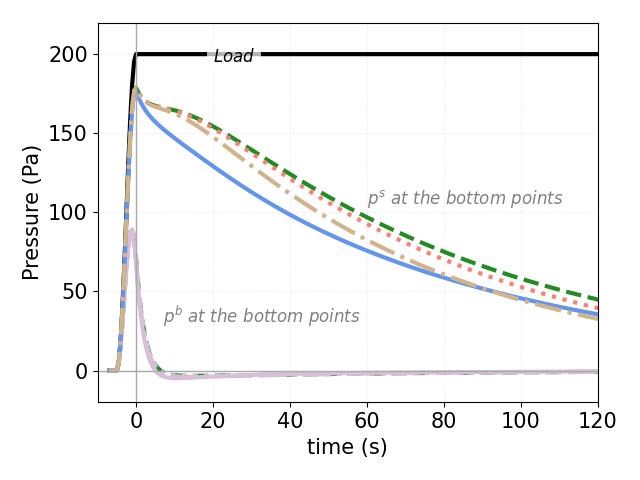}
    \includegraphics[width=0.4\linewidth]{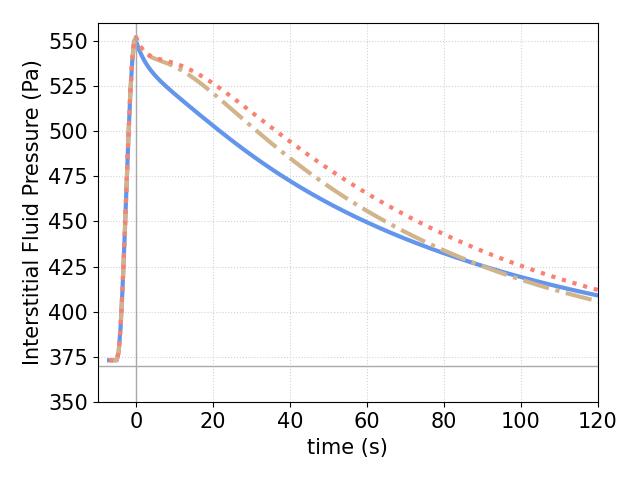}
    \includegraphics[width=0.4\linewidth]{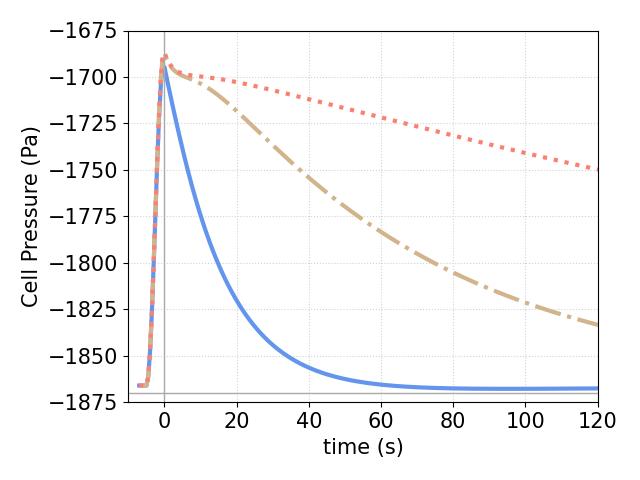}
    \includegraphics[width=0.4\linewidth]{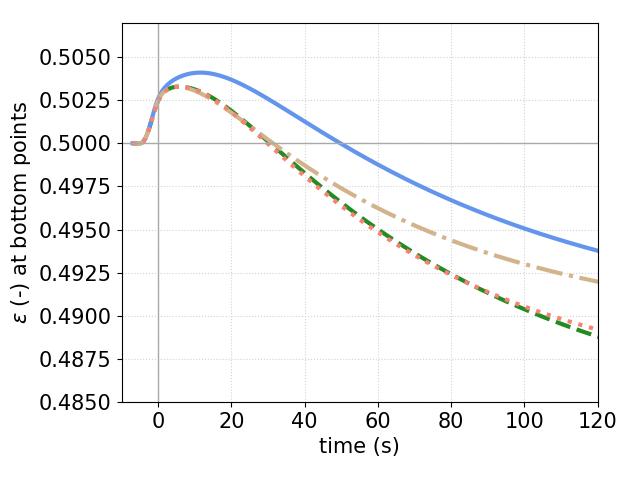}
    \includegraphics[width=0.4\linewidth]{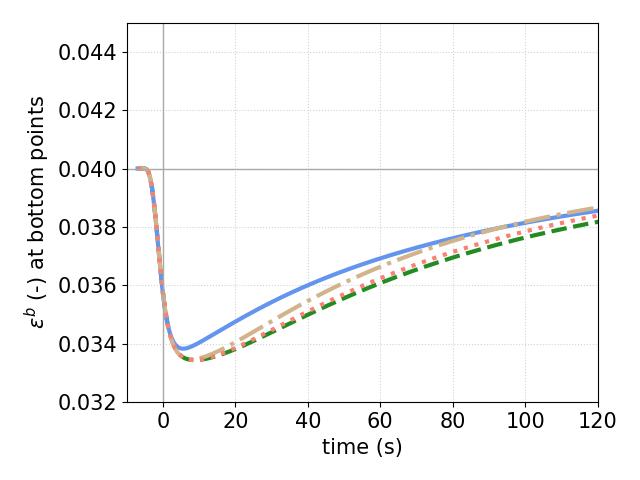}
    \includegraphics[width=0.4\linewidth,trim={3cm 2cm 3cm 2cm},clip]{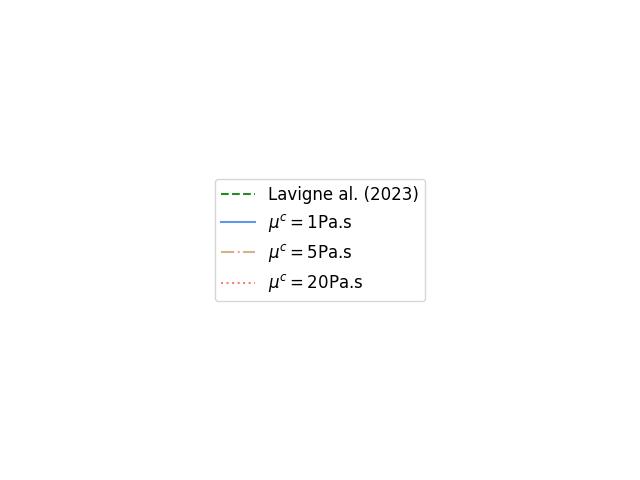}
    \caption{Results of the consolidation test on a 1D 2-compartment column including a biphasic interstitium. The interstitial fluid is the non wetting phase occupying $85~\%$ of the interstitium and the cells represent the wetting phase accounting for $15~\%$ of the interstitium. Different cell viscosities are tested $(\mu^c=1,\,5,\,20)$. The results include the phase as well as the vascular and extra-vascular porosities evaluated at the bottom points of the column. The mono-phasic computation corresponds to the test presented in \citet{Lavigne2023}. }
    \label{fig:1D:column:results}
\end{figure}

Figure \ref{fig:1D:column:results} illustrates the model response for the 1D consolidation examples described in \citet{Scium2021,Lavigne2023}, with an initial blood volume fraction of 0.04. The model exhibits behaviour consistent with the findings in \citet{Scium2021,Lavigne2023}, but highlights the influence of cell viscosity $\mu^c$ on pressure responses. When the cell viscosity matches that of the interstitial fluid $\mu^l$, the medium behaves as a mono-phasic system dominated by a fluid-filled interstitium (with increased porosity). In this case, both interstitial fluid pressure and cell pressure relax over a relatively short timescale. However, as the viscosity of the cells increases, the relaxation time of the cells lengthens. Consequently, the solid pressure and porosities reveal a time-dependent response characterised by two distinct timescales: a shorter timescale associated with interstitial fluid flow and a longer timescale linked to cell phase flow. When cell viscosity is further elevated to a highly viscous regime, the system's short-term behaviour resembles that of a mono-phasic interstitium, as the cell phase effectively behaves like a solid phase.

\newpage

\subsection{Mesh Convergence}

\begin{figure}[ht!]
    \centering
    \includegraphics[width=0.8\linewidth]{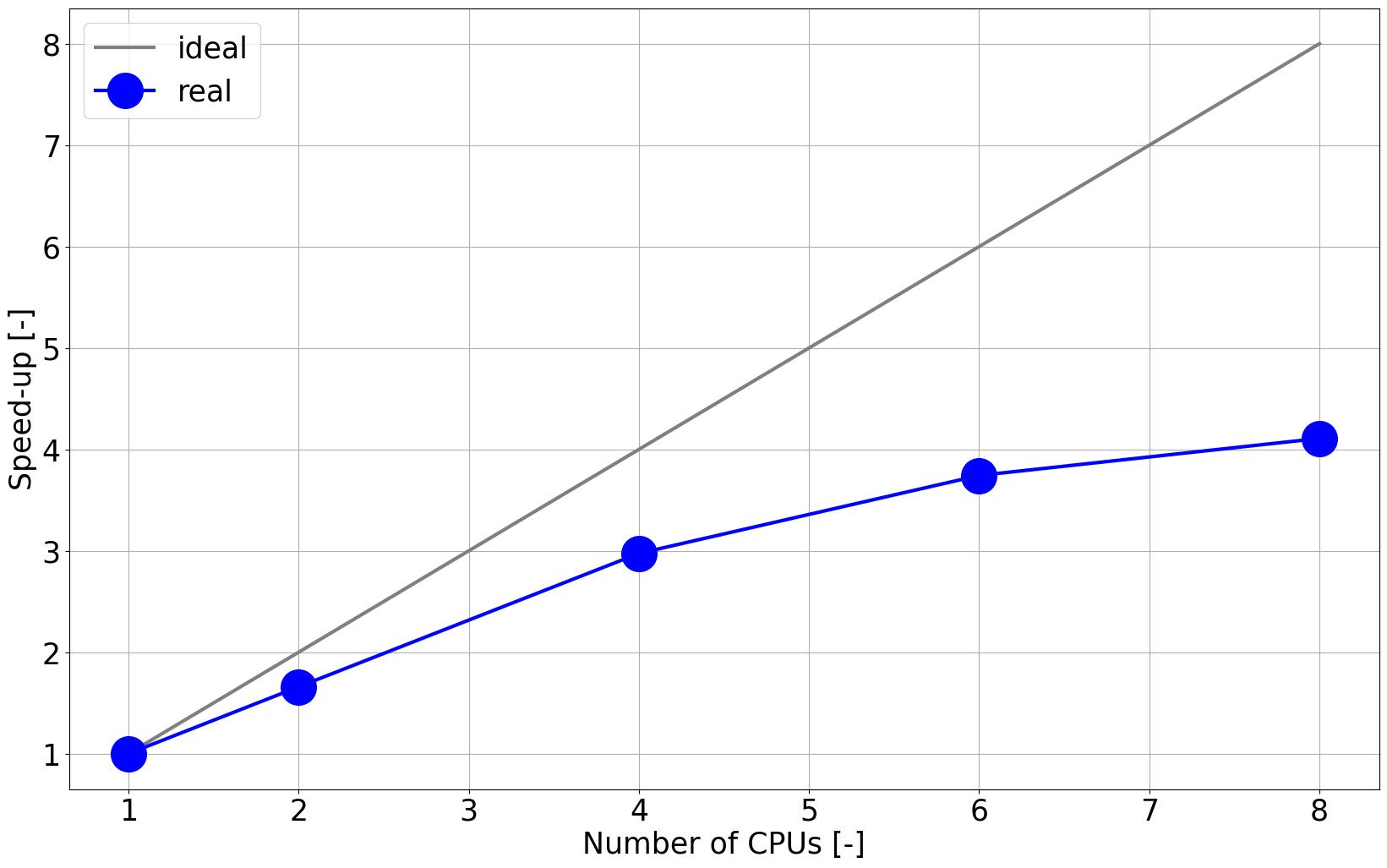}
    \caption{Speed-up for the 250k element mesh as the CPU count increases.}
    \label{XX}
\end{figure}

The analysis, carried out over the first 150 seconds of the experiment, revealed RMSE values of 5.5~\%, 2.2~\%, and 0.5~\% for the 25k, 44k, and 125k element meshes, respectively. Based on these findings, the 44k element mesh, demonstrating a reasonable balance between computational efficiency and accuracy, was selected for subsequent simulations. Furthermore, the computational codes were validated for parallel execution using the FEniCSX framework. 
A substantial performance enhancement was achieved, demonstrating a five-fold speed-up when utilising 4 CPUs for the 125k element mesh and 8 CPUs for the 250k element mesh. Figure \ref{XX} illustrates the speed-up for the 250k element mesh as the CPU count increases. Due to the limited degrees of freedom, performance gains plateau beyond 8 CPUs. To ensure accurate speed-up analysis, all writing and post-processing were disabled. For details on HPC installation procedures, the reader is referred to \ref{parallel}.

\subsection{Sensitivity analysis}

\ref{sec:appendix:sens:curves} presents the first-order sensitivity curves, and Table \ref{tab:sobol:01} provides the first-order Sobol indices. The Sobol indices suggest two dominant parameters, the Young modulus and the exponent $\alpha$ that account for 92.7~\% of the variance. The analysis of the curves reveals that an increase in Young's modulus leads to a reduction in displacement, thereby decreasing the ischaemic response. In contrast, increasing the hydraulic permeability of the interstitial fluid and the compressibility of the vessel tends to alleviate ischaemia and enhance the hyperaemic response. An increase in blood hydraulic permeability results in a decreased hyperaemic response. Both ischaemic and hyperaemic responses are further amplified by an increase in the exponent $\alpha$. Interestingly, the Young modulus mainly impacts the displacement magnitude, whereas the other parametric factor played an important role in the hysteresis response of the model in the flux-displacement curves.

\begin{table}[ht!]
\centering
{%
\begin{tabular}{ccc}
\hline
Parameter                          & $\theta_i\, \text{or}\,\theta_{ij}$ [IU] & $S_i\,\text{or}\,S_{ij}$ [\%] \\ \hline
Young Modulus (E)                  & $2.00\times10^{-1}$             & $52.78$               \\
$\alpha$ (Eq. \ref{eq:twocomp:vascpermea})     & $-1.74\times10^{-1}$            & $39.87$               \\
Vessel Compressibility ($K$)         & $5.83\times10^{-2}$             & $4.46$               \\
Hydraulic permeability ISF ($L^l$) & $4.68\times10^{-2}$             & $2.88$               \\
Hydraulic permeability Blood ($L^b_0$) & $-2.07\times10^{-3}$            & $5.64\times10^{-3}$  \\ \hline
\end{tabular}%
}
\caption{First-order sobol indices sorted from the largest to the lowest impact on the solution.}
\label{tab:sobol:01}
\end{table}

The results of the second-order sensitivity analysis, including $\theta_i,\,\theta_{ij},\,S_i,\,S_{ij}$, are provided in Table \ref{tab:sobol} and are sorted in increasing order of $S_i$ and $S_{ij}$.

\begin{table}[ht!]
\centering
{%
\begin{tabular}{ccc}
\hline
Parameter                          & $\theta_i\, \text{or}\,\theta_{ij}$ [IU] & $S_i\,\text{or}\,S_{ij}$ [\%] \\ \hline
$(E , \alpha)$                 & $-7.90\times10^{-1}$            & $50.9$               \\
$(\alpha , K)$                 & $-6.90\times10^{-1}$            & $38.9$              \\ 
Young Modulus (E)                  & $2.00\times10^{-1}$             & $3.28$               \\
$(E , K)$                      & $1.86\times10^{-1}$             & $2.81$               \\
$\alpha$ (Eq. \ref{eq:twocomp:vascpermea})     & $-1.74\times10^{-1}$            & $2.48$               \\
$(E , L^l)$                    & $1.08\times10^{-1}$             & $0.94$               \\
Vessel Compressibility ($K$)         & $5.83\times10^{-2}$             & $0.27$               \\
$(L^b_0 , \alpha)$               & $-4.76\times10^{-2}$            & $0.19$               \\
Hydraulic permeability ISF ($L^l$) & $4.68\times10^{-2}$             & $0.17$               \\
$(L^b_0 , K)$                    & $-8.94\times10^{-3}$            & $6.53\times10^{-3}$  \\
$(L^l , L^b_0)$                  & $-5.80\times10^{-3}$            & $2.75\times10^{-3}$  \\
$(L^l , K)$                    & $3.83\times10^{-3}$             & $1.19\times10^{-3}$  \\
$(L^l , \alpha)$               & $2.22\times10^{-3}$             & $4.03\times10^{-4}$  \\
Hydraulic permeability Blood ($L^b_0$) & $-2.07\times10^{-3}$            & $3.50\times10^{-4}$  \\
($E , L^b_0)$                    & $-1.19\times10^{-3}$            & $1.16\times10^{-4}$  \\ \hline
\end{tabular}%
}
\caption{Second-order sobol indices sorted from the largest to the lowest impact on the solution.}
\label{tab:sobol}
\end{table}

Sensitivity analysis reveals that the interaction between parameters influences the variance of the solution, accounting for almost $94~\%$ of this variance. However, three main parameters $(E,\,\alpha,\,K)$ now predominantly impact the variance of the solution, both independently and through their interactions, accounting for $98.7~\%$ of this variance. Therefore, blood flow is mainly influenced by the choice of the Young modulus, the exponent $\alpha$ describing the non-linearity of the permeability, but also $K$ the compressibility of the vessel.

\subsection{Model capacity to reproduce the flow}

This section aims at analysing the 3D blood flow obtained in the generic FE model considering the experimental procedure described in Section \ref{sec:experimental}. For the sake of comparison, the LDF signal has been post-processed to show the relative variation to the basal skin blood flow at each time point. Three parameter sets have been considered (considering the international standard units):
\begin{itemize}
    \item Set$_1$: $[E,\;L^l,\;L^b,\;\alpha,\;K]\,=\,[2\times10^5,\; 10^{-14},\;	10^{-9},\;	3,\; 10^{3}]$ 
    \item Set$_2$: $[E,\;L^l,\;L^b,\;\alpha,\;K]\,=\,[2\times10^5,\; 10^{-13},\;	10^{-9},\;	5,\; 10^{3}]$
    \item Set$_3$: $[E,\;L^l,\;L^b,\;\alpha,\;K]\,=\,[8\times10^4,\; 10^{-13},\;	10^{-9},\;	6,\; 10^{3}]$
\end{itemize}

\begin{figure}[ht!]
    \centering
    \includegraphics[width=1\linewidth,trim={0 0 0 0},clip]{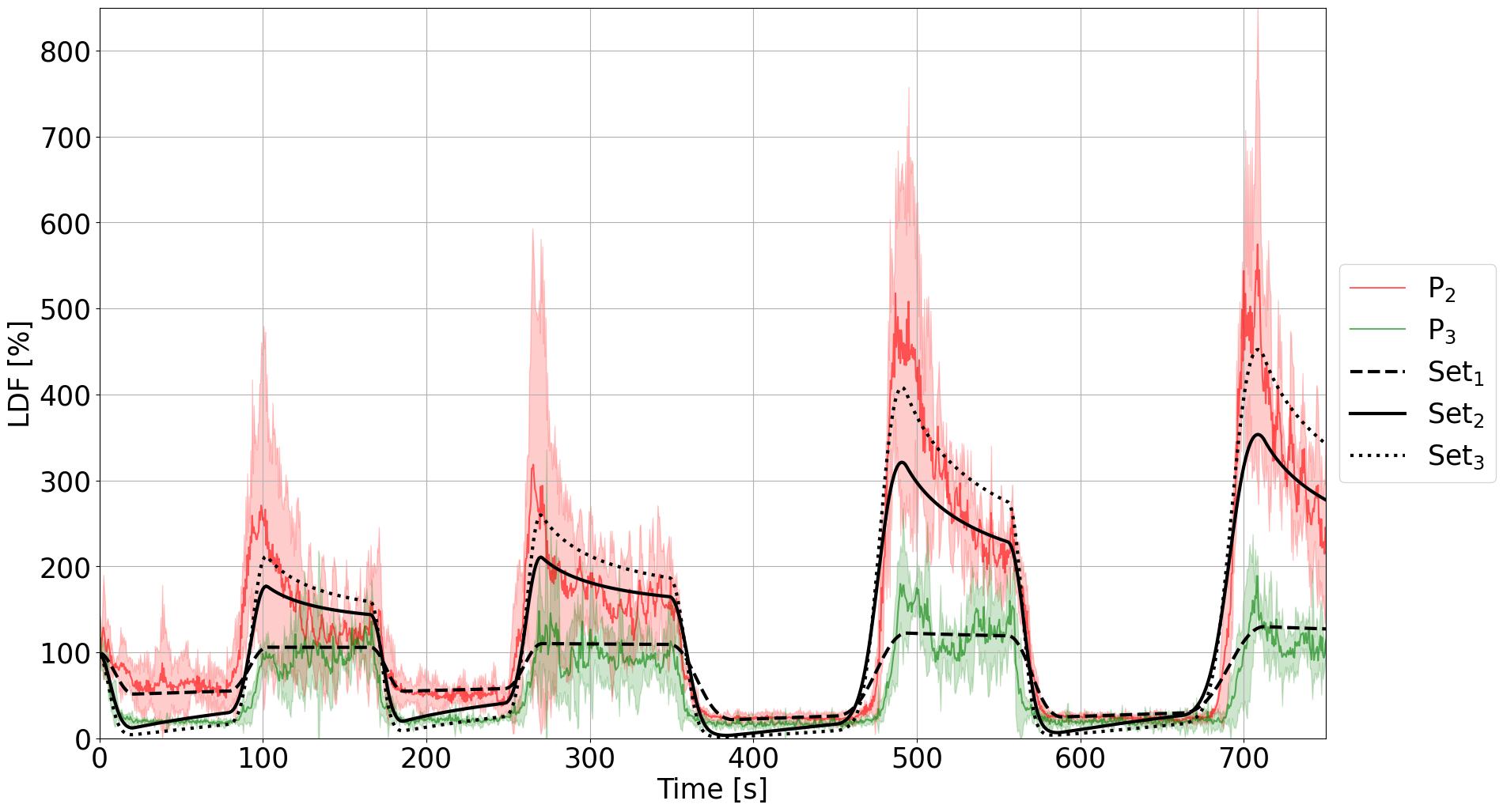}
    \caption{Three different sets of parameters were computed to cover different possible outputs of the model. The results are superimposed on the IQR experimental corridor of the maximal and minimal observed responses, highlighting the capacity of the model to qualitatively reproduce both the ischaemic and hyperaemic responses.}
    \label{fig:superpmodel}
\end{figure}

\begin{figure}
    \centering
    \vspace{-2cm}
    \includegraphics[width=\linewidth]{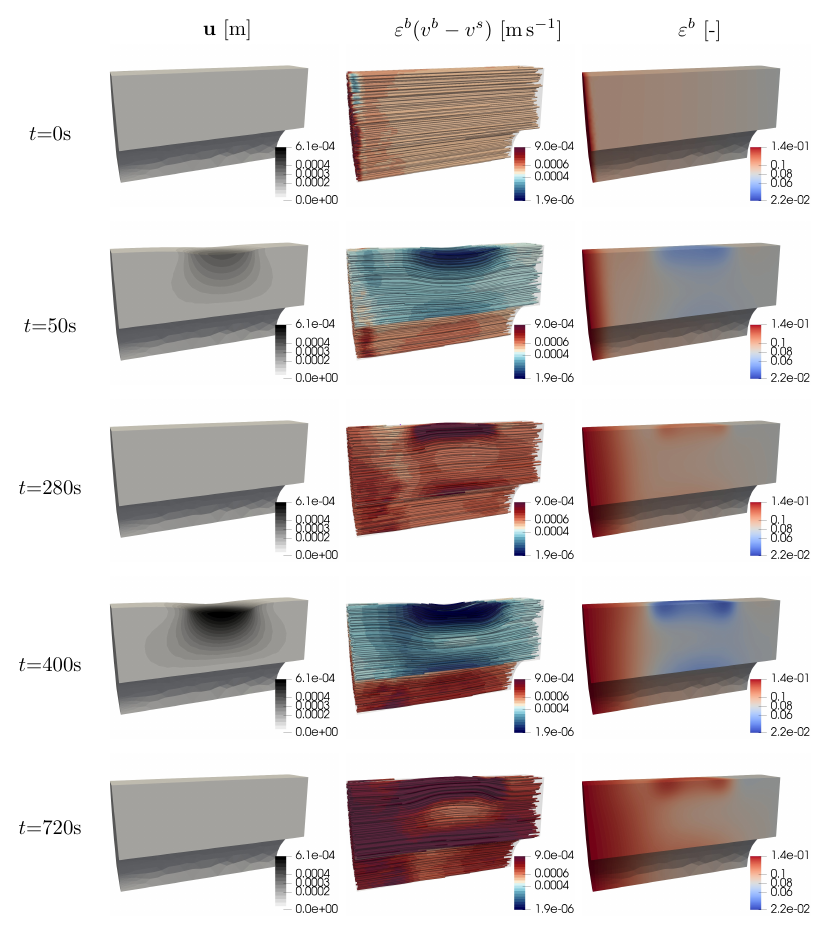}
    \caption{From left to right: displacement magnitude, blood flow, vascular porosity maps. The results represent the initial state, the first cycle ischaemia ($t=50$~\si{\second}), the second hyperaemic peak ($t=280$~\si{\second}), a maximum ischaemia ($t=400$~\si{\second}) and the maximum hyperaemia ($t=720$~\si{\second}). When indenting the skin, the flow is locally reduced and the vascular porosity decreases. Conversely, vasodilation is observed with an increase blood flow during and directly after the release of the load. During the indented phase, vasodilations seems to occur before the indented location.}
    \label{fig:modele}
\end{figure}

The resulting micro-circulatory response (computed LDF signal) is superimposed on the experimental corridor Figure \ref{fig:superpmodel}. 
We observe that two of the three models provide post-occlusive reactive hyperaemia. These are the two sets with the highest values of $\alpha$. The other one, Set$_1$ allows for a more accurate reproduction of the ischaemic plateau, but no hyperaemia is observed. A deeper analysis of the Set$_2$ is provided in Figure \ref{fig:modele}.

Initially, the boundary condition allows for a blood flux of approximately $400 ~\si{\micro\meter\per\second}$. During the establishment of this flux, in the initial five time steps, vascular porosity increases slightly to 9~\%  in the region of interest. During the first two cycles, the application of the load induces a compressive displacement, reducing the flux directly beneath the load. The overall flux adapts by circumventing this region. A concurrent reduction in vascular porosity is observed, indicating the onset of occlusion processes with increased vascular porosity before the loaded region (vasodilation).

Upon releasing the load, strong reflux is observed beneath the probe, characterised by increased blood velocity and an accompanying increase in vascular porosity. This is followed by a gradual relaxation of the flow, returning it to the initial homeostatic state. When a higher load is applied, the maximum displacement reaches $0.6\si{\milli\meter}$, resulting in an increase in intensity and depth of ischaemia, as well as a more pronounced post-occlusive reactive hyperaemia effect.

\section{Discussion}
\label{sec:disc:skin}

The primary objective of this study was to evaluate the feasibility of employing a two-compartment porous media mechanical model in conjunction with experimental measurements of skin micro-circulatory haemodynamics. This helps in the reproduction of the mechanical response of the skin under indentation loading \textit{in vivo} in the perspective of injury prevention. Laser Doppler flowmetry (LDF) was used as the experimental method to assess the micro-circulatory response of the skin under load conditions. The load was applied using the LDF probe itself.

LDF is a well-established, non-invasive, real-time technique for measuring micro-circulatory blood flow of the skin (\citet{Poffo2014, Varghese2009317, FolgosiCorrea2012}). However, the method has known limitations. For example, the LDF technique provides measurements from a limited penetration depth of 0.5-1~\si{\milli\meter} (\citet{Heurtier2013, FREDRIKSSON20094, Poffo2014}), which may not fully capture deeper vascular responses. Furthermore, the precision of LDF is sensitive to various external factors, such as probe position, contact force, skin temperature, and patient-specific characteristics such as tobacco use (\citet{Nogami2019287, Ajan2024, Kouadio201858, Petrofsky2012}). Therefore, stringent calibration procedures and the control of experimental conditions are critical to ensuring the reproducibility and reliability of LDF measurements. To overcome these limitations, smokers were excluded from the study and the temperature was measured to limit bias in the analysis while ensuring a thermal stabilisation period prior to the start of the experiment. This ensures the reproducibility and reliability of the LDF measurements. 

Although LDF has been used to diagnose micro-circulatory disorders, it is essential to consider the reference range and variability of the measured parameters to validly interpret LDF values in individual patients (\citet{Mrowietz2017347}). However, the LDF measurements allowed for the identification of values consistent with those reported in the literature, with an initial basal skin blood flow of $21.1 \pm 20.5$~AU. \citet{Borchardt2017} reported similar basal skin blood flow values for the forearm. In their results, pressure-induced effects did not result in significant changes compared to basal blood flow from the skin. However, a significant increase of 105.5\% was observed, from $32.7 \pm 12.6$~AU to $67.2 \pm 20.6$~AU, over a period of 1 hour after direct cold atmospheric plasma treatment for 270~s. \citet{Fei2018656} reported forearm blood flow basal skin blood flow values of $33 \pm 6.5$~AU measured by Laser Speckle Contrast Imaging (LSCI) and $60 \pm 11.5$~AU measured by Laser Doppler Perfusion Imaging (LDPI). Changes in blood flow during iontophoresis showed an increase of $724 \pm 412\%$ (LDPI) and $259 \pm 87\%$ (LSCI) from basal skin blood flow.

In terms of fingertip blood perfusion, studies such as \citet{Saha2020} reported basal skin blood flow values around $20$~AU. 
\citet{Volkov2017} described procedures including the recording of background perfusion and velocity ($0.3-0.5$~min), occlusion testing ($1-1.5$~min), and post-occlusion recordings ($0.1-1$~min), with occlusion pressure set at approximately $29.3~\si{\kilo\pascal}$. LDF amplitudes were also reported to vary between 10 and 60~AU for the fingers (\citet{Mrowietz2019129}).

The experiment also aligns with the literature on the observed behaviour. Similarly to \citet{Volkov2017}, the load application induces a reduced LDF signal. When the load is released, strong reflux is observed. This reflux is called post-occlusive reactive hyperaemia (PORH) (\citet{Fuchs2017, Balasubramanian2021}). \citet{ostergren1986skin} identified during the PORH a capillary blood velocity that reached a maximum of $1.2 \pm 0.7~\si{\milli\meter\per\second}$ at $7.8 \pm 2.4$~seconds after cuff release.

PORH is a subcategory of reactive hyperaemia (RH), which is a physiological response that leads to increased microvascular blood flow to the skin. Optical coherence tomography (OCT) has been used to visualise and quantify the microvascular changes of the skin in response to RH in humans, demonstrating an increase in microvascular diameter, flow rate, speed and density after a stimulus (\citet{Argarini2020, WangEvers2021}). Four major factors have been proposed to be involved in the hyperaemic response: metabolic vasodilators, endothelial vasodilators, myogenic response, and sensory nerves (\citet{Cracowski2006}). This hyperaemic state also explains the increase in temperature (thermal hyperaemia), since increased local blood flow to the skin leads to thermal hyperaemia (\citet{Golay2004, Choi2014, Wong2006, Rosenberry2020, Basiladze201583, Leo2020, Horn2022}). 

No significant differences were found between male and female populations for relative LDF variations with basal blood flow of the skin. This suggests that despite potential differences in absolute perfusion levels, the microvascular response to compressive loading and subsequent hyperaemia remains comparable between the sexes.  
The lack of significant differences (\( p > 0.3 \) for all conditions) indicates that sex-related factors do not substantially influence relative ischaemic and hyperaemic responses, at least within the population studied. This finding is consistent with previous research suggesting that while baseline skin blood flow may vary due to differences in skin temperature, vascular tone, or hormonal influences, the fundamental regulatory mechanisms of reactive hyperaemia and ischaemic adaptation are preserved between sexes.  
Furthermore, the observed normal distribution of the results supports the robustness of the experimental design, strengthening the reliability of the findings. These results further highlight the importance of normalising the LDF values to basal blood flow, as absolute perfusion measurements alone may introduce variability that is not related to the actual microvascular response. This normalisation approach reduces inter-individual differences and allows for a more accurate comparison of vascular reactivity across different populations. 
Future studies with larger sample sizes and additional physiological parameters (e.g., hormone levels, vascular compliance) could provide further insight into possible subtle differences that were not detectable in the present study. However, within the current experimental framework, our findings suggest that the physiological response to compressive loading is largely independent of sex, reinforcing the validity of using mixed-sex data in similar studies.

To reproduce the mechanical behaviour and micro-circulatory response of human skin \textit{ in vivo}, a two-compartment model inspired by \citet{Scium2021} was developed, calculated in FEniCSx 0.9.0 and compared to the experiment. This model is based on the Thermodynamically Constrained Average Theory (TCAT) and accounts for both the fluid behaviour of the interstitial fluid and the vascular network, paving the way for the integration of biological exchange as proposed in \ref{appendix:02}. Powered by the characteristic times of the experiment and a preliminary 1D consolidation problem in a column, the theory was simplified to provide a consistent comparison with the experiment. In addition, in the absence of data on the displacement field, no calibration was achievable, so parameters from the literature were used and a sensitivity analysis was performed.

Sensitivity analysis revealed that three main parameters governed the response to blood flow. Young's modulus directly affected the displacement of the sample and, therefore, the level of occlusion. The other two parameters are related to the choice of exponent in the constitutive permeability law and the choice of compressibility of the vessel in the constitutive law of vascular porosity. This suggests that the choice and definition of the vascular porosity state law and the permeability state law are crucial for a physically relevant interpretation of the model. This can be explained in more detail by the non-Newtonian behaviour of blood. In fact, when blood is subjected to compressive loads, the deformation rates and the resulting shear stress can lead to changes in viscosity. At low deformation rates, cell aggregations increase viscosity, while at higher rates, these structures break down, leading to a decrease in viscosity (\citet{Javadi2023775}). The compressive load can also affect the formation of the cell-free layer (CFL) in micro-vessels, which influences blood flow and viscosity. Higher compressive loads can enhance the deformability of red blood cells, affecting CFL and overall blood rheology (\citet{ZHANG2011, Yalcin2015}).

In the proposed study, three different simulations were further compared to the experiment to qualitatively assess the capacity of the model to reproduce the experiment. One of them allowed for a precise reproduction of the ischaemic plateau, whereas the two others highlighted the capacity to capture both ischaemia and hyperaemia, supporting the capacity of this physically based model to account for skin micro-circulation. However, this study also supports the need to develop experimental procedures that monitor both mechanical fields (force/displacement) and fluid flows to further validate such models and inform their boundary conditions. Therefore, experiment should concentrate on the explicit evaluation of skin fluids flow \textit{in vivo}. Then, biochemistry and nutrient exchanges could be introduced and evaluated as proposed in the proof of concept of \ref{appendix:02}.

\section{Conclusion}

This study introduces a hierarchical modelling framework to integrate micro-circulation dynamics within a porous medium. The proposed theory conceptualises the interstitium as a biphasic system, distinguishing between the characteristic timescales of cells and the interstitial fluid. Appendix A extends the model to account for biological transport processes, such as oxygen diffusion.

To evaluate the model, a one-dimensional column consolidation test was performed. Initial results indicated that cell viscosity introduces an additional characteristic timescale, separate from that governed by interstitial fluid flow. Notably, when the cell viscosity is high and the timescales are short, the cells behave similarly to a solid.

Building on these insights, the model was simplified to correspond with an experimental study focused on healthy skin. A sensitivity analysis, based on material properties from the existing literature, identified three key parameters with strong interdependence: Young's modulus, the exponent of the permeability law, and the assumed compressibility of the vessel. Since displacement data was unavailable, parameter calibration was not performed, highlighting the need for further experimental studies to refine mechanical property estimates. However, the model successfully captured the qualitative features of both ischaemia and post-occlusive reactive hyperaemia within the experimental scope.

All numerical simulations were performed using the open-source software FEniCSx v0.9.0. To promote transparency and reproducibility, anonymised experimental data and associated finite element code have been openly shared on GitHub.


\section*{Acknowledgments}
This research was funded in whole or in part by the Luxembourg National Research Fund (FNR), grant reference No. 17013182. For the purpose of open access, the author has applied a Creative Commons Attribution 4.0 International (CC BY 4.0) licence to any Author-Accredited manuscript version arising from this submission.

The authors also acknowledge the use of the Cassiopee Arts et Métiers Institute of Technology HPC Centre made available for conducting the research reported in this article.


\section*{Declarations}
\textbf{Competing interests:} The authors declare that they have no known competing financial interests or personal relationships that could have appeared to influence the work reported in this article.

\textbf{Supplementary material:} The Python codes corresponding to the FEniCSx models and the experimental data of this article are made available on GitHub.

\appendix

\section{Mono-phasic interstitium: Simplified Variational Form}
\label{appendix:mono:var}
The variational form is obtained considering ($q^{c}$,$q^{l}$, $q^b$,$\mathbf{w}$) the test functions defined in the mixed space $L_0^2(\Omega)\times L_0^2(\Omega)\times[H^1(\Omega)]^3$. The solutions of the problem are the capillary pressure, the cell pressure, the blood pressure, and displacement of the solid: ($\pl,\,\pb,\,\us$).The problem to be solved reads:
\begin{align}
\us&=\mathbf{u}_{\text{imposed}}\,\text{on}\, \partial\Omega_u \label{eq:mono:twocomp:dirichlet}\\
{\color{black} p^\alpha} &= p_{\text{imposed}}\,\text{on}\, \partial\Omega_p^\alpha \label{eq:mono:twocomp:bc_pressure}
\end{align}
\begin{align}
\begin{split}
\int_\Omega \frac{k^{b}}{\mu^{b}} \nabla \pb \nabla q^l \ddroit \Omega 
+  \int_\Omega \frac{k^{l}}{\mu^{l}} \nabla \pl \nabla q^l \ddroit \Omega \\
 + \int_\Omega \nabla \cdot \left(\frac{\mathrm{D}^s \us}{\mathrm{D}t} \right) q^l \ddroit \Omega = 0,\,\forall~q^l\in {L}_0^2(\Omega)
\end{split}  \label{eq:mono:twocomp:variationalmassIF}
\end{align}
\begin{align}
\begin{split}
\int_\Omega \tilde{C}_{e,p} \frac{\mathrm{D}^s \pl}{\mathrm{D}t} q^b \ddroit \Omega 
- \int_\Omega \tilde{C}_{e,p} \frac{\mathrm{D}^s \pb}{\mathrm{D}t} q^b \ddroit \Omega \\
+ \int_\Omega \frac{k^{b}}{\mu^{b}}\mathbf{\nabla}\pb \nabla q^b \ddroit \Omega 
+ \int_\Omega \varepsilon^b \nabla \cdot \left(\frac{\mathrm{D}^s \us}{\mathrm{D}t} \right) q^b \ddroit \Omega = 0,\,\forall~q^b\in {L}_0^2(\Omega)
\end{split}  \label{eq:mono:twocomp:variationalmassblood}
\end{align}
\begin{align}
\begin{split}
\int_\Omega \mathbf{t}^{\text{eff}}(\us) : \nabla \mathbf{w} \ddroit \Omega  
- \int_\Omega \left(1-\varepsilon^b\right) \pl \nabla \cdot \mathbf{w} \ddroit \Omega \\ 
- \int_\Omega \varepsilon^b \pb \nabla \cdot \mathbf{w} \ddroit \Omega 
- \int_{\Gamma_s} \mathbf{t}^{\text{imposed}} \cdot \mathbf{w} \ddroit \Gamma_s =0, \forall~\mathbf{w}\in[H^1(\Omega)]^3 \label{eq:mono:twocomp:variationalmomentum}
\end{split}
\end{align}

\section{Oxygen Biochemistry}
\label{appendix:02}

Consider the exchange of oxygen between the blood compartment and cells in the extravascular space through the interstitial fluid. The idea is to establish the mass conservation Equation of the blood with a mass exchange between the blood and the interstitial fluid and a consumption law by the cells as an extension of the model proposed by \citet{Scium2021}.

\subsection{Strong form}

Let $\omega^{O2,l}$ be the mass fraction of oxygen in the IF and $\omega^{O2,b}$ the mass fraction of oxygen in the blood. $\omega^{O2,l}$ is a new unknown in the problem to be computed. According to the formulas in Sciumè et al. 2021, the mass conservation of the species reads:
\begin{align}
\begin{split}
&\underbrace{\frac{\partial}{\partial t}(S^{l}\varepsilon \rho^{l} \omegaodeux)}_{\text{Accumulation rate}}
+ \underbrace{\nabla \cdot (S^{l}\varepsilon \rho^{l} \omegaodeux \mathbf{v}^{l})}_{\text{Outward of species advective transport}}\\
&+ \underbrace{\nabla \cdot (S^{l}\varepsilon \rho^{l} \omegaodeux \mathbf{u}^{O2,l})}_{\text{Outward of species diffusive transport}}
- \underbrace{S^{l}\varepsilon r^{O2,l} \mathbf{u}^{O2,l})}_{\text{Intraphase reactive exchange of mass}}\\
&+ \underbrace{\Sigma_{\kappa}\overset{O2,l\rightarrow O2,\kappa}{M}}_{\text{Intraphase mass transport of the species}}
+ \underbrace{\omegaodeux\Sigma_{\kappa}\overset{\kappa\rightarrow IF}{M}}_{\text{Interphase mass transport}} = 0 \label{eq:o2_conservation}
\end{split}
\end{align}

In our case, this Equation can be re-written:
\begin{align}
\begin{split}
&\underbrace{\frac{\mathrm{D}^s}{\mathrm{D}t}(S^{l}\varepsilon \rho^{l} \omegaodeux)}_{\text{Accumulation rate}}
+ \underbrace{\nabla \cdot (S^{l}\varepsilon \rho^{l} \omegaodeux (\mathbf{v}^{l}-\vs))}_{\text{Infiltration}}\\
&+ \underbrace{\nabla \cdot (S^{l}\varepsilon \rho^{l} \omegaodeux \mathbf{u}^{O2,l})}_{Diffusion}
+ \underbrace{S^{l}\varepsilon \rho^{l} \omegaodeux \nabla \cdot \vs}_{\text{ECM deformation}}\\
&=\underbrace{\overset{O2,b\rightarrow O2,l}{M}}_{\text{Blood to IF transport}}
 - \underbrace{\overset{O2,l\rightarrow O2,c}{M}}_{\text{O2 consumption from the cells}} \label{eq:o2_conservation2}
\end{split}
\end{align}

where the increase of oxygen from blood is driven by (exchange proportional to the vessel wall area):

\begin{align} 
	\overset{O2,b\rightarrow O2,l}{M} = h_v \varepsilon^b (\omega^{O2,b} - \omegaodeux)\label{eq:o2apportblood}
\end{align}

where $\omega^{O2,b}$ is the mass fraction of oxygen within the blood and is assumed constant. The coefficient $h_v$ is representative of the vessel wall permeability.

The consumption of oxygen from the cells is proportional to the fluid saturation in cells and is given by ($\omega_{crit}$ is the hypoxia threshold obtained from Henry's law and $\gamma_0$ relates to the cell metabolism):

\begin{itemize}
\item if $\omegaodeux\geq \omega_{crit}$:
\begin{align} 
	\overset{O2,l \rightarrow O2,c}{M} = \gamma_0 \, S^c \, \varepsilon 
\label{eq:o2cellconsoa}
\end{align}
\item if $\omegaodeux\leq \omega_{crit}$:
\begin{align}
	\overset{O2,l\rightarrow O2,c}{M} = \gamma_0 \, S^c \, \varepsilon \, \left[\frac{1}{2}\left(1- \cos \pi \frac{\omegaodeux}{\omega_{crit}} \right) \right]\label{eq:o2cellconsob}
\end{align}
\end{itemize}

Introducing Fick's law, the diffusive flux of oxygen reads:

\begin{align} 
	\omegaodeux\mathbf{u}^{O2,l} = - D_{eff}^{O2,l}\nabla \omegaodeux      \label{eq:o2_diff_flux}
\end{align}

where $D_{eff}^{O2,l}$ is the effective diffusion coefficient of oxygen: 

\begin{align} 
	D_{eff}^{O2,l}=D_{0}^{O2,l}(\varepsilon^{l})^{\delta}=D_{0}^{O2,l}(S^{l}\varepsilon)^{\delta}      
\label{eq:o2_diff_effective_coeff}
\end{align}

where $D_{0}^{O2,IF}$ is the oxygen diffusion coefficient in the bulk interstitial fluid and $\delta$ is a coefficient related to the tortuosity of the medium.

Considering a constant density $\rho^l$, the mass conservation becomes:
\begin{align}
\begin{split}
&S^{l}\varepsilon\frac{\mathrm{D}^s}{\mathrm{D}t}(\omegaodeux)+S^{l} \omegaodeux\frac{\mathrm{D}^s}{\mathrm{D}t}(\varepsilon)+\varepsilon  \omegaodeux\frac{\mathrm{D}^s}{\mathrm{D}t}(S^{l}) \\
& - \nabla \cdot ( \frac{k^{l}}{\mu^{l}}  \omegaodeux \nabla \pl )
- \nabla \cdot (S^{l}\varepsilon D_{eff}^{O2,l}\nabla \omegaodeux ) \\
&+ S^{l}\varepsilon  \omegaodeux \nabla \cdot \vs = \frac{1}{\rho^{l}}\left(\overset{O2,b\rightarrow O2,l}{M} - \overset{O2,l\rightarrow O2,c}{M}\right) 
\end{split}
\label{eq:o2_conservation3}\end{align}

Using the divergence of a product rule, we get:
\begin{align}
\begin{split}
\nabla \cdot ( \frac{k^{l}}{\mu^{l}}  \omegaodeux \nabla \pl ) =  \omegaodeux \nabla\cdot\left(\frac{k^{l}}{\mu^{l}} \nabla \pl\right) + \frac{k^{l}}{\mu^{l}} \nabla \omegaodeux \nabla \pl 
\end{split}
\label{eq:o2_conservation5}\end{align}

Then, Equation \ref{eq:o2_conservation3} becomes:
\begin{align}
\begin{split}
&(1-S^{c})\varepsilon\frac{\mathrm{D}^s}{\mathrm{D}t}(\omegaodeux)\\
&+\omegaodeux\left(\underbrace{\frac{\mathrm{D}^s}{\mathrm{D}t}((1-S^{c})\varepsilon) - \nabla\cdot\left(\frac{k^{l}}{\mu^{l}} \nabla \pl\right) + (1-S^{c})\varepsilon \nabla \cdot \vs}_{(\ref{eq:twocomp:IFmass0b})\implies =0} \right) \\
& - \frac{k^{l}}{\mu^{l}} \nabla \omegaodeux \nabla \pl 
- \nabla \cdot ((1-S^{c})\varepsilon D_{eff}^{O2,l}\nabla \omegaodeux ) \\
& = \frac{1}{\rho^{l}}\left(\overset{O2,b\rightarrow O2,l}{M} - \overset{O2,l\rightarrow O2,c}{M}\right) 
\end{split}
\label{eq:o2_conservation3bb}\\
\begin{split}
&(1-S^{c})\varepsilon\frac{\mathrm{D}^s}{\mathrm{D}t}(\omegaodeux)\\
& - \frac{k^{l}}{\mu^{l}} \nabla \omegaodeux \nabla \pl 
- \nabla \cdot ((1-S^{c})\varepsilon D_{eff}^{O2,l}\nabla \omegaodeux ) \\
& = \frac{1}{\rho^{l}}\left(\overset{O2,b\rightarrow O2,l}{M} - \overset{O2,l\rightarrow O2,c}{M}\right) 
\end{split}
\label{eq:o2_conservation3b}
\end{align}

\subsection{Variational form}


Consider ($w_o$) the test function defined in the space $L_0^2(\Omega)$. The following system of Equations completes the system of Equation of the previous section. 

Consider ($q^{c}$,$q^{l}$, $q^b$,$\mathbf{w}$, $w_o$) the test functions defined in the mixed space $L_0^2(\Omega)\times L_0^2(\Omega)\times[H^1(\Omega)]^3\times L_0^2(\Omega)$. The solutions of the problem are the capillary pressure, the cell pressure, the blood pressure, the displacement of the solid, and the fraction of the oxygen mass: ($\pl,\,\plc,\,\pb,\,\us,\,\omegaodeux$).

Using Equations \ref{eq:twocomp:momentum}, \ref{eq:massblood2}, \ref{eq:twocomp:cellmass4c}, \ref{eq:twocomp:IFmass2}, and \ref{eq:o2_conservation3b}, the problem to be solved reads:

\begin{align}
\us&=\mathbf{u}_{\text{imposed}}\,\text{on}\, \partial\Omega_u \label{eq:oxy:dirichlet}\\
{\color{black} p^\alpha} &= p_{\text{imposed}}\,\text{on}\, \partial\Omega_p^\alpha \label{eq:oxy:bc_pressure}
\end{align}
\begin{align}
\begin{split}
&\int_\Omega C_{m,c} \frac{\mathrm{D}^s \plc}{\mathrm{D}t} q^c \ddroit \Omega \\
&+ \int_\Omega  S^{c} \frac{k^{b}}{\mu^{b}} \nabla \pb \nabla q^c \ddroit \Omega \\
& + \int_\Omega \frac{k^{c}}{\mu^{c}}\mathbf{\nabla}\pl \nabla q^c \ddroit \Omega \\ 
& - \int_\Omega \frac{k^{c}}{\mu^{c}}\mathbf{\nabla}\plc \nabla q^c \ddroit \Omega \\ 
&+ \int_\Omega S^c \nabla \cdot \left(\frac{\mathrm{D}^s \us}{\mathrm{D}t} \right) q^c \ddroit \Omega = 0,\,\forall~q^c\in {L}_0^2(\Omega)
\end{split}  \label{eq:oxy:variationalmasscell}
\end{align}
\begin{align}
\begin{split}
&\int_\Omega \frac{k^{b}}{\mu^{b}} \nabla \pb \nabla q^l \ddroit \Omega \\
&+  \int_\Omega \left[\frac{k^{c}}{\mu^{c}} + \frac{k^{l}}{\mu^{l}} \right] \nabla \pl \nabla q^l \ddroit \Omega \\
&- \int_\Omega \frac{k^{c}}{\mu^{c}} \nabla \plc \nabla q^l \ddroit \Omega \\
& + \int_\Omega \nabla \cdot \left(\frac{\mathrm{D}^s \us}{\mathrm{D}t} \right) q^l \ddroit \Omega = 0,\,\forall~q^l\in {L}_0^2(\Omega)
\end{split}  \label{eq:oxy:variationalmassIF}
\end{align}
\begin{align}
\begin{split}
&\int_\Omega C_{e,p} \frac{\mathrm{D}^s \pl}{\mathrm{D}t} q^b \ddroit \Omega \\
&- \int_\Omega C_{e,p}C_{state} \frac{\mathrm{D}^s \plc}{\mathrm{D}t} q^b \ddroit \Omega \\
&- \int_\Omega C_{e,p} \frac{\mathrm{D}^s \pb}{\mathrm{D}t} q^b \ddroit \Omega \\
&+ \int_\Omega \frac{k^{b}}{\mu^{b}}\mathbf{\nabla}\pb \nabla q^b \ddroit \Omega \\ 
 &+ \int_\Omega \varepsilon^b \nabla \cdot \left(\frac{\mathrm{D}^s \us}{\mathrm{D}t} \right) q^b \ddroit \Omega = 0,\,\forall~q^b\in {L}_0^2(\Omega)
\end{split}  \label{eq:oxy:variationalmassblood}
\end{align}
\begin{align}
\begin{split}
&\int_\Omega \mathbf{t}^{\text{eff}}(\us) : \nabla \mathbf{w} \ddroit \Omega  \\
&- \int_\Omega \left(1-\varepsilon^b\right) \left(\pl - S^c \plc \right) \nabla \cdot \mathbf{w} \ddroit \Omega \\ 
&- \int_\Omega \varepsilon^b \pb \nabla \cdot \mathbf{w} \ddroit \Omega \\
&- \int_{\Gamma_s} \mathbf{t}^{\text{imposed}} \cdot \mathbf{w} \ddroit \Gamma_s =0, \forall~\mathbf{w}\in[H^1(\Omega)]^3 \label{eq:oxy:o2_variational}
\end{split}
\end{align}
\begin{align}
\begin{split}
&\int_\Omega (1-S^{c})\varepsilon\frac{\mathrm{D}^s}{\mathrm{D}t}(\omegaodeux) w_o \ddroit \Omega \\
&- \int_\Omega \frac{k^{l}}{\mu^{l}} \nabla \omegaodeux \nabla \pl  w_o \ddroit \Omega \\
&+ \int_\Omega (1-S^{c})\varepsilon D_{eff}^{O2,l} \nabla \omegaodeux  \nabla w_o \ddroit \Omega\\
&= \int_\Omega \frac{1}{\rho^{l}}\left(\overset{O2,b\rightarrow O2,l}{M} 
- \overset{O2,l\rightarrow O2,c}{M}\right)  w_o \ddroit \Omega,\,\forall~w_o\in {L}_0^2(\Omega)
\end{split}
\label{eq:o2_variational}\end{align}

\section{Individual LDF signals for the 10 subjects}
\label{appendix:LDF}

This appendix provides the median LDF and temperature signals with the IQR corridor for each of the patients. 

\begin{figure}[ht!]
    \centering
    \includegraphics[width=0.495\linewidth]{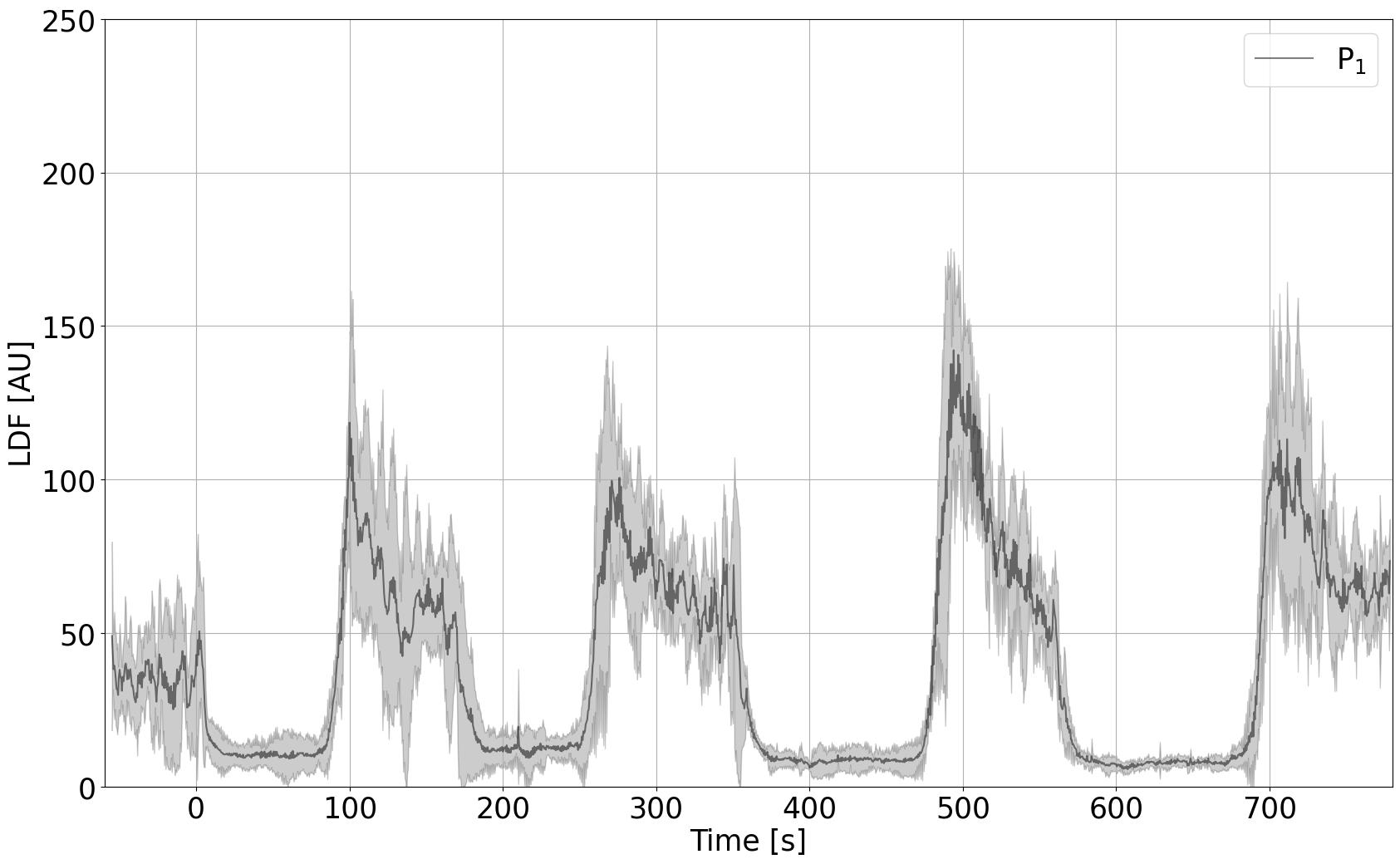}
    \includegraphics[width=0.495\linewidth]{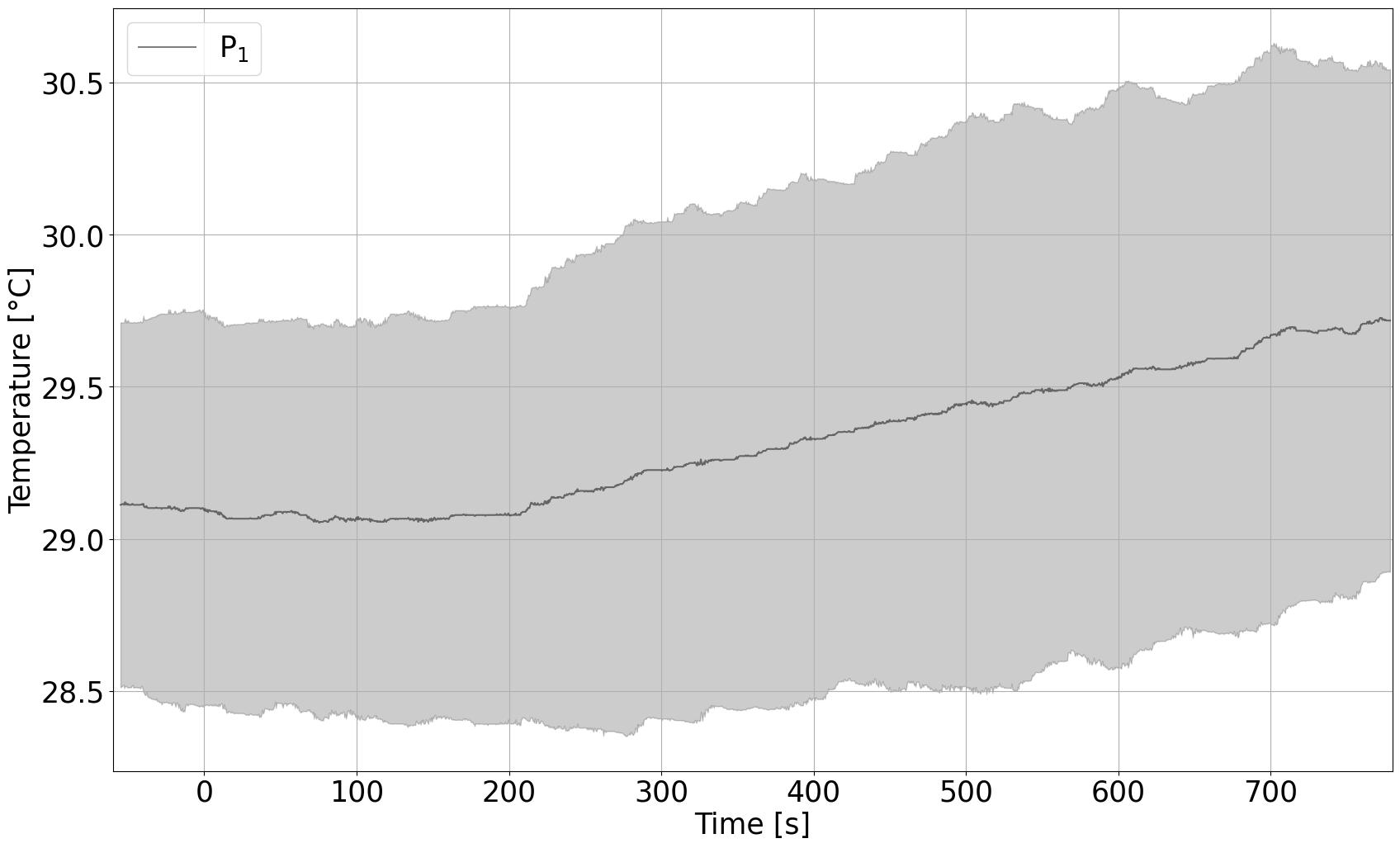}
\end{figure}

\begin{figure}[ht!]
    \centering
    \includegraphics[width=0.495\linewidth]{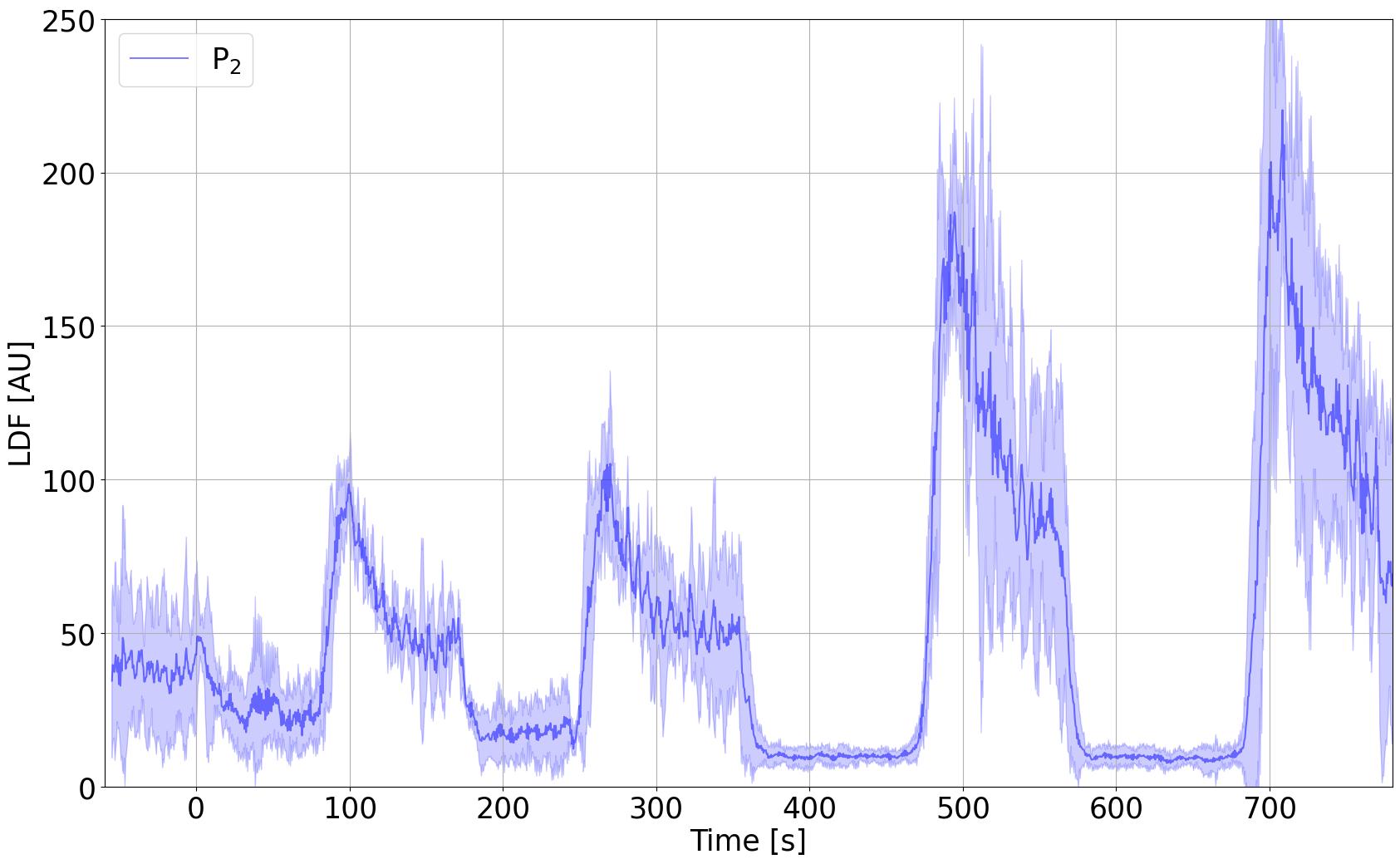}
    \includegraphics[width=0.495\linewidth]{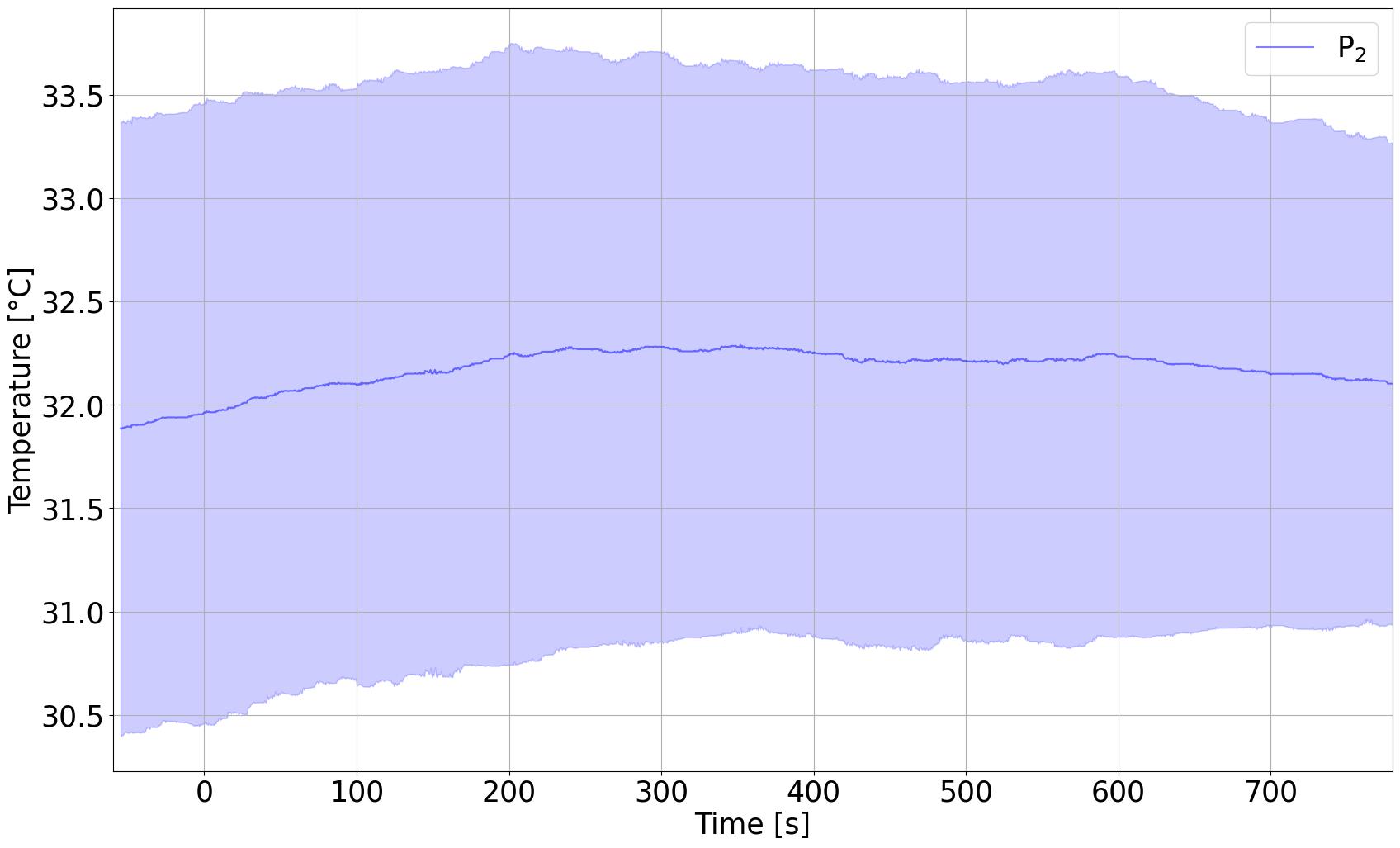}
\end{figure}

\begin{figure}[ht!]
    \centering
    \includegraphics[width=0.495\linewidth]{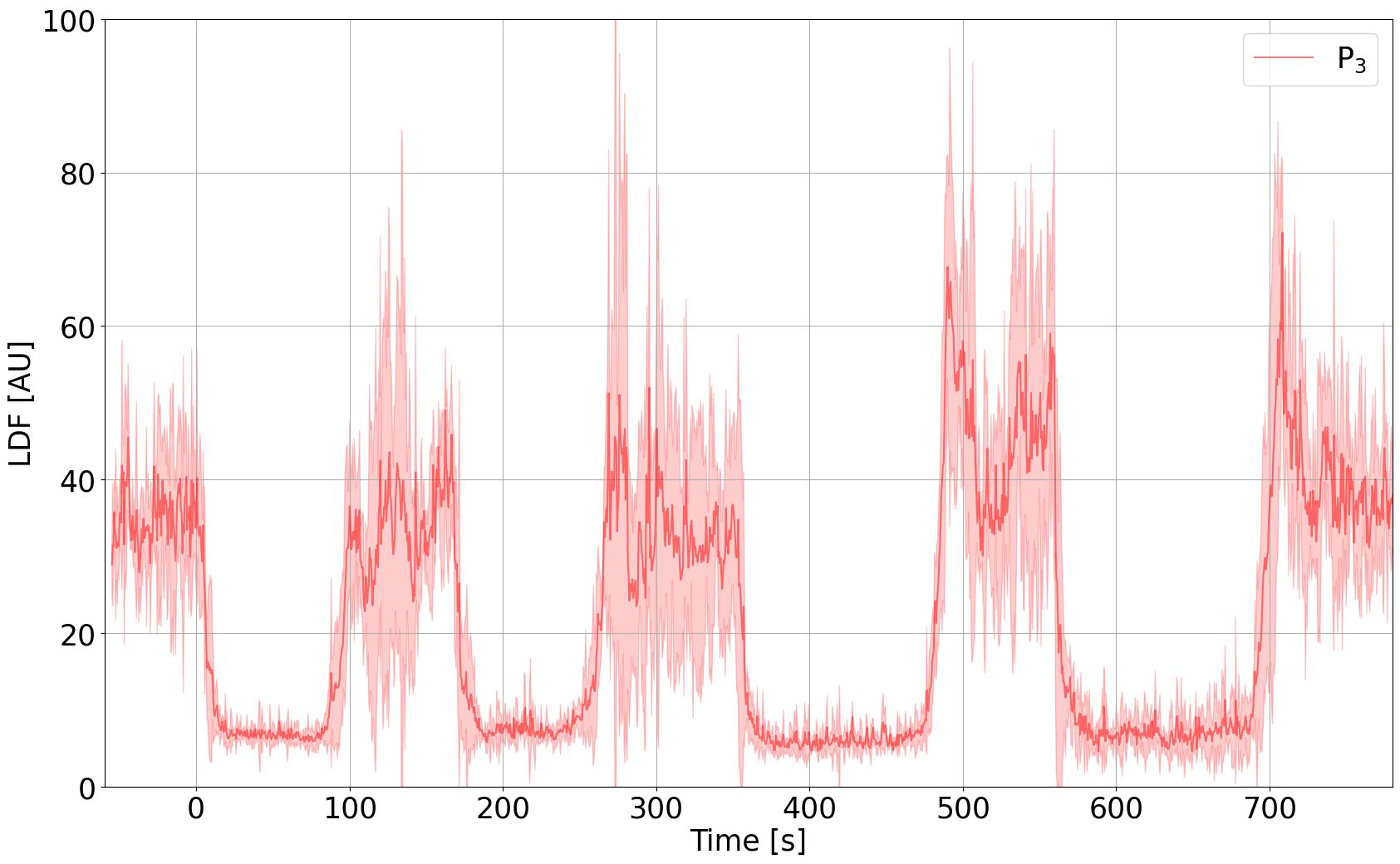}
    \includegraphics[width=0.495\linewidth]{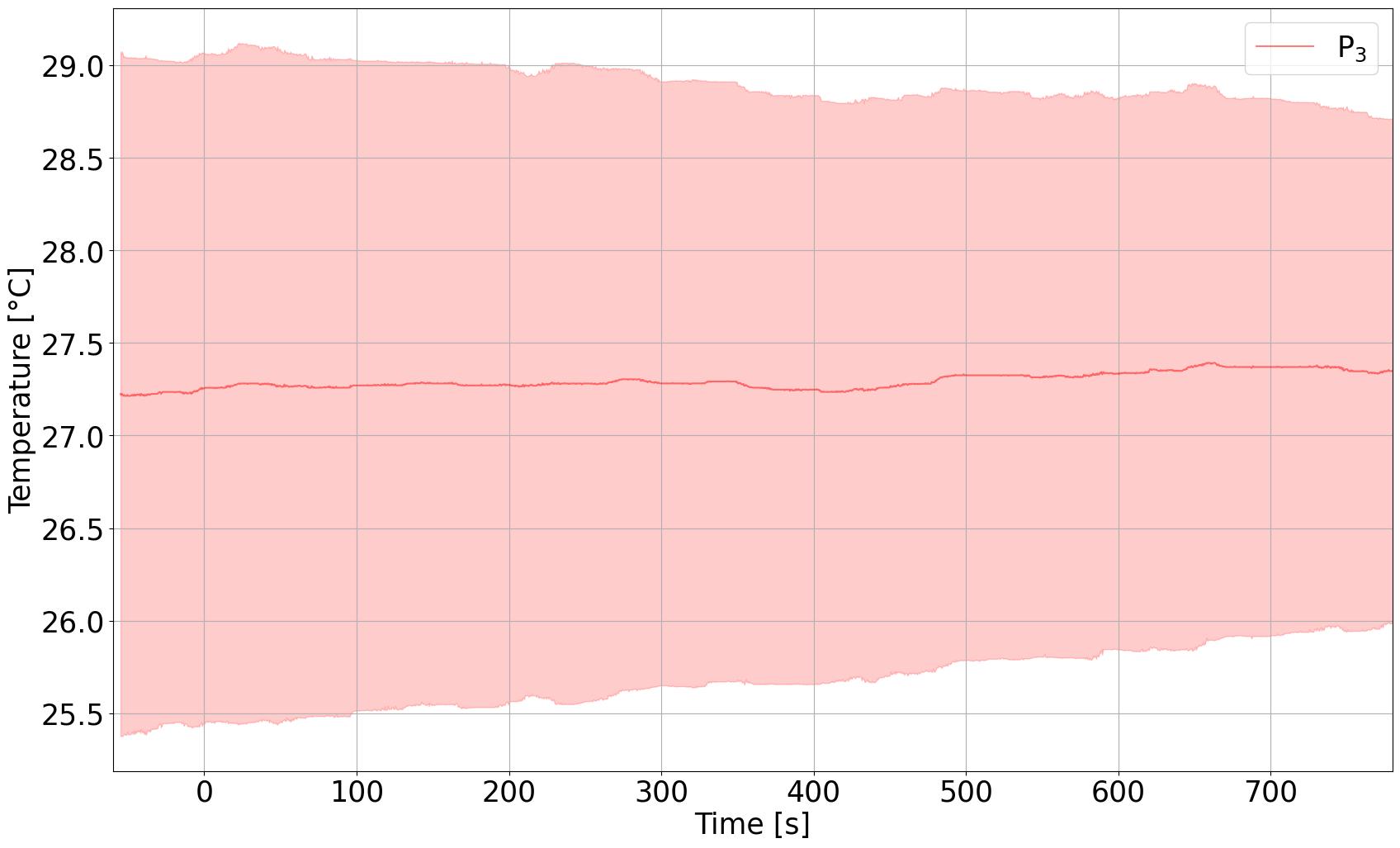}
\end{figure}

\begin{figure}[ht!]
    \centering
    \includegraphics[width=0.495\linewidth]{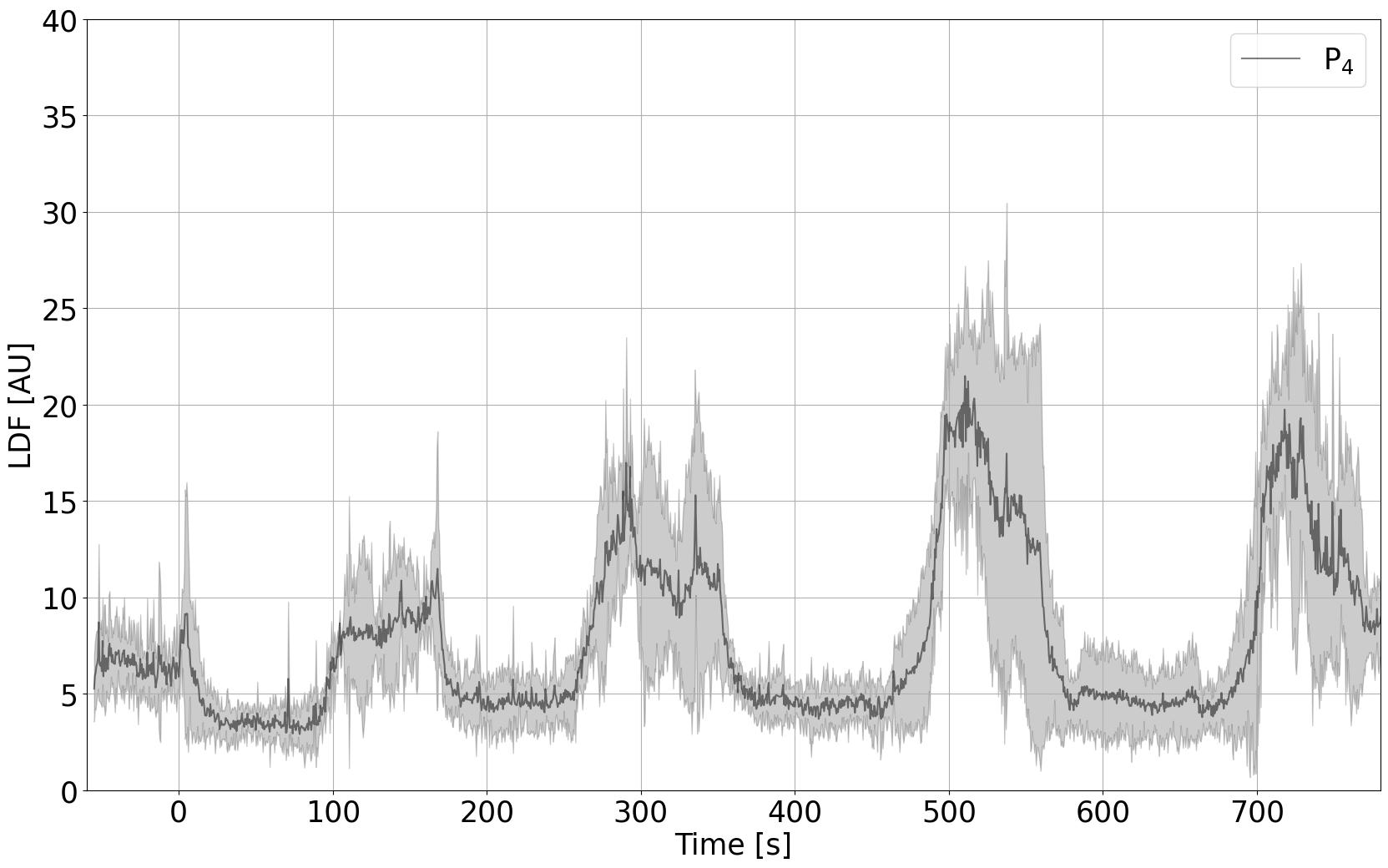}
    \includegraphics[width=0.495\linewidth]{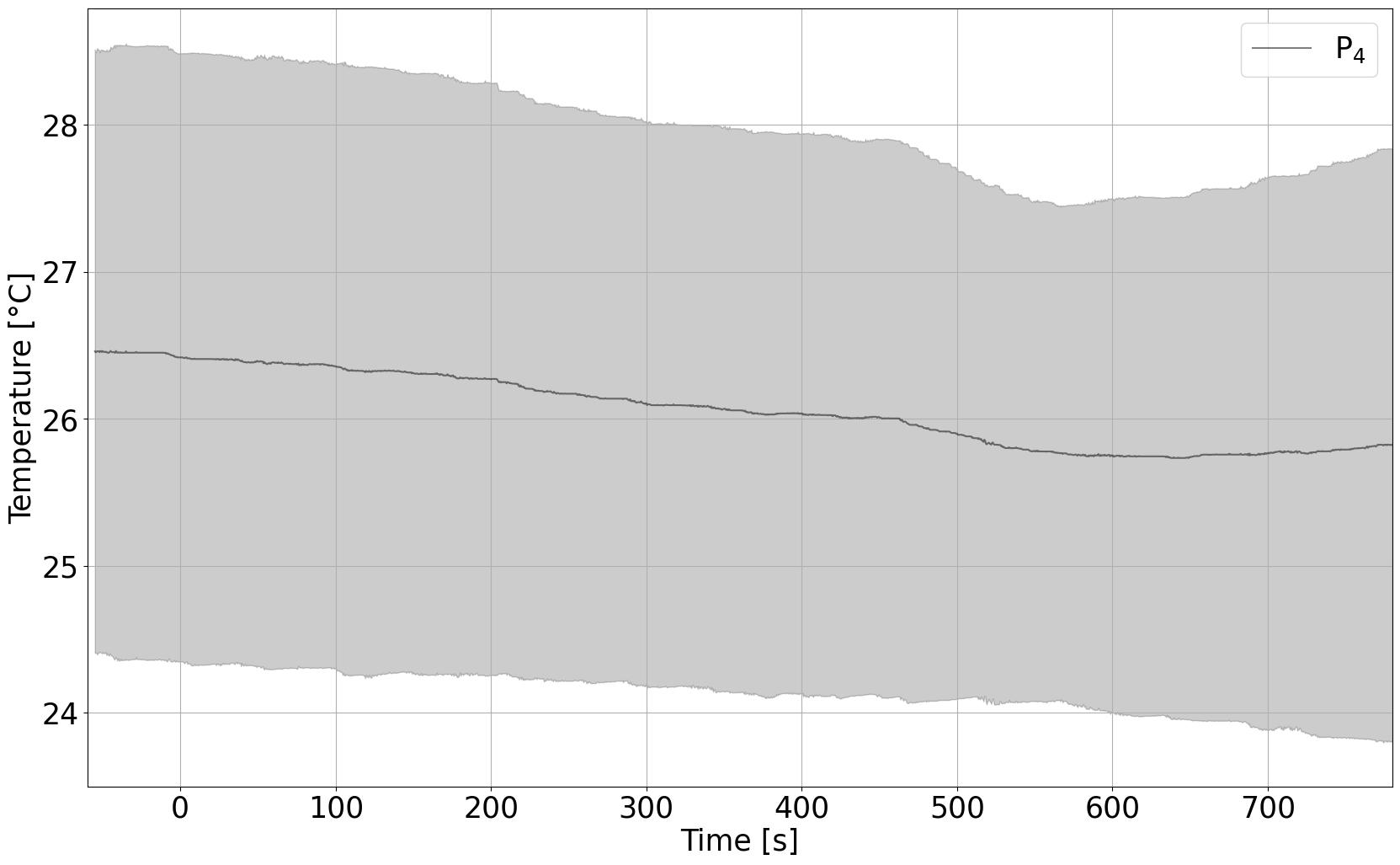}
\end{figure}

\begin{figure}[ht!]
    \centering
    \includegraphics[width=0.495\linewidth]{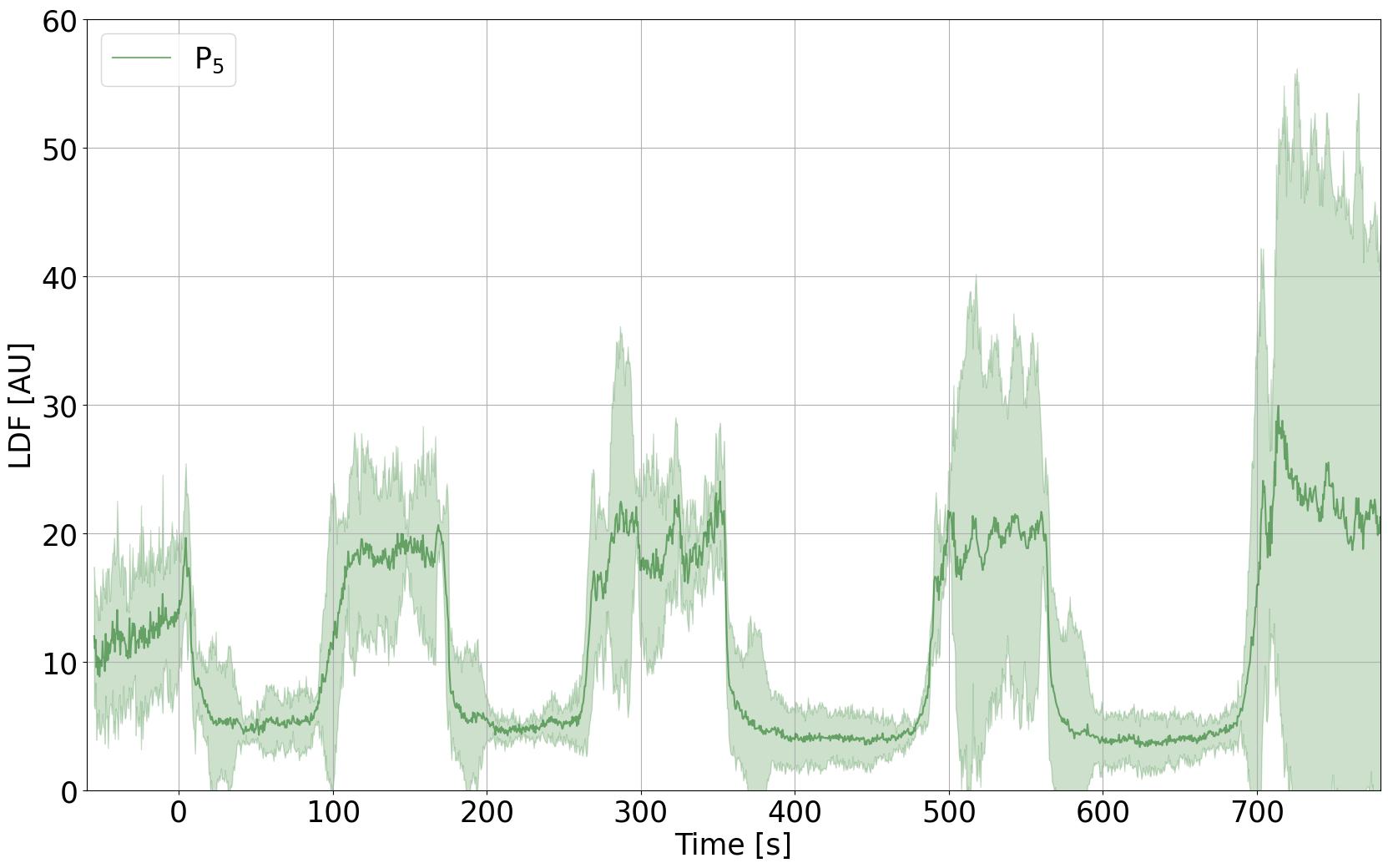}
    \includegraphics[width=0.495\linewidth]{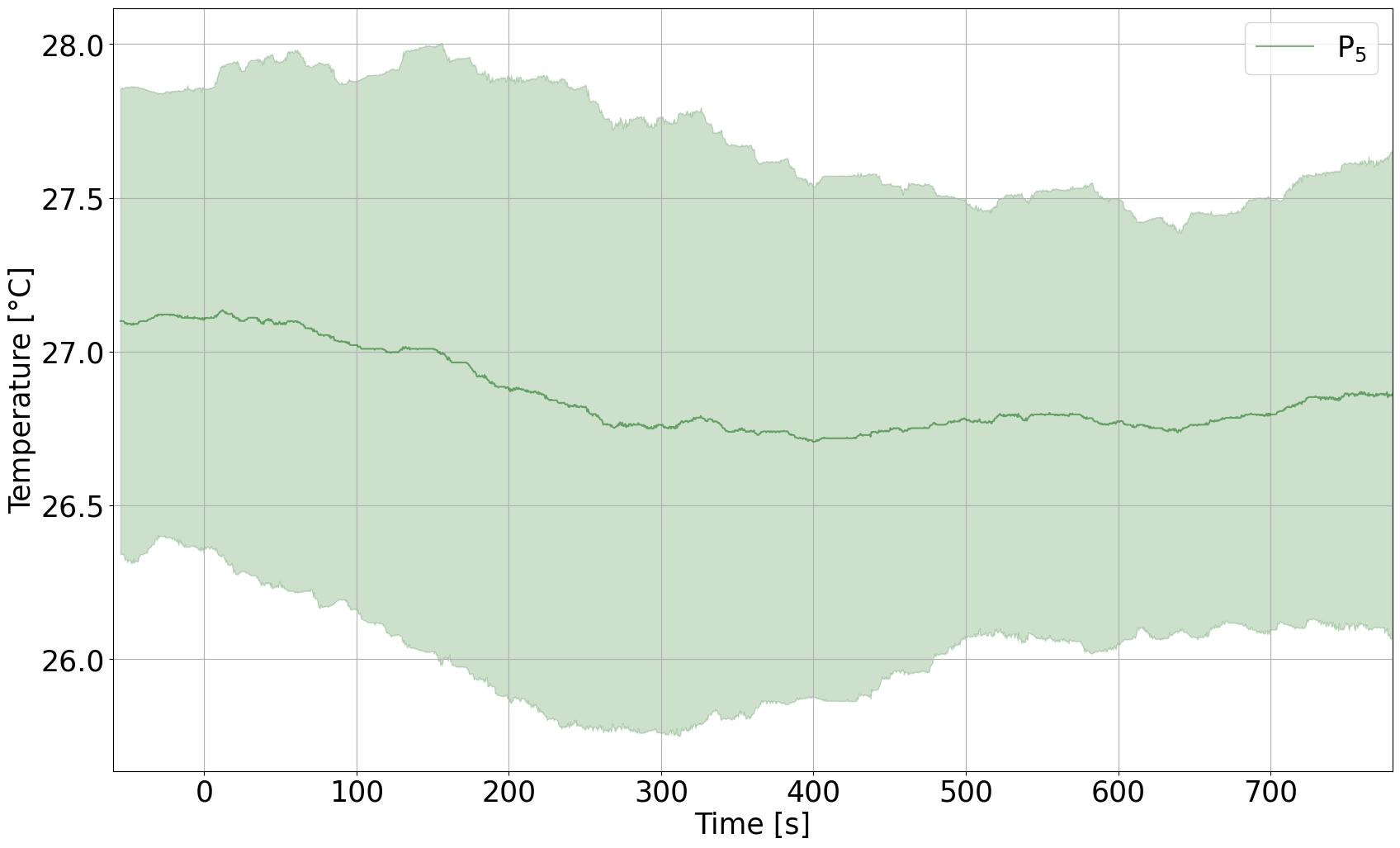}
\end{figure}

\begin{figure}[ht!]
    \centering
    \includegraphics[width=0.495\linewidth]{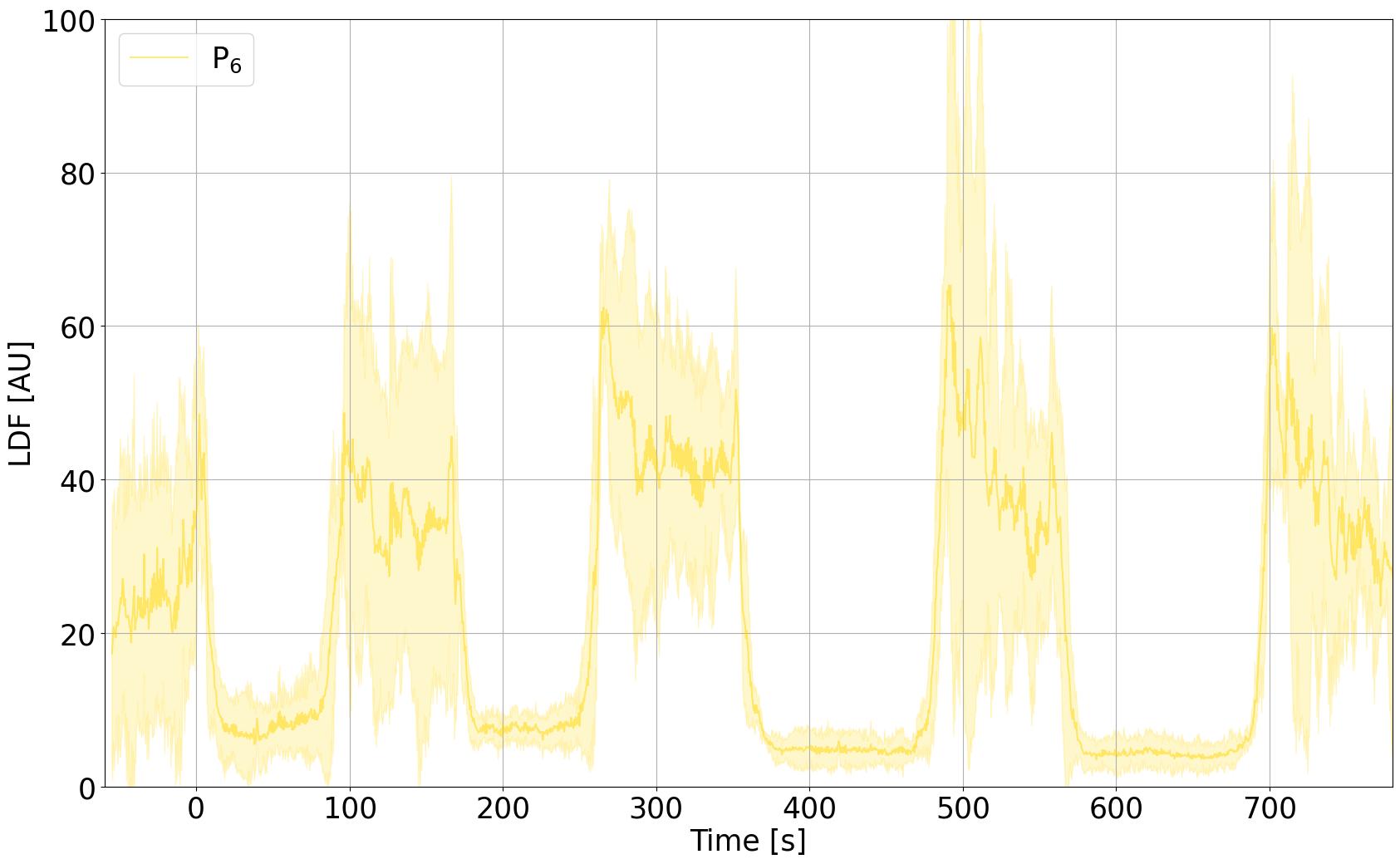}
    \includegraphics[width=0.495\linewidth]{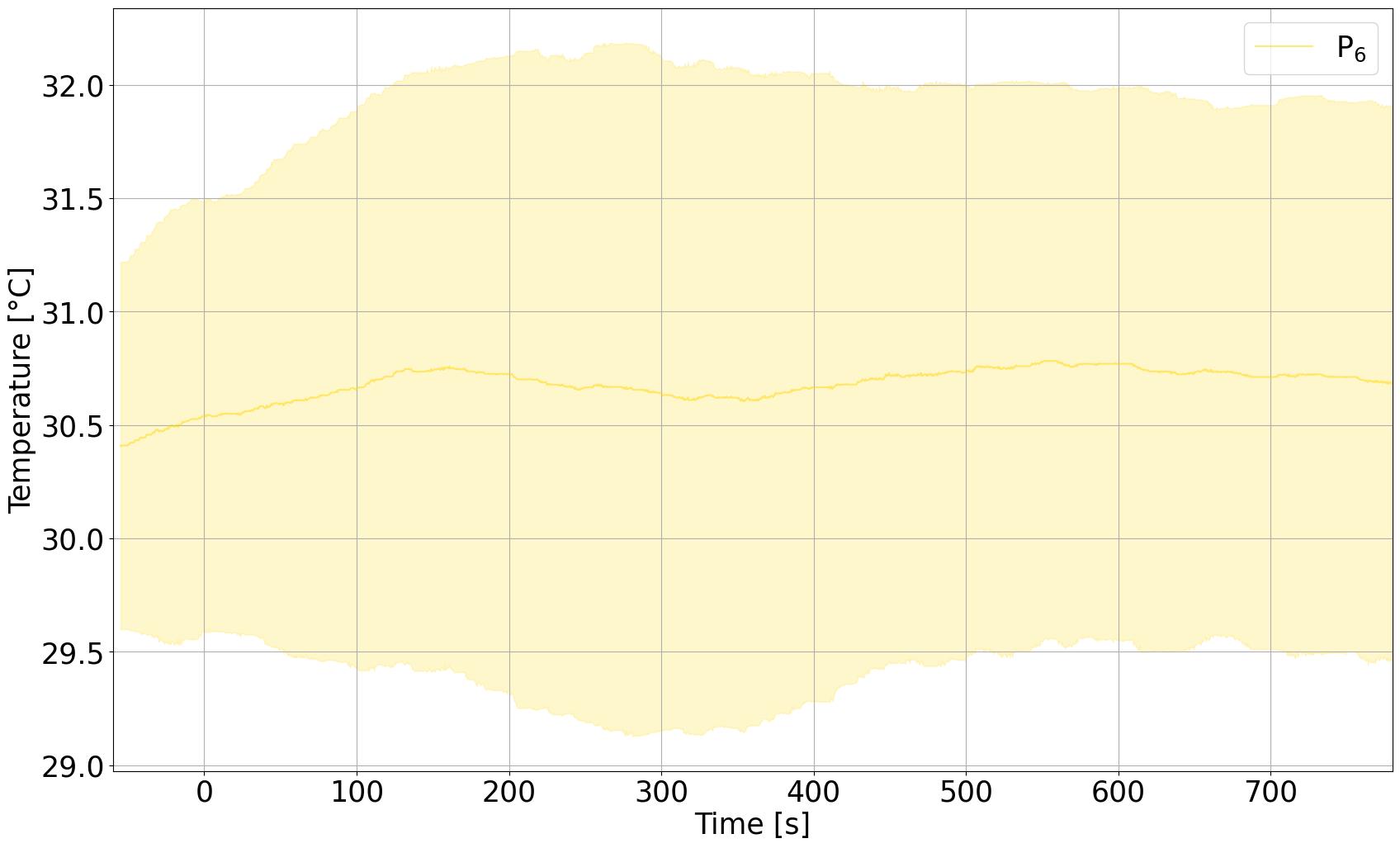}
\end{figure}

\begin{figure}[ht!]
    \centering
    \includegraphics[width=0.495\linewidth]{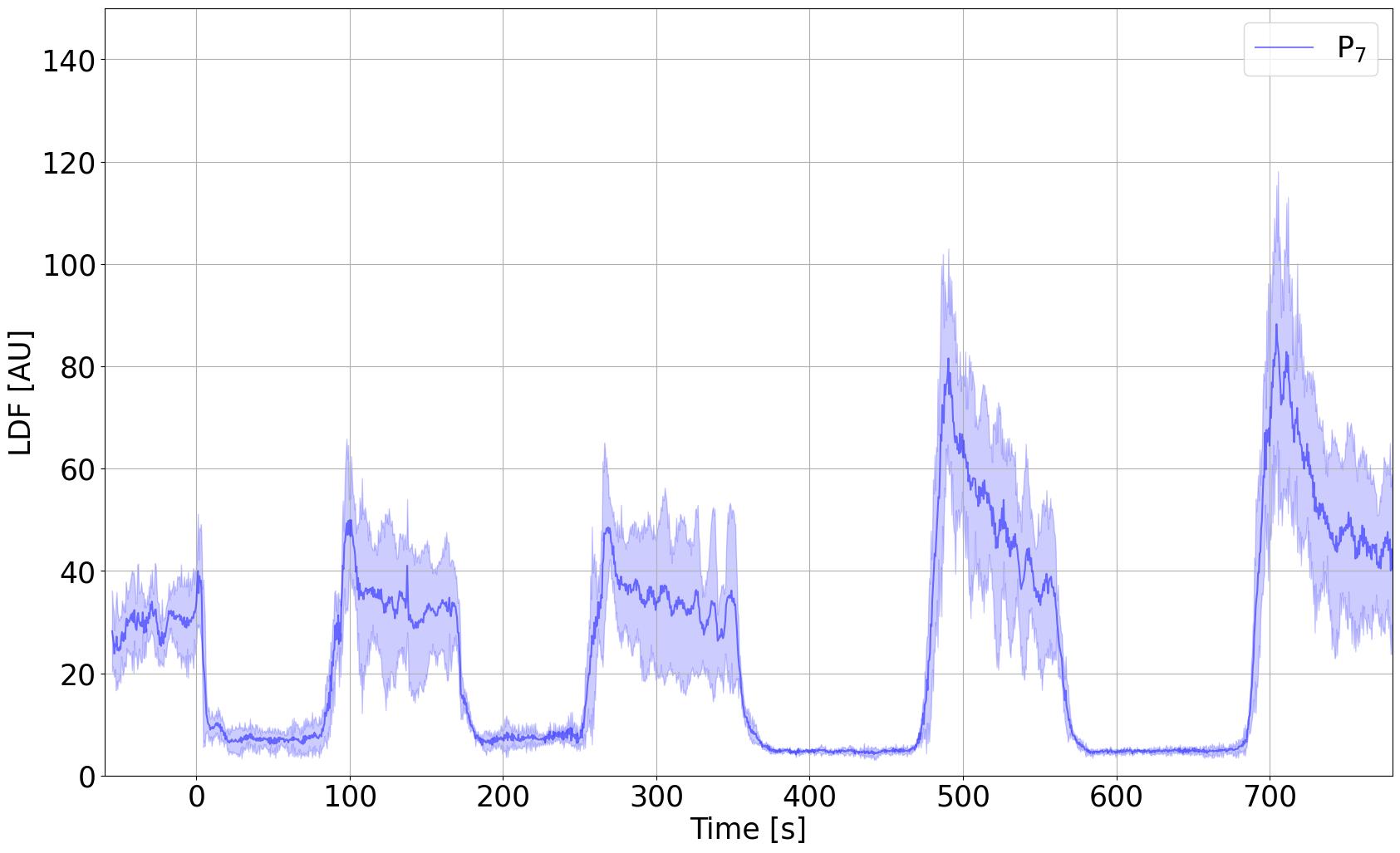}
    \includegraphics[width=0.495\linewidth]{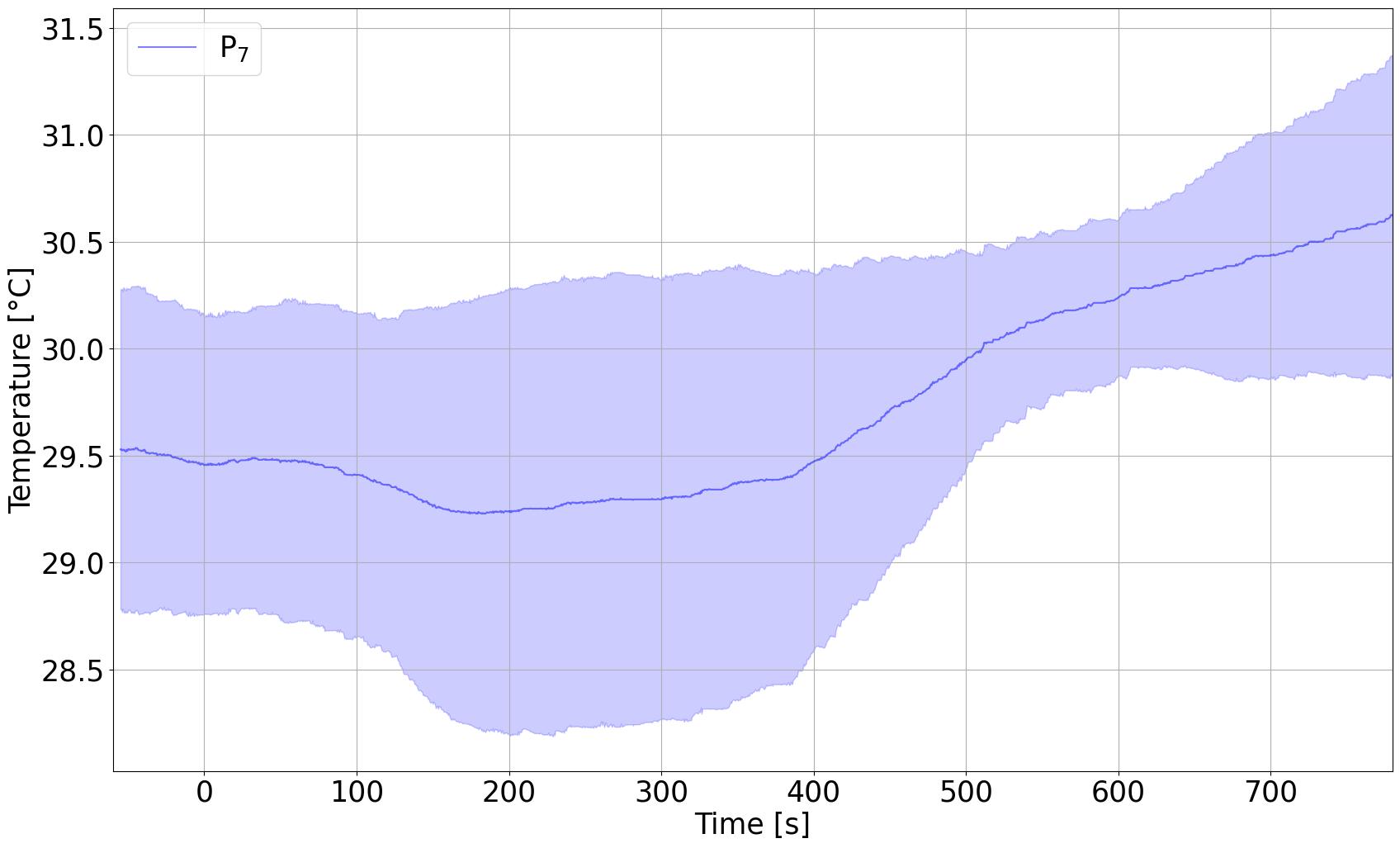}
\end{figure}

\begin{figure}[ht!]
    \centering
    \includegraphics[width=0.495\linewidth]{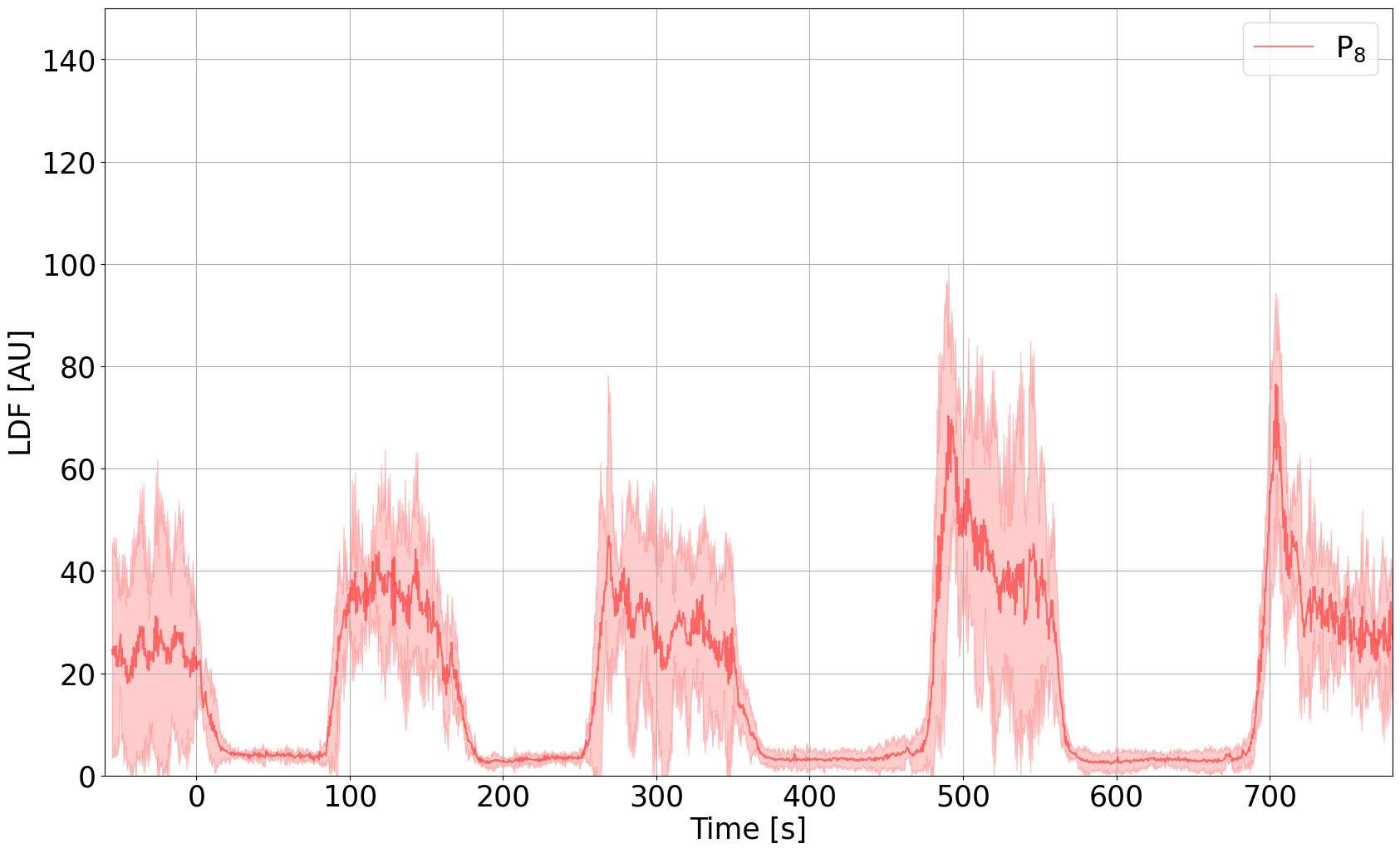}
    \includegraphics[width=0.495\linewidth]{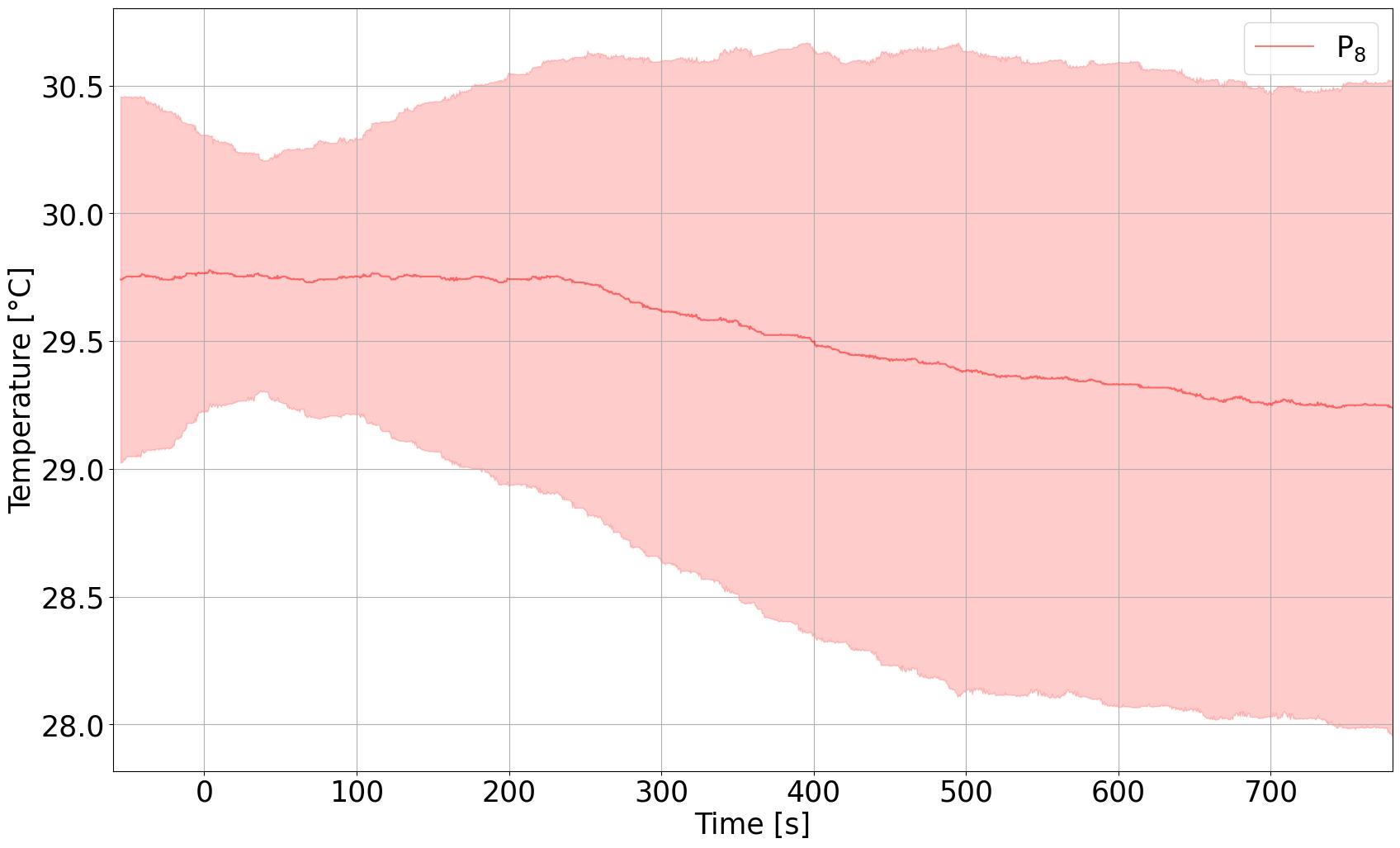}
\end{figure}

\begin{figure}[ht!]
    \centering
    \includegraphics[width=0.495\linewidth]{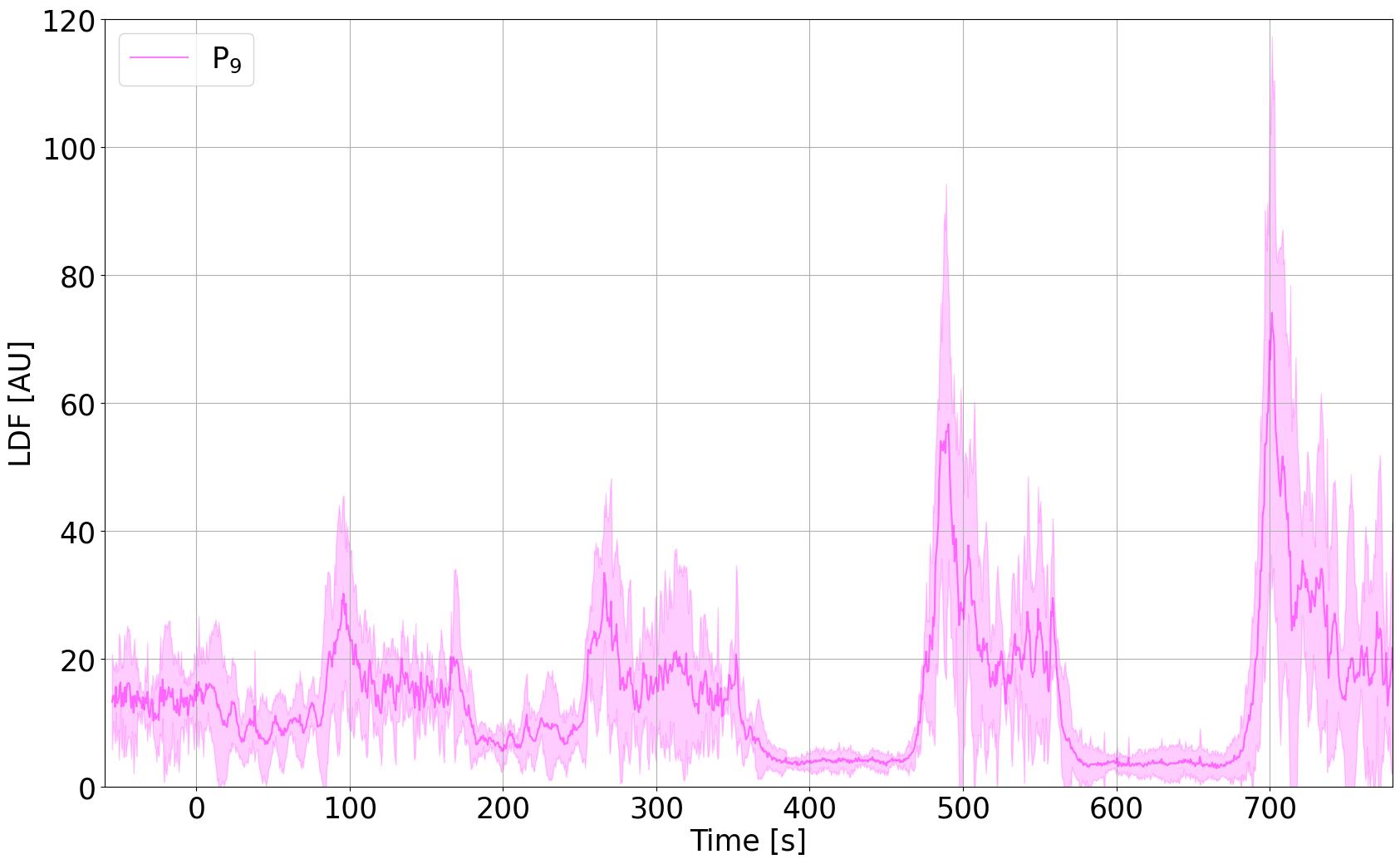}
    \includegraphics[width=0.495\linewidth]{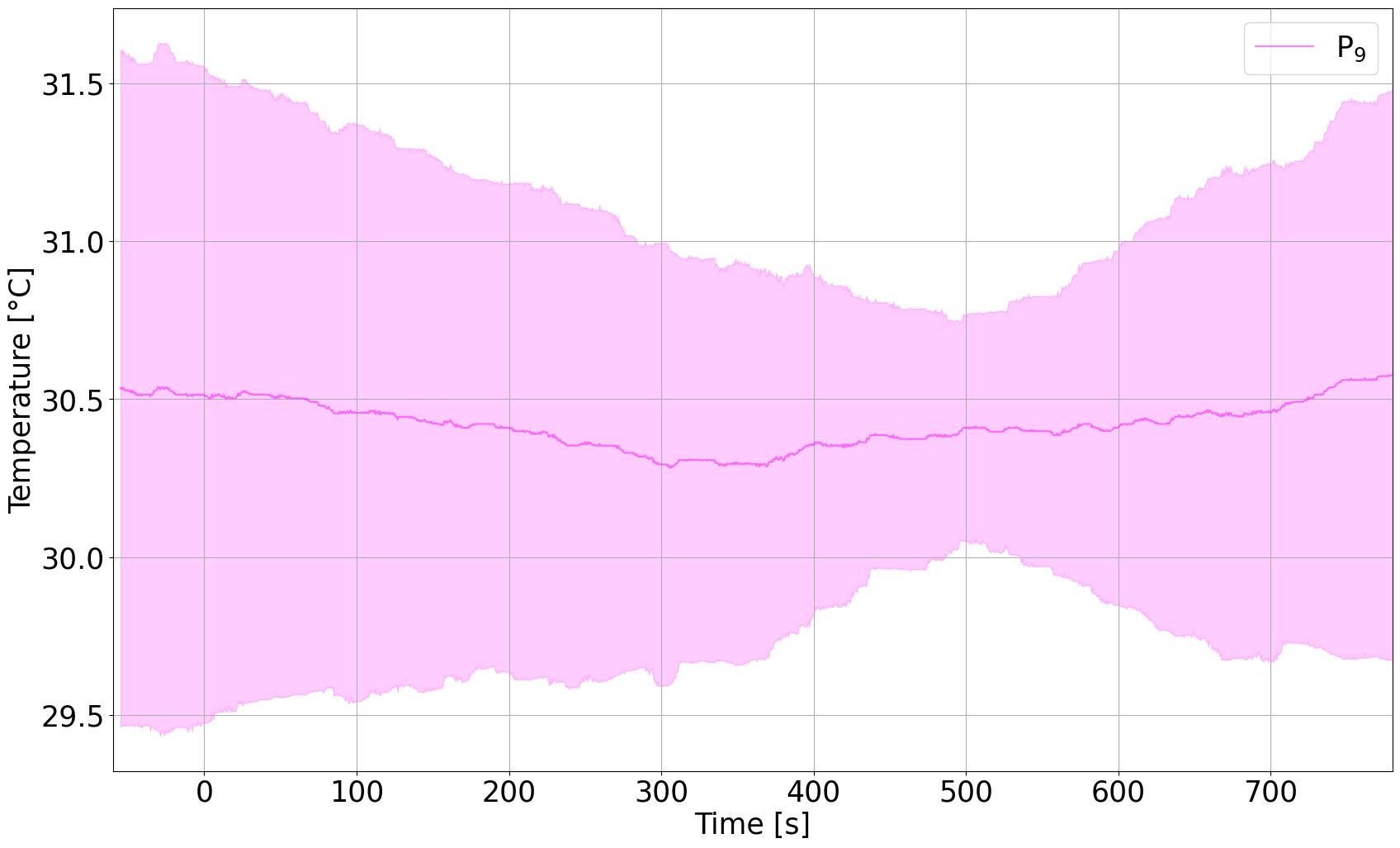}
\end{figure}

\begin{figure}[ht!]
    \centering
    \includegraphics[width=0.495\linewidth]{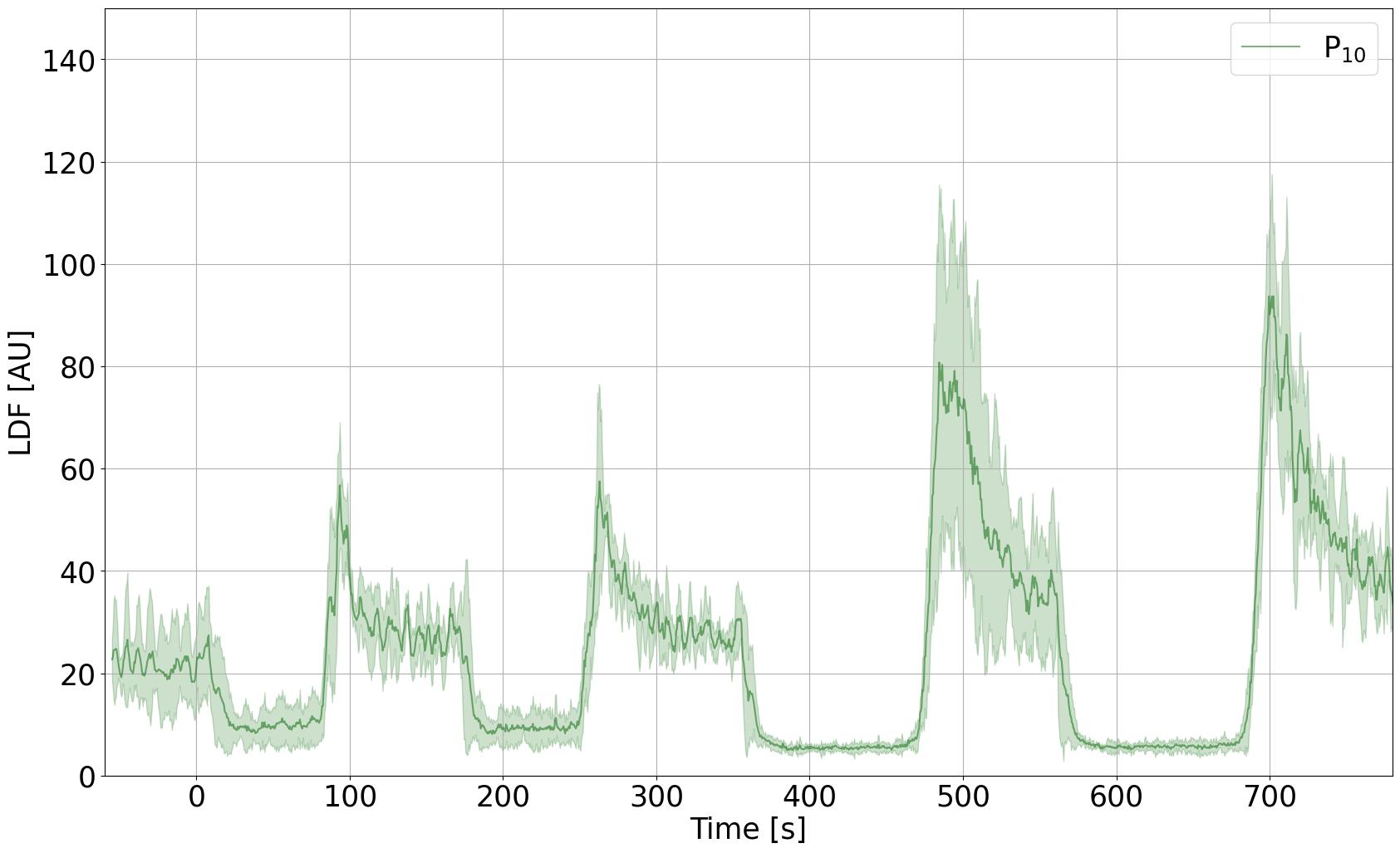}
    \includegraphics[width=0.495\linewidth]{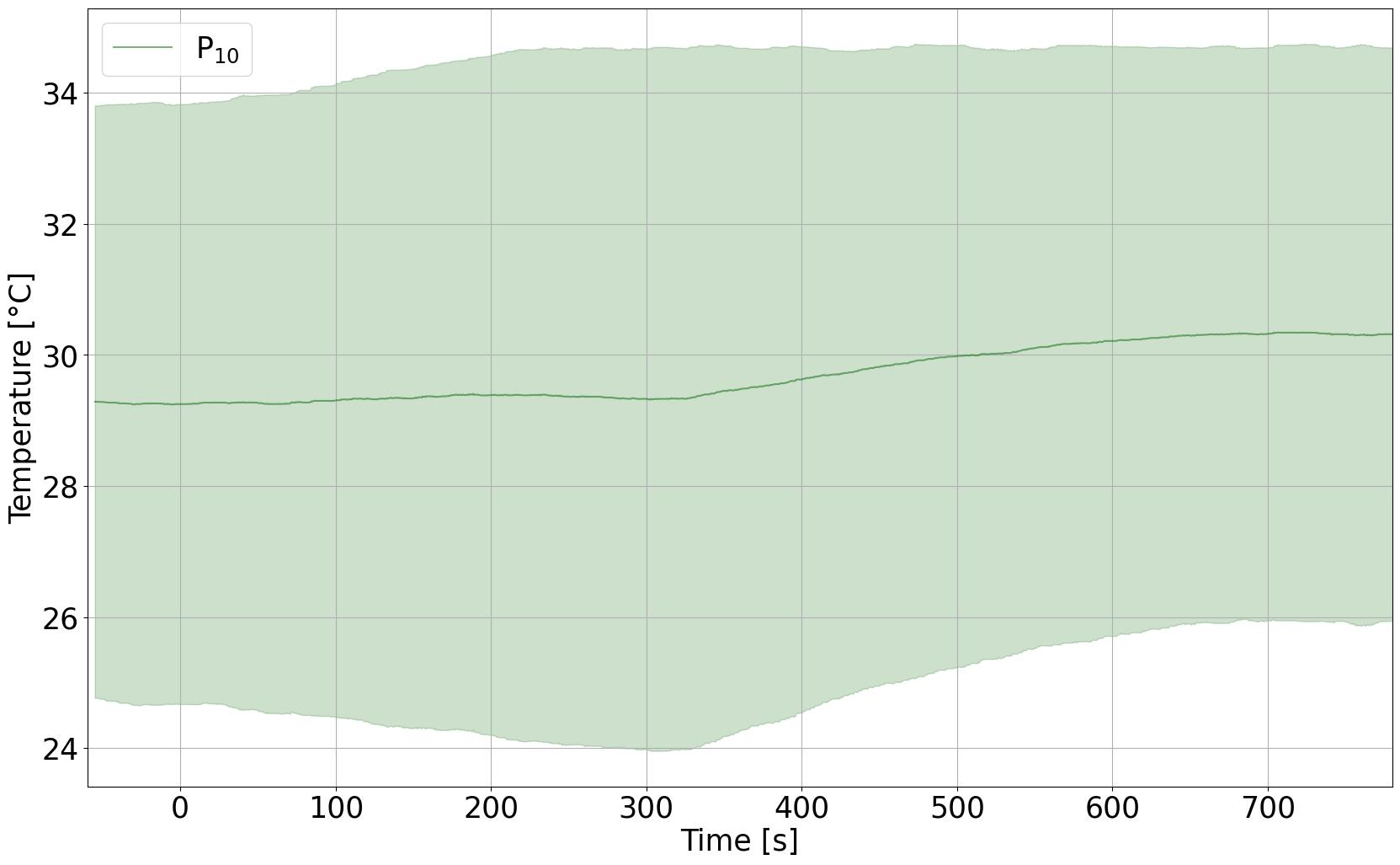}
\end{figure}

\begin{figure}[ht!]
    \centering
    \includegraphics[width=0.495\linewidth]{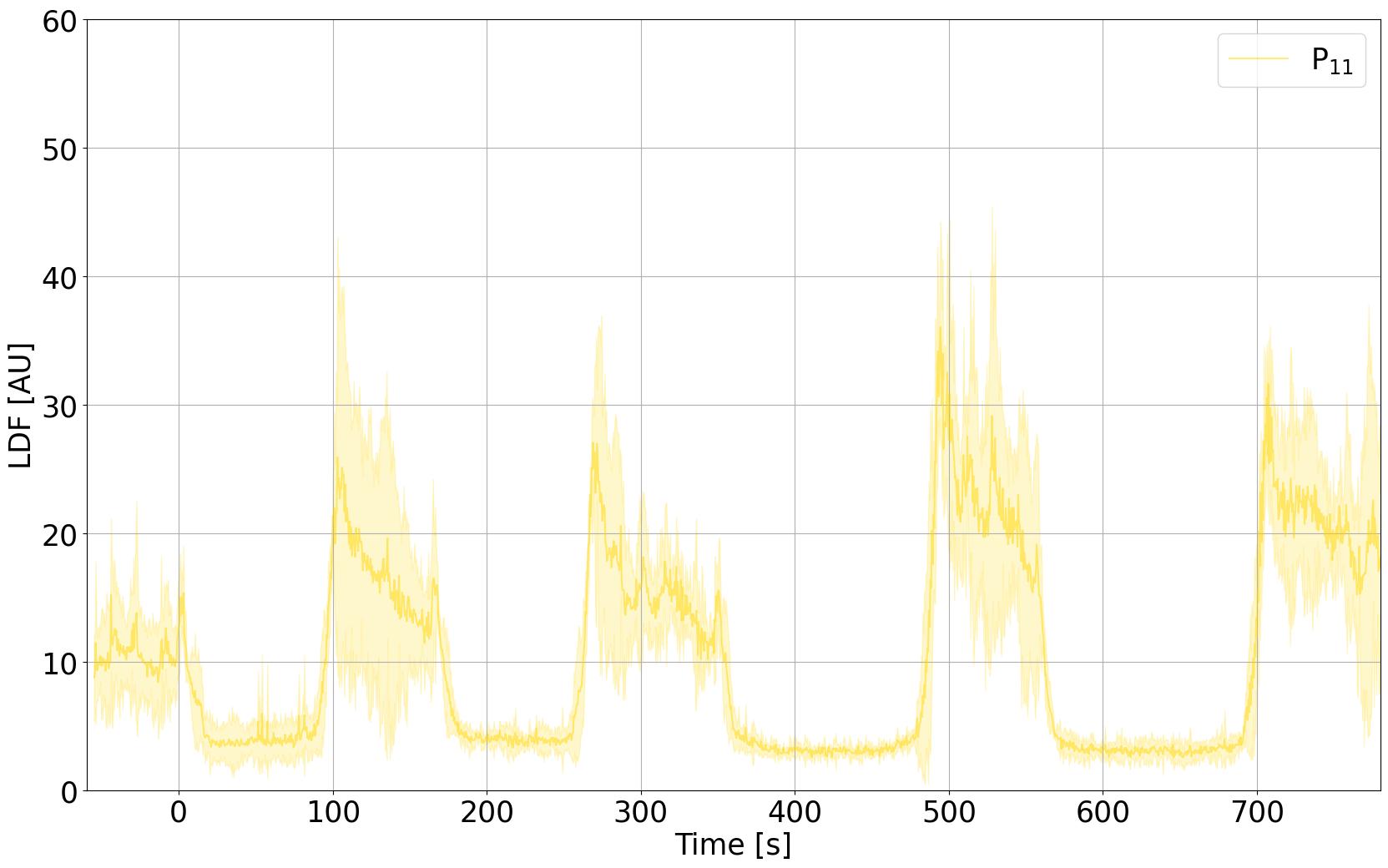}
    \includegraphics[width=0.495\linewidth]{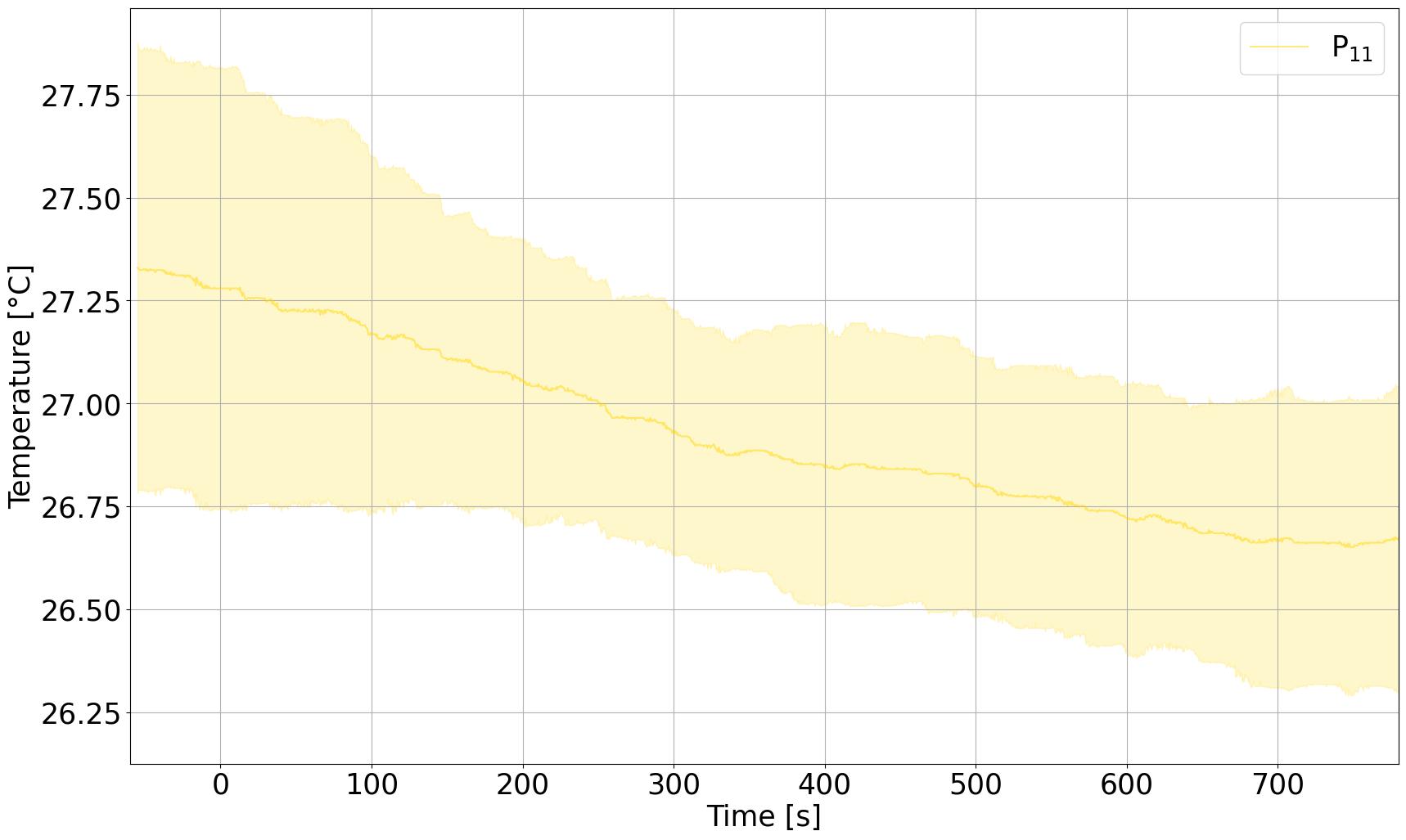}
\end{figure}

\newpage

\section{Sensitivity Curves}
\label{sec:appendix:sens:curves}
\begin{figure}[ht!]
    \centering
    \includegraphics[width=0.4\linewidth]{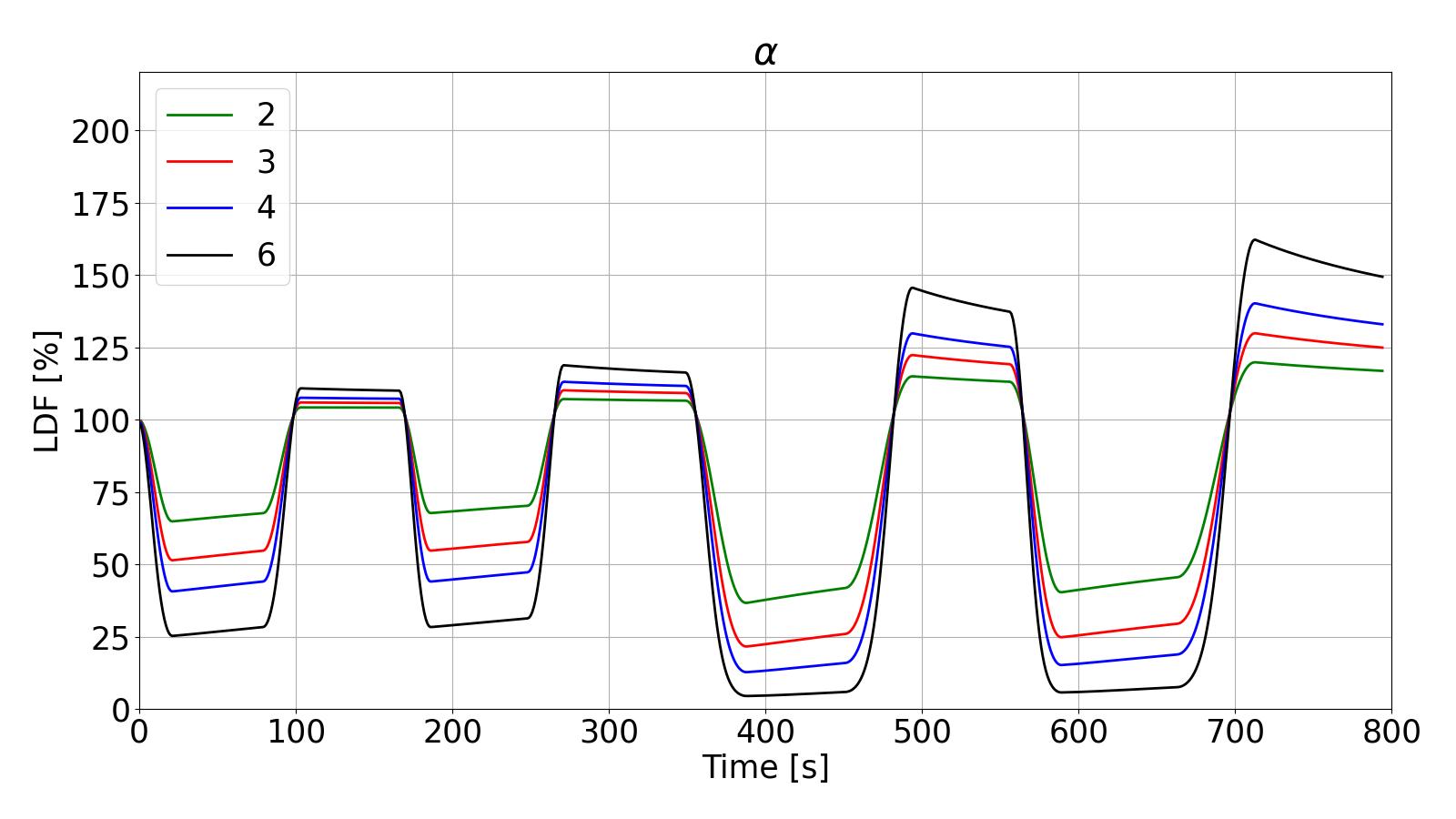}
    \includegraphics[width=0.4\linewidth]{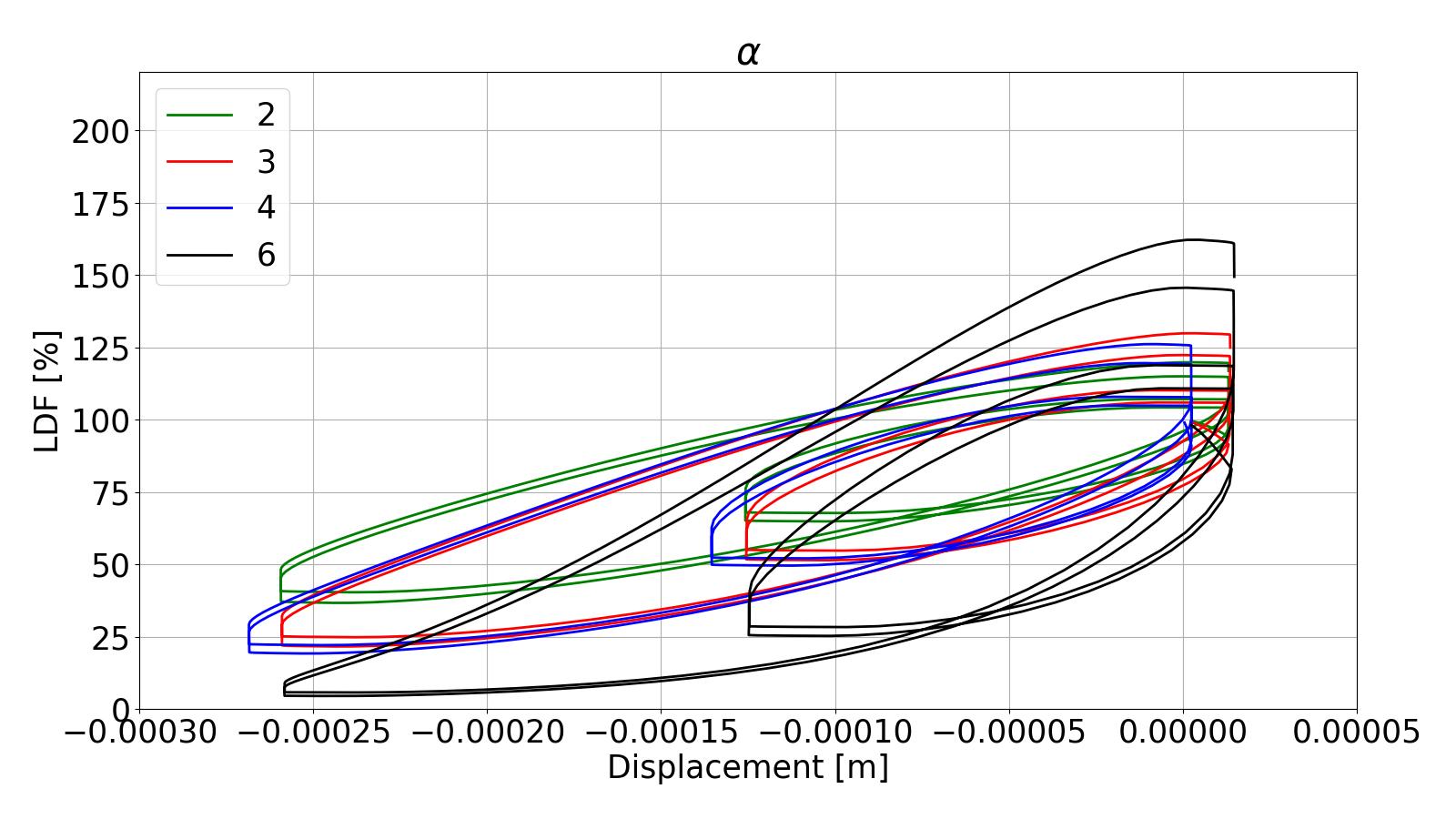}
    \includegraphics[width=0.4\linewidth]{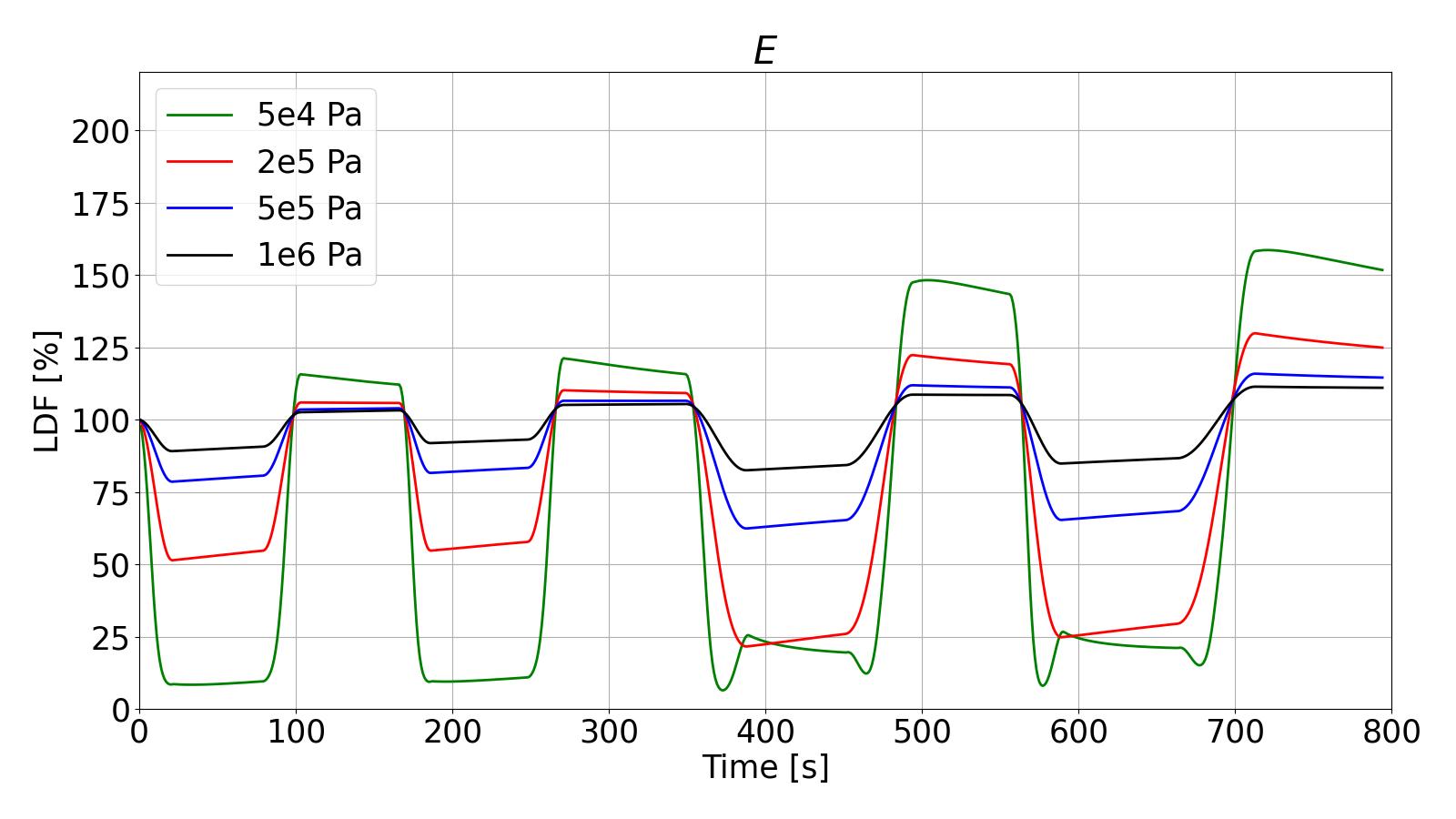}
    \includegraphics[width=0.4\linewidth]{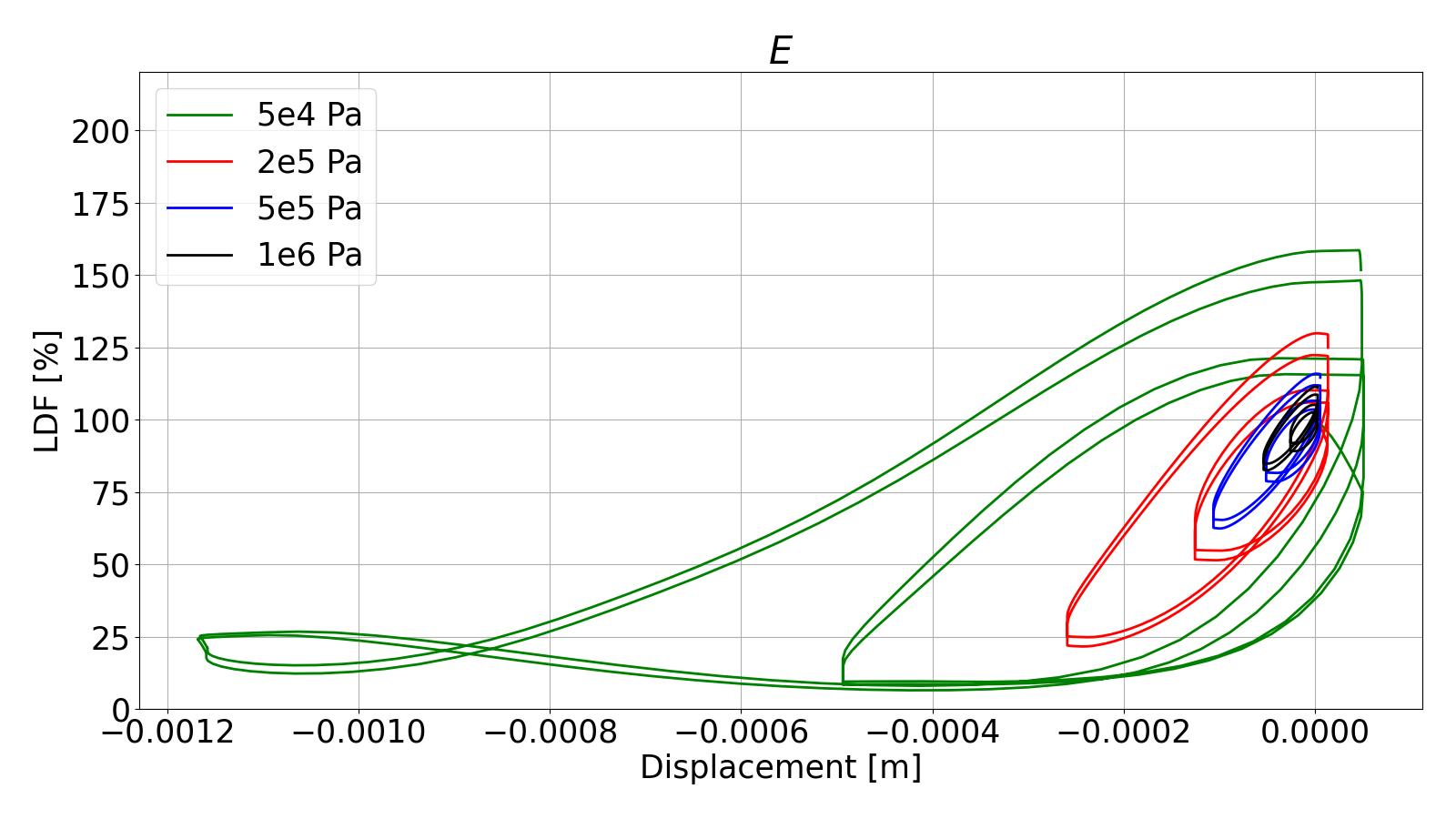}
    \includegraphics[width=0.4\linewidth]{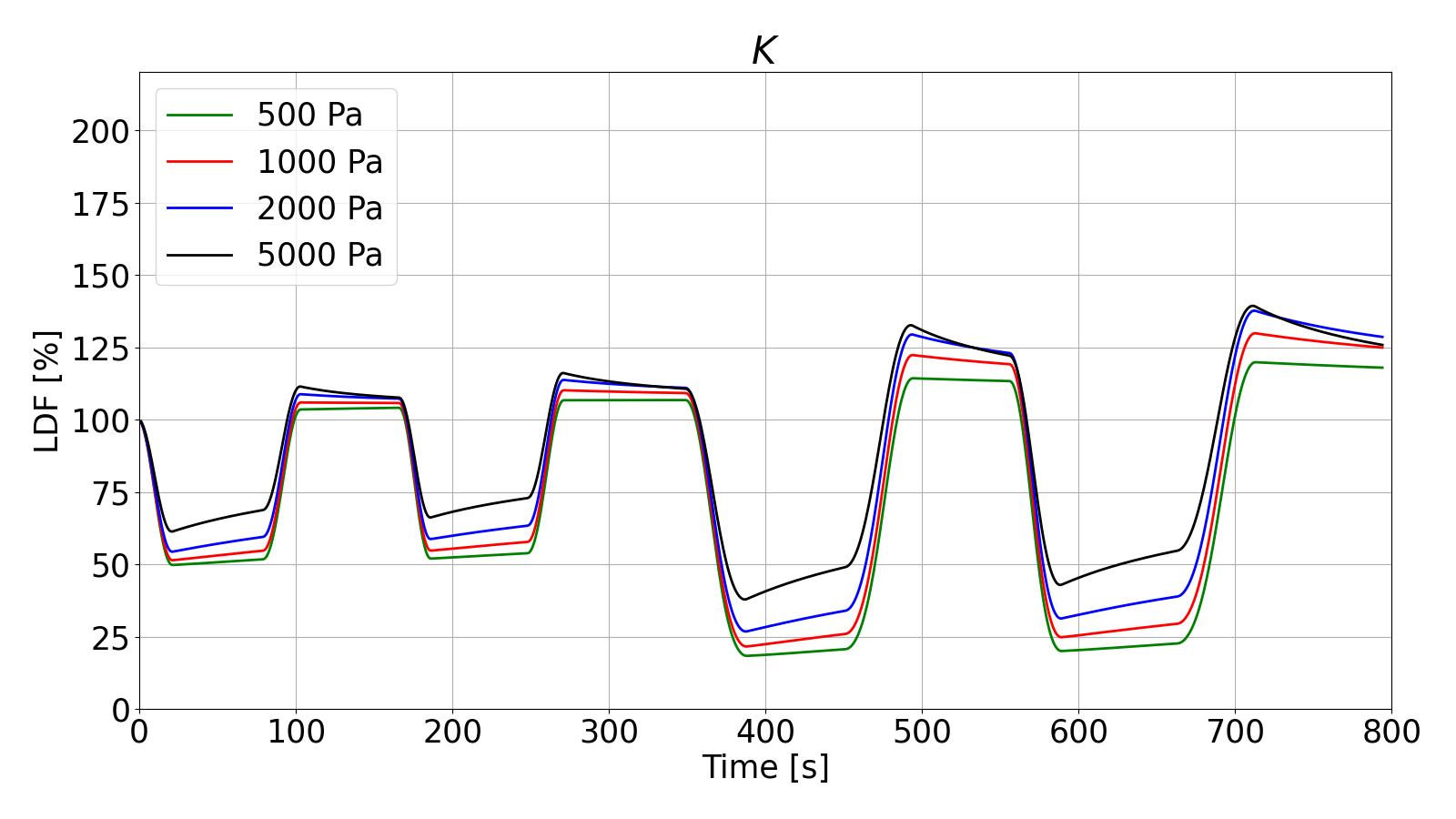}
    \includegraphics[width=0.4\linewidth]{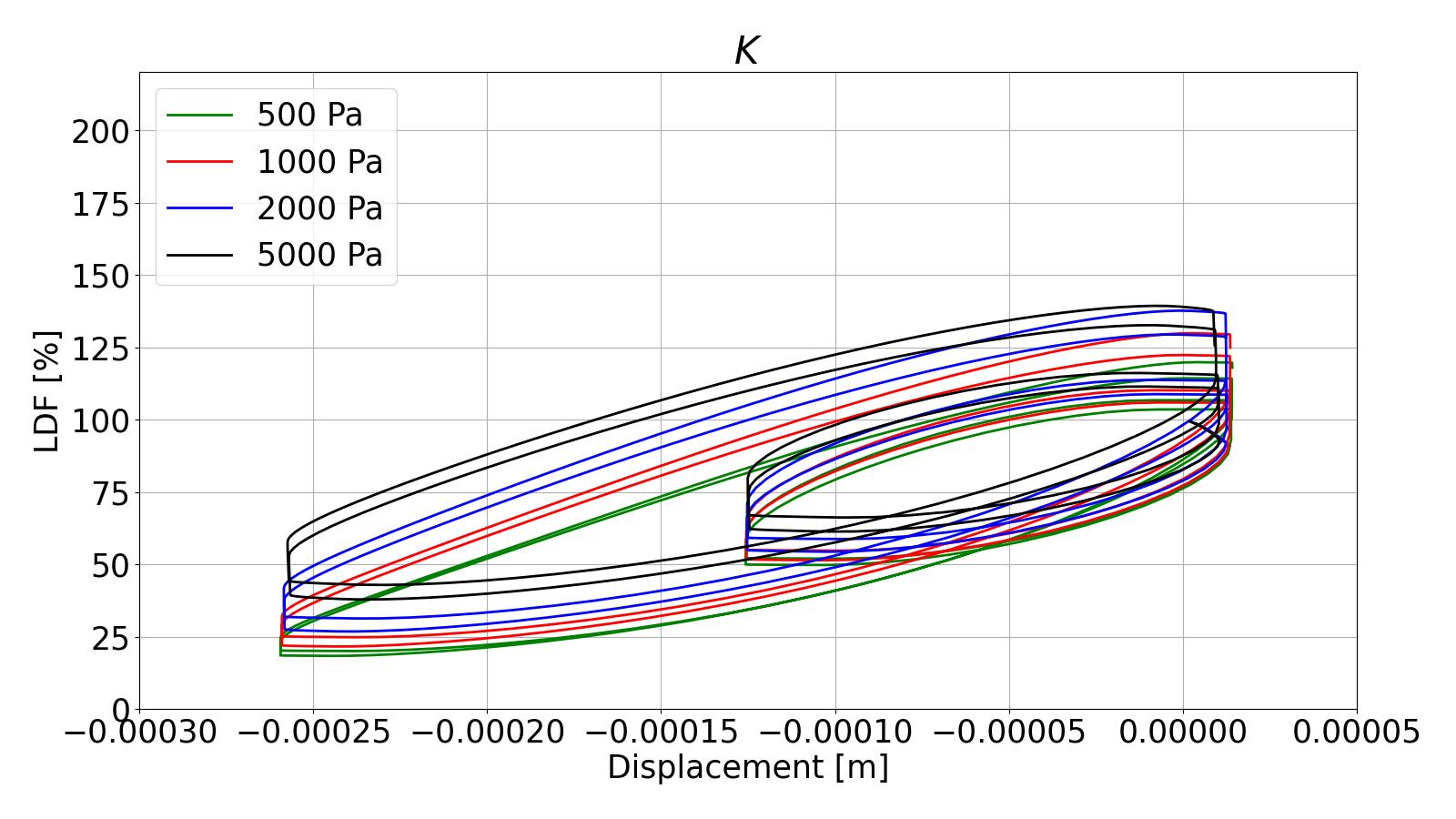}
    \includegraphics[width=0.4\linewidth]{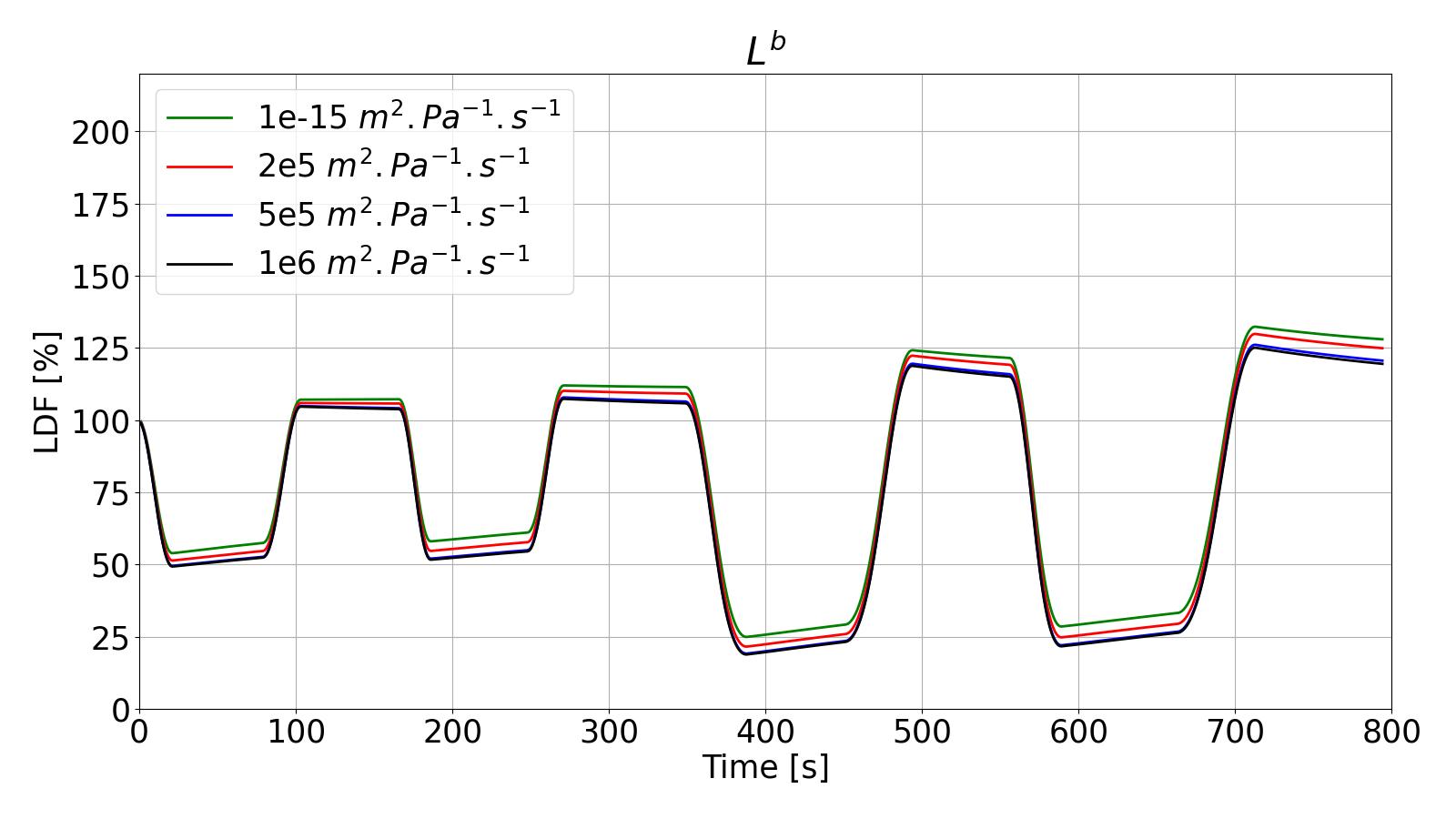}
    \includegraphics[width=0.4\linewidth]{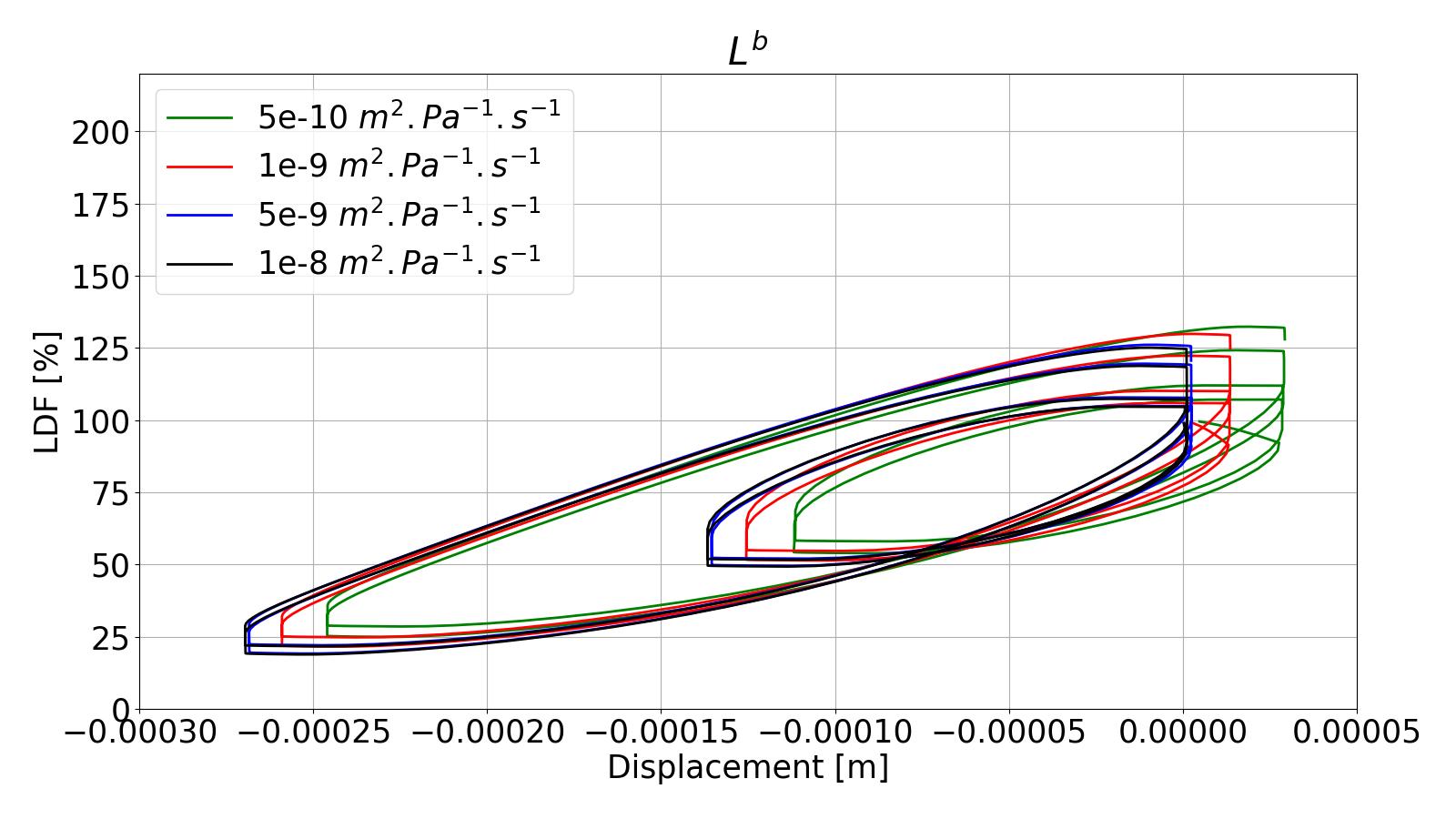}
    \includegraphics[width=0.4\linewidth]{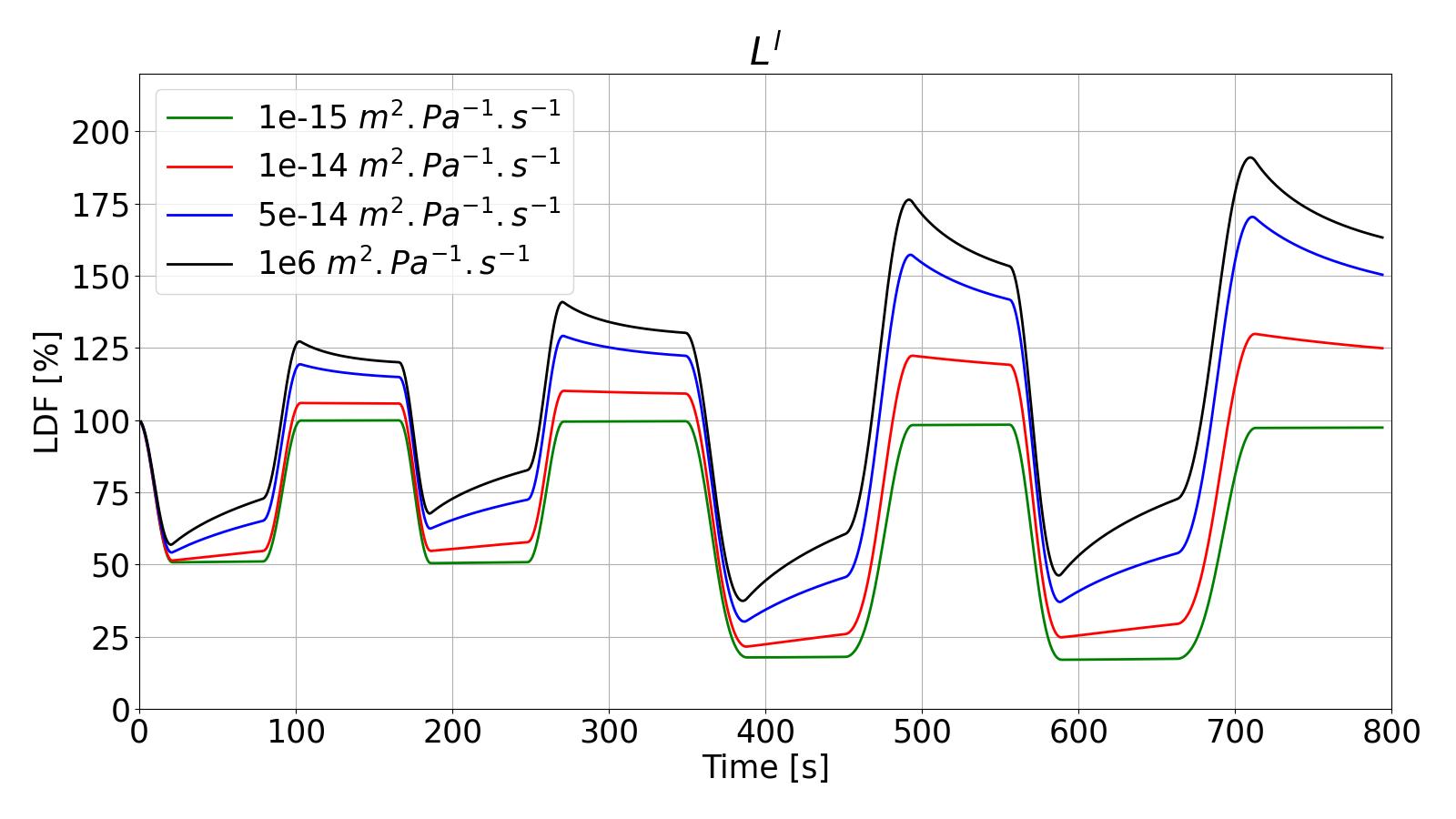}
    \includegraphics[width=0.4\linewidth]{Figure_D_Lb_b.jpg}
    \caption{Impact of the parameters on the model output: first-order sensitivity. Each parameter was independently varied to study the impact on the model response.}
    \label{fig:sensitivity:appendix}
\end{figure}

This appendix provides the sensitivity curves for each parameter (first-order sensitivity) in Figure \ref{fig:sensitivity:appendix}.

The analysis reveals that an increase in Young's modulus leads to a reduction in displacement, thereby decreasing the ischaemic response. In contrast, increasing the hydraulic permeability of the interstitial fluid and the compressibility of the vessel tends to alleviate ischaemia and enhance the hyperaemic response. In contrast, an increase in blood hydraulic permeability results in a diminished hyperaemic response. Both ischaemic and hyperaemic responses are further amplified by an increase in the exponent $\alpha$. Interestingly, the Young modulus mainly impacts the displacement magnitude, whereas the other parametric factor played an important role in the hysteresis response of the model in the flux-displacement curves.

\newpage

\section{Parallel environment install}
\label{parallel}

This procedure was used to install FEniCSx on the HPC Cassiopee (Red Hat Enterprise Linux 8.10).

\subsection{Install}
\begin{itemize}
\item Get version 0.23.0 of spack:
\begin{lstlisting}[language=bash,linewidth=.99\textwidth]
wget https://github.com/spack/spack/releases/download/
                            v0.23.0/spack-0.23.0.tar.gz
\end{lstlisting}

\item untar the archive:
\begin{lstlisting}[language=bash]
tar -xf spack-0.23.0.tar.gz
\end{lstlisting}

\item Optional: disable the local configuration of spack (current user home directory). Useful if you have multiple versions of spack
\begin{lstlisting}[language=bash]
export SPACK_DISABLE_LOCAL_CONFIG=true
\end{lstlisting}

\item activate spack:
\begin{lstlisting}[language=bash]
source spack/share/spack/setup-env.sh
\end{lstlisting}

\item add system compiler to spack (need to be installed in your os):
\begin{lstlisting}[language=bash]
spack compiler find 
\end{lstlisting}

\item optional: add a decent compiler (useful if you have an old distribution) and add it as compiler to spack:
\begin{lstlisting}[language=bash]
spack install gcc@12
spack compiler find \$(spack location -i gcc@12)
\end{lstlisting}

\item create a spack environment (useful if you want to load several packages at one time):
\begin{lstlisting}[language=bash]
spack env create MyFenicsxEnv
\end{lstlisting}

\item activate the environment, add some packages, and install them:
\begin{lstlisting}[language=bash]
spack env activate MyFenicsxEnv
spack add petsc +fortran +mumps +trilinos +superlu-dist
spack add python
spack add py-fenics-dolfinx@0.9.0
spack add py-numpy
spack add py-matplotlib
spack add py-pandas
spack add py-gmsh
spack install
\end{lstlisting}
Congratulation, you have installed fenicsx 0.9.0, working with mpi!

\item create the porous\_fenicsx package:
\begin{lstlisting}[language=bash]
cd $PATH/porous_fenicsx
python -m pip install .
\end{lstlisting}

\item deactivate your environment:
\begin{lstlisting}[language=bash]
spack env deactivate
\end{lstlisting}
\end{itemize}

\subsection{Use}

\begin{itemize}
\item activate spack in your session (can be added to your .bashrc):
\begin{lstlisting}[language=bash]
source FullPathToSpack/share/spack/setup-env.sh
\end{lstlisting}
\item activate your environment:
\begin{lstlisting}[language=bash]
spack env activate MyFenicsxEnv
\end{lstlisting}
\item launch your script in parallel (replace NCPUs with the number of CPUs you want):
\begin{lstlisting}[language=bash]
mpirun -n NCPUs python myscript.py
\end{lstlisting}
\item deactivate your environment:
\begin{lstlisting}[language=bash]
spack env deactivate
\end{lstlisting}
\end{itemize}

\newpage

\bibliographystyle{apalike} 
\bibliography{references,reference_besancon,reference_tuto,references_ldf}

\begin{thebibliography}{}

\bibitem[Abdo et~al., 2020]{Abdo2020}
Abdo, J.~M., Sopko, N.~A., and Milner, S.~M. (2020).
\newblock The applied anatomy of human skin: A model for regeneration.
\newblock {\em Wound Medicine}, 28:100179.

\bibitem[Ajan et~al., 2024]{Ajan2024}
Ajan, A., Roberg, K., Fredriksson, I., and Abtahi, J. (2024).
\newblock Reproducibility of laser doppler flowmetry in gingival microcirculation. a study on six different protocols.
\newblock {\em Microvascular Research}, 153.
\newblock Cited by: 0.

\bibitem[Alnæs et~al., 2015]{FEniCS}
Alnæs, M., Blechta, J., Hake, J., Johansson, A., Kehlet, B., Logg, A., Richardson, C., Ring, J., Rognes, M.~E., and Wells, G.~N. (2015).
\newblock The fenics project version 1.5.
\newblock {\em <p>Archive of Numerical Software}, Vol 3:<strong>Starting Point and Frequency: </strong>Year: 2013</p>.

\bibitem[An et~al., 2010]{An2010}
An, L., Qin, J., and Wang, R.~K. (2010).
\newblock Ultrahigh sensitive optical microangiography for in vivo imaging of microcirculations within human skin tissue beds.
\newblock {\em Optics Express}, 18(8):8220.

\bibitem[Anthony et~al., 2008]{Anthony2008}
Anthony, D., Parboteeah, S., Saleh, M., and Papanikolaou, P. (2008).
\newblock Norton, waterlow and braden scores: a review of the literature and a comparison between the scores and clinical judgement.
\newblock {\em Journal of Clinical Nursing}, 17(5):646–653.

\bibitem[Argarini et~al., 2020]{Argarini2020}
Argarini, R., Smith, K.~J., Carter, H.~H., Naylor, L.~H., McLaughlin, R.~A., and Green, D.~J. (2020).
\newblock Visualizing and quantifying the impact of reactive hyperemia on cutaneous microvessels in humans.
\newblock {\em Journal of Applied Physiology}, 128(1):17–24.

\bibitem[Balasubramanian et~al., 2021]{Balasubramanian2021}
Balasubramanian, G.~V., Chockalingam, N., and Naemi, R. (2021).
\newblock The role of cutaneous microcirculatory responses in tissue injury, inflammation and repair at the foot in diabetes.
\newblock {\em Frontiers in Bioengineering and Biotechnology}, 9.

\bibitem[Baran et~al., 2015]{Baran2015}
Baran, U., Choi, W.~J., and Wang, R.~K. (2015).
\newblock Potential use of <scp>oct</scp>‐based microangiography in clinical dermatology.
\newblock {\em Skin Research and Technology}, 22(2):238–246.

\bibitem[Baratta et~al., 2023]{barratta}
Baratta, I.~A., Dean, J.~P., Dokken, J.~S., Habera, M., Hale, J.~S., Richardson, C.~N., Rognes, M.~E., Scroggs, M.~W., Sime, N., and Wells, G.~N. (2023).
\newblock Dolfinx: The next generation fenics problem solving environment.

\bibitem[Basiladze et~al., 2015]{Basiladze201583}
Basiladze, T., Bekaia, G., Gongadze, N., and Mitagvaria, N. (2015).
\newblock Possible mechanism of hyperemia in the skin caused by non-painful mechanical pressure.
\newblock {\em Georgian medical news}, page 83 – 88.

\bibitem[Bauer et~al., 2005]{Bauer2005}
Bauer, D., Grebe, R., and Ehrlacher, A. (2005).
\newblock First phase microcirculatory reaction to mechanical skin irritation: a three layer model of a compliant vascular tree.
\newblock {\em Journal of Theoretical Biology}, 232(2):249–260.

\bibitem[Borchardt et~al., 2017]{Borchardt2017}
Borchardt, T., Ernst, J., Helmke, A., Tanyeli, M., Schilling, A.~F., Felmerer, G., and Vi\"{o}l, W. (2017).
\newblock Effect of direct cold atmospheric plasma (di<scp>cap</scp>) on microcirculation of intact skin in a controlled mechanical environment.
\newblock {\em Microcirculation}, 24(8).

\bibitem[Braverman, 1989]{Braverman1989}
Braverman, I.~M. (1989).
\newblock Ultrastructure and organization of the cutaneous microvasculature in normal and pathologic states.
\newblock {\em Journal of Investigative Dermatology}, 93(2):S2–S9.

\bibitem[Breuls et~al., 2003]{Breuls2003}
Breuls, R. G.~M., Bouten, C. V.~C., Oomens, C. W.~J., Bader, D.~L., and Baaijens, F. P.~T. (2003).
\newblock A theoretical analysis of damage evolution in skeletal muscle tissue with reference to pressure ulcer development.
\newblock {\em Journal of Biomechanical Engineering}, 125(6):902–909.

\bibitem[Breuls et~al., 2002]{Breuls2002}
Breuls, R. G.~M., Sengers, B.~G., Oomens, C. W.~J., Bouten, C. V.~C., and Baaijens, F. P.~T. (2002).
\newblock Predicting local cell deformations in engineered tissue constructs: A multilevel finite element approach.
\newblock {\em Journal of Biomechanical Engineering}, 124(2):198–207.

\bibitem[Budday et~al., 2019]{Budday2019}
Budday, S., Ovaert, T.~C., Holzapfel, G.~A., Steinmann, P., and Kuhl, E. (2019).
\newblock Fifty shades of brain: A review on the mechanical testing and modeling of brain tissue.
\newblock {\em Archives of Computational Methods in Engineering}, 27(4):1187--1230.

\bibitem[Bulle, 2022]{bulle:tel-03652547}
Bulle, R. (2022).
\newblock {\em A posteriori error estimation for finite element approximations of fractional Laplacian problems and applications to poro-elasticity}.
\newblock Theses, {Universit{\'e} Bourgogne Franche-Comt{\'e} ; Universit{\'e} du Luxembourg}.

\bibitem[Bulle et~al., 2021]{Bulle2021}
Bulle, R., Alotta, G., Marchiori, G., Berni, M., Lopomo, N.~F., Zaffagnini, S., Bordas, S. P.~A., and Barrera, O. (2021).
\newblock The human meniscus behaves as a functionally graded fractional porous medium under confined compression conditions.
\newblock {\em Applied Sciences}, 11(20):9405.

\bibitem[Burns et~al., 2010]{Burns2010}
Burns, T., Breathnach, S.~M., Cox, N.~H., and Griffiths, C., editors (2010).
\newblock {\em Rook's textbook of dermatology}.
\newblock Wiley-Blackwell, Chichester, England, 8 edition.

\bibitem[Carrasco-Mantis et~al., 2023]{CarrascoMantis2023}
Carrasco-Mantis, A., Randelovic, T., Castro-Abril, H., Ochoa, I., Doblaré, M., and Sanz-Herrera, J.~A. (2023).
\newblock A mechanobiological model for tumor spheroid evolution with application to glioblastoma: A continuum multiphysics approach.
\newblock {\em Computers in Biology and Medicine}, 159:106897.

\bibitem[Ceelen et~al., 2008]{Ceelen2008}
Ceelen, K., Stekelenburg, A., Loerakker, S., Strijkers, G., Bader, D., Nicolay, K., Baaijens, F., and Oomens, C. (2008).
\newblock Compression-induced damage and internal tissue strains are related.
\newblock {\em Journal of Biomechanics}, 41(16):3399--3404.

\bibitem[Choi et~al., 2014]{Choi2014}
Choi, P.~J., Brunt, V.~E., Fujii, N., and Minson, C.~T. (2014).
\newblock New approach to measure cutaneous microvascular function: an improved test of no-mediated vasodilation by thermal hyperemia.
\newblock {\em Journal of Applied Physiology}, 117(3):277–283.

\bibitem[Coleman et~al., 2013]{Coleman2013}
Coleman, S., Gorecki, C., Nelson, E.~A., Closs, S.~J., Defloor, T., Halfens, R., Farrin, A., Brown, J., Schoonhoven, L., and Nixon, J. (2013).
\newblock Patient risk factors for pressure ulcer development: Systematic review.
\newblock {\em International Journal of Nursing Studies}, 50(7):974--1003.

\bibitem[Cracowski et~al., 2006]{Cracowski2006}
Cracowski, J.-L., Minson, C.~T., Salvat-Melis, M., and Halliwill, J.~R. (2006).
\newblock Methodological issues in the assessment of skin microvascular endothelial function in humans.
\newblock {\em Trends in Pharmacological Sciences}, 27(9):503–508.

\bibitem[de~Lucio et~al., 2023]{deLucio2023}
de~Lucio, M., Leng, Y., Hans, A., Bilionis, I., Brindise, M., Ardekani, A.~M., Vlachos, P.~P., and Gomez, H. (2023).
\newblock Modeling large-volume subcutaneous injection of monoclonal antibodies with anisotropic porohyperelastic models and data-driven tissue layer geometries.
\newblock {\em Journal of the Mechanical Behavior of Biomedical Materials}, 138:105602.

\bibitem[de~Lucio et~al., 2024]{deLucio2024}
de~Lucio, M., Leng, Y., Wang, H., Vlachos, P.~P., and Gomez, H. (2024).
\newblock Modeling drug transport and absorption in subcutaneous injection of monoclonal antibodies: Impact of tissue deformation, devices, and physiology.
\newblock {\em International Journal of Pharmaceutics}, 661:124446.

\bibitem[Dwivedi et~al., 2022]{Dwivedi2022}
Dwivedi, K.~K., Lakhani, P., Kumar, S., and Kumar, N. (2022).
\newblock Effect of collagen fibre orientation on the poisson's ratio and stress relaxation of skin: an ex vivo and in vivo study.
\newblock {\em Royal Society Open Science}, 9(3).

\bibitem[Fagrell et~al., 1977]{Fagrell1977-ws}
Fagrell, B., Fronek, A., and Intaglietta, M. (1977).
\newblock A microscope-television system for studying flow velocity in human skin capillaries.
\newblock {\em Am. J. Physiol. Heart Circ. Physiol.}, 233(2):H318--H321.

\bibitem[Fang et~al., 2024]{Fang2024}
Fang, Y., Cui, J., Xu, J., Ma, J., Liu, Z., and Zhang, H. (2024).
\newblock Research progress and clinical application of laser doppler blood flow measurement technology.
\newblock {\em Proceedings of SPIE - The International Society for Optical Engineering}, 13182.

\bibitem[Farkas et~al., 2004]{Farkas2004}
Farkas, K., Kolossv{\'a}ry, E., J{\'a}rai, Z., Nemcsik, J., and Farsang, C. (2004).
\newblock Non-invasive assessment of microvascular endothelial function by laser doppler flowmetry in patients with essential hypertension.
\newblock {\em Atherosclerosis}, 173(1):97--102.

\bibitem[Fedorovich et~al., 2018]{Fedorovich2018}
Fedorovich, A., Drapkina, O., Pronko, K., Sinopalnikov, V., and Zemskov, V. (2018).
\newblock Telemonitoring of capillary blood flow in the human skin: New opportunities and prospects.
\newblock {\em Clinical Practice}.

\bibitem[Fei et~al., 2018]{Fei2018656}
Fei, W., Xu, S., Ma, J., Zhai, W., Cheng, S., Chang, Y., Wang, X., Gao, J., Tang, H., Yang, S., and Zhang, X. (2018).
\newblock Fundamental supply of skin blood flow in the chinese han population: Measurements by a full-field laser perfusion imager.
\newblock {\em Skin Research and Technology}, 24(4):656 – 662.
\newblock Cited by: 1.

\bibitem[Folgosi-Correa and Nogueira, 2012]{FolgosiCorrea2012}
Folgosi-Correa, M.~S. and Nogueira, G. E.~C. (2012).
\newblock Time of correlation of low-frequency fluctuations in the regional laser doppler flow signal from human skin.
\newblock In Popp, J., Drexler, W., Tuchin, V.~V., and Matthews, D.~L., editors, {\em Biophotonics: Photonic Solutions for Better Health Care III}, volume 8427, page 84272D. SPIE.

\bibitem[Fredriksson et~al., 2009]{FREDRIKSSON20094}
Fredriksson, I., Larsson, M., and Strömberg, T. (2009).
\newblock Measurement depth and volume in laser doppler flowmetry.
\newblock {\em Microvascular Research}, 78(1):4--13.

\bibitem[Fromy et~al., 1998]{fromy1998}
Fromy, B., Abraham, P., and Saumet, J.-L. (1998).
\newblock Non-nociceptive capsaicin-sensitive nerve terminal stimulation allows for an original vasodilatory reflex in the human skin.
\newblock {\em Brain Research}, 811(1-2):166--168.

\bibitem[Fromy et~al., 2000]{fromy2000}
Fromy, B., Abraham, P., and Saumet, J.-L. (2000).
\newblock Progressive calibrated pressure device to measure cutaneous blood flow changes to external pressure strain.
\newblock {\em Brain Research Protocols}, 5(2):198--203.

\bibitem[Fuchs et~al., 2017]{Fuchs2017}
Fuchs, D., Dupon, P.~P., Schaap, L.~A., and Draijer, R. (2017).
\newblock The association between diabetes and dermal microvascular dysfunction non-invasively assessed by laser doppler with local thermal hyperemia: a systematic review with meta-analysis.
\newblock {\em Cardiovascular Diabetology}, 16(1).

\bibitem[Fukumura and Jain, 2007]{FUKUMURA200772}
Fukumura, D. and Jain, R.~K. (2007).
\newblock Tumor microvasculature and microenvironment: Targets for anti-angiogenesis and normalization.
\newblock {\em Microvascular Research}, 74(2):72--84.
\newblock Therapeutic Applications of Angiogenesis and Anti-angiogenesis.

\bibitem[Geuzaine and Remacle, 2009]{Geuzaine2009}
Geuzaine, C. and Remacle, J. (2009).
\newblock Gmsh: A 3‐d finite element mesh generator with built‐in pre‐ and post‐processing facilities.
\newblock {\em International Journal for Numerical Methods in Engineering}, 79(11):1309–1331.

\bibitem[Golay et~al., 2004]{Golay2004}
Golay, S., Haeberli, C., Delachaux, A., Liaudet, L., Kucera, P., Waeber, B., and Feihl, F. (2004).
\newblock Local heating of human skin causes hyperemia without mediation by muscarinic cholinergic receptors or prostanoids.
\newblock {\em Journal of Applied Physiology}, 97(5):1781–1786.

\bibitem[Greiner et~al., 2021]{Greiner2021}
Greiner, A., Reiter, N., Paulsen, F., Holzapfel, G.~A., Steinmann, P., Comellas, E., and Budday, S. (2021).
\newblock Poro-viscoelastic effects during biomechanical testing of human brain tissue.
\newblock {\em Frontiers in Mechanical Engineering}, 7.

\bibitem[Han et~al., 2023]{Han2023}
Han, D., Huang, Z., Rahimi, E., and Ardekani, A.~M. (2023).
\newblock Solute transport across the lymphatic vasculature in a soft skin tissue.
\newblock {\em Biology}, 12(7):942.

\bibitem[Hervas-Raluy et~al., 2023]{HervasRaluy2023}
Hervas-Raluy, S., Wirthl, B., Guerrero, P.~E., Robalo~Rei, G., Nitzler, J., Coronado, E., Font~de Mora~Sainz, J., Schrefler, B.~A., Gomez-Benito, M.~J., Garcia-Aznar, J.~M., and Wall, W.~A. (2023).
\newblock Tumour growth: An approach to calibrate parameters of a multiphase porous media model based on in vitro observations of neuroblastoma spheroid growth in a hydrogel microenvironment.
\newblock {\em Computers in Biology and Medicine}, 159:106895.

\bibitem[Horn et~al., 2022]{Horn2022}
Horn, A.~G., Schulze, K.~M., Weber, R.~E., Barstow, T.~J., Musch, T.~I., Poole, D.~C., and Behnke, B.~J. (2022).
\newblock Post-occlusive reactive hyperemia and skeletal muscle capillary hemodynamics.
\newblock {\em Microvascular Research}, 140:104283.

\bibitem[Hosseini-Farid et~al., 2020]{HosseiniFarid2020}
Hosseini-Farid, M., Ramzanpour, M., McLean, J., Ziejewski, M., and Karami, G. (2020).
\newblock A poro-hyper-viscoelastic rate-dependent constitutive modeling for the analysis of brain tissues.
\newblock {\em Journal of the Mechanical Behavior of Biomedical Materials}, 102:103475.

\bibitem[Hsu and Fuchs, 2021]{Hsu2021}
Hsu, Y.-C. and Fuchs, E. (2021).
\newblock Building and maintaining the skin.
\newblock {\em Cold Spring Harbor Perspectives in Biology}, 14(7):a040840.

\bibitem[Humeau-Heurtier et~al., 2013]{Heurtier2013}
Humeau-Heurtier, A., Guerreschi, E., Abraham, P., and Mahé, G. (2013).
\newblock Relevance of laser doppler and laser speckle techniques for assessing vascular function: State of the art and future trends.
\newblock {\em IEEE Transactions on Biomedical Engineering}, 60(3):659--666.

\bibitem[Javadi et~al., 2023]{Javadi2023775}
Javadi, E., Armstrong, M.~J., and Jamali, S. (2023).
\newblock A fully physiologically-informed time- and rate-dependent hemorheological constitutive model.
\newblock {\em Journal of Rheology}, 67(3):775 – 788.
\newblock Cited by: 1; All Open Access, Bronze Open Access.

\bibitem[Joodaki and Panzer, 2018]{Joodaki2018}
Joodaki, H. and Panzer, M.~B. (2018).
\newblock Skin mechanical properties and modeling: A review.
\newblock {\em Proceedings of the Institution of Mechanical Engineers, Part H: Journal of Engineering in Medicine}, 232(4):323--343.

\bibitem[Kallepalli et~al., 2022]{Kallepalli2022}
Kallepalli, A., Halls, J., James, D.~B., and Richardson, M.~A. (2022).
\newblock An ultrasonography‐based approach for tissue modelling to inform photo‐therapy treatment strategies.
\newblock {\em Journal of Biophotonics}, 15(4).

\bibitem[Kalra and Lowe, 2016]{Kalra2016}
Kalra, A. and Lowe, A. (2016).
\newblock Mechanical behaviour of skin: A review.
\newblock {\em Journal of Material Science {\&} Engineering}, 5(4).

\bibitem[Kazemi et~al., 2013]{Kazemi2013}
Kazemi, M., Dabiri, Y., and Li, L.~P. (2013).
\newblock Recent advances in computational mechanics of the human knee joint.
\newblock {\em Computational and Mathematical Methods in Medicine}, 2013:1–27.

\bibitem[Kelly et~al., 1995]{Kelly1995}
Kelly, R.~I., Pearse, R., Bull, R.~H., Leveque, J.-L., de~Rigal, J., and Mortimer, P.~S. (1995).
\newblock The effects of aging on the cutaneous microvasculature.
\newblock {\em Journal of the American Academy of Dermatology}, 33(5):749–756.

\bibitem[Kim et~al., 2020]{Kim2020}
Kim, J., Lee, J., and Lee, E. (2020).
\newblock Risk factors for newly acquired pressure ulcer and the impact of nurse staffing on pressure ulcer incidence.
\newblock {\em Journal of Nursing Management}, 30(5).

\bibitem[Kouadio et~al., 2018]{Kouadio201858}
Kouadio, A.~A., Jordana, F., Koffi, N.~J., Le~Bars, P., and Soueidan, A. (2018).
\newblock The use of laser doppler flowmetry to evaluate oral soft tissue blood flow in humans: A review.
\newblock {\em Archives of Oral Biology}, 86:58 – 71.
\newblock Cited by: 37.

\bibitem[Kubli et~al., 2000]{Kubli2000}
Kubli, S., Waeber, B., Dalle-Ave, A., and Feihl, F. (2000).
\newblock Reproducibility of laser doppler imaging of skin blood flow as a tool to assess endothelial function.
\newblock {\em Journal of cardiovascular pharmacology}, 36(5):640--648.

\bibitem[Lakhani et~al., 2021]{Lakhani2021}
Lakhani, P., Dwivedi, K.~K., Parashar, A., and Kumar, N. (2021).
\newblock Non-invasive in vivo quantification of directional dependent variation in mechanical properties for human skin.
\newblock {\em Frontiers in Bioengineering and Biotechnology}, 9.

\bibitem[Lavigne et~al., 2022]{Lavigne2022}
Lavigne, T., Sciumè, G., Laporte, S., Pillet, H., Urcun, S., Wheatley, B., and Rohan, P.-Y. (2022).
\newblock Société de biomécanique young investigator award 2021: Numerical investigation of the time-dependent stress–strain mechanical behaviour of skeletal muscle tissue in the context of pressure ulcer prevention.
\newblock {\em Clinical Biomechanics}, 93:105592.

\bibitem[Lavigne et~al., 2023]{Lavigne2023}
Lavigne, T., Urcun, S., Rohan, P.-Y., Scium{\`{e}}, G., Baroli, D., and Bordas, S.~P. (2023).
\newblock Single and bi-compartment poro-elastic model of perfused biological soft tissues: {FEniCSx} implementation and tutorial.
\newblock {\em Journal of the Mechanical Behavior of Biomedical Materials}, 143:105902.

\bibitem[Leng et~al., 2021]{Leng2021}
Leng, Y., de~Lucio, M., and Gomez, H. (2021).
\newblock Using poro-elasticity to model the large deformation of tissue during subcutaneous injection.
\newblock {\em Computer Methods in Applied Mechanics and Engineering}, 384:113919.

\bibitem[Leo et~al., 2020]{Leo2020}
Leo, F., Krenz, T., Wolff, G., Weidenbach, M., Heiss, C., Kelm, M., Isakson, B., and Cortese-Krott, M.~M. (2020).
\newblock Assessment of tissue perfusion and vascular function in mice by scanning laser doppler perfusion imaging.
\newblock {\em Biochemical Pharmacology}, 176:113893.

\bibitem[Liao et~al., 2013]{Liao2013}
Liao, F., Burns, S., and Jan, Y.-K. (2013).
\newblock Skin blood flow dynamics and its role in pressure ulcers.
\newblock {\em Journal of Tissue Viability}, 22(2):25--36.

\bibitem[Loerakker et~al., 2011]{Loerakker2011}
Loerakker, S., Manders, E., Strijkers, G.~J., Nicolay, K., Baaijens, F. P.~T., Bader, D.~L., and Oomens, C. W.~J. (2011).
\newblock The effects of deformation, ischemia, and reperfusion on the development of muscle damage during prolonged loading.
\newblock {\em Journal of Applied Physiology}, 111(4):1168–1177.

\bibitem[Loerakker et~al., 2010]{Loerakker2010}
Loerakker, S., Stekelenburg, A., Strijkers, G.~J., Rijpkema, J. J.~M., Baaijens, F. P.~T., Bader, D.~L., Nicolay, K., and Oomens, C. W.~J. (2010).
\newblock Temporal effects of mechanical loading on deformation-induced damage in skeletal muscle tissue.
\newblock {\em Annals of Biomedical Engineering}, 38(8):2577--2587.

\bibitem[Lustig et~al., 2021]{Lustig2021}
Lustig, A., Margi, R., Orlov, A., Orlova, D., Azaria, L., and Gefen, A. (2021).
\newblock The mechanobiology theory of the development of medical device-related pressure ulcers revealed through a cell-scale computational modeling framework.
\newblock {\em Biomechanics and Modeling in Mechanobiology}, 20(3):851–860.

\bibitem[Millet et~al., 2011]{Millet2011}
Millet, C., Roustit, M., Blaise, S., and Cracowski, J. (2011).
\newblock Comparison between laser speckle contrast imaging and laser doppler imaging to assess skin blood flow in humans.
\newblock {\em Microvascular research}, 82(2):147--151.

\bibitem[Mithraratne et~al., 2012]{Mithraratne20121071}
Mithraratne, K., Ho, H., Hunter, P., and Fernandez, J. (2012).
\newblock Mechanics of the foot part 2: A coupled solid-fluid model to investigate blood transport in the pathologic foot.
\newblock {\em International Journal for Numerical Methods in Biomedical Engineering}, 28(10):1071 – 1081.
\newblock Cited by: 17.

\bibitem[Moore et~al., 2019]{Moore2019}
Moore, Z., Avsar, P., Conaty, L., Moore, D.~H., Patton, D., and O’Connor, T. (2019).
\newblock The prevalence of pressure ulcers in europe, what does the european data tell us: a systematic review.
\newblock {\em Journal of Wound Care}, 28(11):710–719.

\bibitem[Mrowietz et~al., 2019]{Mrowietz2019129}
Mrowietz, C., Franke, R., Pindur, G., Sternitzky, R., Jung, F., and Wolf, U. (2019).
\newblock Evaluation of laser-doppler-fluxmetry for the diagnosis of microcirculatory disorders.
\newblock {\em Clinical Hemorheology and Microcirculation}, 71(2):129 – 135.
\newblock Cited by: 12.

\bibitem[Mrowietz et~al., 2017]{Mrowietz2017347}
Mrowietz, C., Franke, R., Pindur, G., Wolf, U., and Jung, F. (2017).
\newblock Reference range and variability of laser-doppler-fluxmetry.
\newblock {\em Clinical Hemorheology and Microcirculation}, 67(3-4):347 – 353.
\newblock Cited by: 8.

\bibitem[Mukhina et~al., 2020]{Mukhina2020}
Mukhina, E., Rohan, P.-Y., Connesson, N., and Payan, Y. (2020).
\newblock Calibration of the fat and muscle hyperelastic material parameters for the assessment of the internal tissue deformation in relation to pressure ulcer prevention.
\newblock {\em Computer Methods in Biomechanics and Biomedical Engineering}, 23(sup1):S197--S199.

\bibitem[Nakagawa et~al., 2010]{Nakagawa2010}
Nakagawa, N., Matsumoto, M., and Sakai, S. (2010).
\newblock In vivo measurement of the water content in the dermis by confocal raman spectroscopy.
\newblock {\em Skin Research and Technology}, 16(2):137--141.

\bibitem[Nguyen-Tu et~al., 2013]{NguyenTu2013}
Nguyen-Tu, M.-S., Begey, A.-L., Decorps, J., Boizot, J., Sommer, P., Fromy, B., and Sigaudo-Roussel, D. (2013).
\newblock Skin microvascular response to pressure load in obese mice.
\newblock {\em Microvascular Research}, 90:138–143.

\bibitem[Nogami et~al., 2019]{Nogami2019287}
Nogami, H., Komatsutani, K., Hirata, T., and Sawada, R. (2019).
\newblock Integrated laser doppler blood flowing combining optical contact force.
\newblock {\em 2019 International Conference on Electronics Packaging, ICEP 2019}, page 287 – 290.

\bibitem[Oftadeh et~al., 2023]{Oftadeh2023}
Oftadeh, R., Azadi, M., Donovan, M., Langer, J., Liao, I.-C., Ortiz, C., Grodzinsky, A.~J., and Luengo, G.~S. (2023).
\newblock Poroelastic behavior and water permeability of human skin at the nanoscale.
\newblock {\em PNAS Nexus}, 2(8).

\bibitem[Oftadeh et~al., 2018]{Oftadeh2018}
Oftadeh, R., Connizzo, B.~K., Nia, H.~T., Ortiz, C., and Grodzinsky, A.~J. (2018).
\newblock Biological connective tissues exhibit viscoelastic and poroelastic behavior at different frequency regimes: Application to tendon and skin biophysics.
\newblock {\em Acta Biomaterialia}, 70:249–259.

\bibitem[Oomens et~al., 2017]{Payan2017}
Oomens, C.~W., {van Vijven}, M., and Peters, G.~W. (2017).
\newblock Chapter 16 - skin mechanics.
\newblock In Payan, Y. and Ohayon, J., editors, {\em Biomechanics of Living Organs}, volume~1 of {\em Translational Epigenetics}, pages 347--357. Academic Press, Oxford.

\bibitem[Ostergren and Fagrell, 1986]{ostergren1986skin}
Ostergren, J. and Fagrell, B. (1986).
\newblock Skin capillary blood cell velocity in man. characteristics and reproducibility of the reactive hyperemia response.
\newblock {\em International journal of microcirculation, clinical and experimental}, 5(1):37--51.

\bibitem[Pailler-Mattei et~al., 2008]{PaillerMattei2008}
Pailler-Mattei, C., Bec, S., and Zahouani, H. (2008).
\newblock In vivo measurements of the elastic mechanical properties of human skin by indentation tests.
\newblock {\em Medical Engineering {\&} Physics}, 30(5):599--606.

\bibitem[Pedanekar et~al., 2018]{Pedanekar2018}
Pedanekar, T., Kedare, R., and Sengupta, A. (2018).
\newblock Monitoring tumor progression by mapping skin microcirculation with laser doppler flowmetry.
\newblock {\em Lasers in Medical Science}, 34(1):61–77.

\bibitem[Petrofsky and Berk, 2012]{Petrofsky2012}
Petrofsky, J.~S. and Berk, L. (2012).
\newblock {\em Skin Moisture and Heat Transfer}, page 561–580.
\newblock Springer Berlin Heidelberg.

\bibitem[Poffo et~al., 2014]{Poffo2014}
Poffo, L., Goujon, J.-M., Le~Page, R., Lemaitre, J., Guendouz, M., Lorrain, N., and Bosc, D. (2014).
\newblock Laser double doppler flowmeter.
\newblock In Popp, J., Tuchin, V.~V., Matthews, D.~L., Pavone, F.~S., and Garside, P., editors, {\em Biophotonics: Photonic Solutions for Better Health Care IV}, volume 9129, page 91290X. SPIE.

\bibitem[Querleux et~al., 2002]{Querleux2002}
Querleux, B., Cornillon, C., Jolivet, O., and Bittoun, J. (2002).
\newblock Anatomy and physiology of subcutaneous adipose tissue by in vivo magnetic resonance imaging and spectroscopy: Relationships with sex and presence of cellulite.
\newblock {\em Skin Research and Technology}, 8(2):118–124.

\bibitem[Raveh~Tilleman et~al., 2004]{raveh2004elastic}
Raveh~Tilleman, T., Tilleman, M., and Neumann, H. (2004).
\newblock The elastic properties of cancerous skin: Poisson’s ratio and young’s modulus.
\newblock {\em Optimization of Incisions in Cutaneous Surgery including Mohs’ Micrographic Surgery}, 105(2).

\bibitem[Ricken and Lambers, 2019]{ricken2019computational}
Ricken, T. and Lambers, L. (2019).
\newblock On computational approaches of liver lobule function and perfusion simulation.
\newblock {\em GAMM-Mitteilungen}, 42(4).

\bibitem[Rosenberry and Nelson, 2020]{Rosenberry2020}
Rosenberry, R. and Nelson, M.~D. (2020).
\newblock Reactive hyperemia: a review of methods, mechanisms, and considerations.
\newblock {\em American Journal of Physiology-Regulatory, Integrative and Comparative Physiology}, 318(3):R605–R618.

\bibitem[Rossi et~al., 2006a]{Rossi2006b}
Rossi, M., Carpi, A., Di~Maria, C., Galetta, F., and Santoro, G. (2006a).
\newblock Spectral analysis of laser doppler skin blood flow oscillations in human essential arterial hypertension.
\newblock {\em Microvascular research}, 72(1-2):34--41.

\bibitem[Rossi et~al., 2006b]{Rossi2006}
Rossi, M., Carpi, A., Galetta, F., Franzoni, F., and Santoro, G. (2006b).
\newblock The investigation of skin blood flowmotion: a new approach to study the microcirculatory impairment in vascular diseases?
\newblock {\em Biomedicine \& Pharmacotherapy}, 60(8):437–442.

\bibitem[Russell et~al., 2015]{Russell2015}
Russell, N.~S., Floot, B., van Werkhoven, E., Schriemer, M., de~Jong-Korlaar, R., Woerdeman, L.~A., Stewart, F.~A., and Scharpfenecker, M. (2015).
\newblock Blood and lymphatic microvessel damage in irradiated human skin: The role of tgf-$\beta$, endoglin and macrophages.
\newblock {\em Radiotherapy and Oncology}, 116(3):455–461.

\bibitem[Saha et~al., 2020]{Saha2020}
Saha, M., Dremin, V., Rafailov, I., Dunaev, A., Sokolovski, S., and Rafailov, E. (2020).
\newblock Wearable laser doppler flowmetry sensor: A feasibility study with smoker and non-smoker volunteers.
\newblock {\em Biosensors}, 10(12):201.

\bibitem[Sanders, 1973]{Sanders1973}
Sanders, R. (1973).
\newblock Torsional elasticity of human skin in vivo.
\newblock {\em Pfl gers Archiv European Journal of Physiology}, 342(3):255--260.

\bibitem[Sciumè, 2021]{Scium2021}
Sciumè, G. (2021).
\newblock Mechanistic modeling of vascular tumor growth: an extension of biot’s theory to hierarchical bi-compartment porous medium systems.
\newblock {\em Acta Mechanica}, 232(4):1445–1478.

\bibitem[Sree et~al., 2019]{Sree2019b}
Sree, V.~D., Rausch, M.~K., and Tepole, A.~B. (2019).
\newblock Linking microvascular collapse to tissue hypoxia in a multiscale model of pressure ulcer initiation.
\newblock {\em Biomechanics and Modeling in Mechanobiology}, 18(6):1947–1964.

\bibitem[Stekelenburg et~al., 2006]{Stekelenburg2006}
Stekelenburg, A., Oomens, C. W.~J., Strijkers, G.~J., Nicolay, K., and Bader, D.~L. (2006).
\newblock Compression-induced deep tissue injury examined with magnetic resonance imaging and histology.
\newblock {\em Journal of Applied Physiology}, 100(6):1946--1954.

\bibitem[Svedman et~al., 1998]{Svedman1998}
Svedman, C., Cherry, G.~W., Strigini, E., and Ryan, T.~J. (1998).
\newblock Laser doppler imaging of skin microcirculation.
\newblock {\em Acta dermato-venereologica}, 78(2):114--118.

\bibitem[Terzaghi, 1943]{Terzaghi_1943}
Terzaghi, K. (1943).
\newblock Theoretical soil mechanics., 1943.
\newblock {\em Published Online}, 19.

\bibitem[Tesselaar et~al., 2017]{Tesselaar2017}
Tesselaar, E., Flejmer, A.~M., Farnebo, S., and Dasu, A. (2017).
\newblock Changes in skin microcirculation during radiation therapy for breast cancer.
\newblock {\em Acta Oncologica}, 56(8):1072–1080.

\bibitem[Traa et~al., 2019]{Traa2019}
Traa, W.~A., van Turnhout, M.~C., Nelissen, J.~L., Strijkers, G.~J., Bader, D.~L., and Oomens, C.~W. (2019).
\newblock There is an individual tolerance to mechanical loading in compression induced deep tissue injury.
\newblock {\em Clinical Biomechanics}, 63:153--160.

\bibitem[Tsuji et~al., 2005]{Tsuji2005}
Tsuji, S., Ichioka, S., Sekiya, N., and Nakatsuka, T. (2005).
\newblock Analysis of ischemia-reperfusion injury in a microcirculatory model of pressure ulcers.
\newblock {\em Wound Repair and Regeneration}, 13(2):209–215.

\bibitem[Urcun et~al., 2023]{urcun_2023}
Urcun, S., Baroli, D., Rohan, P.-Y., Skalli, W., Lubrano, V., Bordas, S.~P., and Sciume, G. (2023).
\newblock Non-operable glioblastoma: proposition of patient-specific forecasting by image-informed poromechanical model.
\newblock {\em Brain Multiphysics}, page 100067.

\bibitem[Urcun et~al., 2022]{Urcun2022a}
Urcun, S., Rohan, P.-Y., Sciumè, G., and Bordas, S.~P. (2022).
\newblock Cortex tissue relaxation and slow to medium load rates dependency can be captured by a two-phase flow poroelastic model.
\newblock {\em Journal of the Mechanical Behavior of Biomedical Materials}, 126:104952.

\bibitem[Urcun et~al., 2021]{Urcun2021}
Urcun, S., Rohan, P.-Y., Skalli, W., Nassoy, P., Bordas, S. P.~A., and Sciumè, G. (2021).
\newblock Digital twinning of cellular capsule technology: Emerging outcomes from the perspective of porous media mechanics.
\newblock {\em PLOS ONE}, 16(7):e0254512.

\bibitem[Uzuner et~al., 2022]{Uzuner2022}
Uzuner, S., Kuntze, G., Li, L., Ronsky, J., and Kucuk, S. (2022).
\newblock Creep behavior of human knee joint determined with high-speed biplanar video-radiography and finite element simulation.
\newblock {\em Journal of the Mechanical Behavior of Biomedical Materials}, 125:104905.

\bibitem[Uzuner et~al., 2020]{Uzuner2020}
Uzuner, S., Li, L., Kucuk, S., and Memisoglu, K. (2020).
\newblock Changes in knee joint mechanics after medial meniscectomy determined with a poromechanical model.
\newblock {\em Journal of Biomechanical Engineering}, 142(10).

\bibitem[van Nierop et~al., 2010]{vanNierop2010}
van Nierop, B.~J., Stekelenburg, A., Loerakker, S., Oomens, C.~W., Bader, D., Strijkers, G.~J., and Nicolay, K. (2010).
\newblock Diffusion of water in skeletal muscle tissue is not influenced by compression in a rat model of deep tissue injury.
\newblock {\em Journal of Biomechanics}, 43(3):570--575.

\bibitem[Vanderwee et~al., 2007]{Vanderwee2007}
Vanderwee, K., Clark, M., Dealey, C., Gunningberg, L., and Defloor, T. (2007).
\newblock Pressure ulcer prevalence in europe: a pilot study.
\newblock {\em Journal of Evaluation in Clinical Practice}, 13(2):227--235.

\bibitem[Varghese et~al., 2009]{Varghese2009317}
Varghese, B., Van~Leeuwen, T.~G., and Steenbergen, W. (2009).
\newblock Coherence domain path length resolved approaches in optical doppler flowmetry.
\newblock {\em Handbook of Interferometers: Research, Technology and Applications}, page 317 – 344.

\bibitem[Volkov et~al., 2017]{Volkov2017}
Volkov, M.~V., Kostrova, D.~A., Margaryants, N.~B., Gurov, I.~P., Erofeev, N.~P., Dremin, V.~V., Zharkikh, E.~V., Zherebtsov, E.~A., Kozlov, I.~O., and Dunaev, A.~V. (2017).
\newblock Evaluation of blood microcirculation parameters by combined use of laser doppler flowmetry and videocapillaroscopy methods.
\newblock In Tuchin, V.~V., Genina, E.~A., Postnov, D.~E., and Derbov, V.~L., editors, {\em Saratov Fall Meeting 2016: Optical Technologies in Biophysics and Medicine XVIII}, volume 10336, page 1033607. SPIE.

\bibitem[Wahlsten et~al., 2019]{Wahlsten2019}
Wahlsten, A., Pensalfini, M., Stracuzzi, A., Restivo, G., Hopf, R., and Mazza, E. (2019).
\newblock On the compressibility and poroelasticity of human and murine skin.
\newblock {\em Biomechanics and Modeling in Mechanobiology}, 18(4):1079–1093.

\bibitem[Wang-Evers et~al., 2021]{WangEvers2021}
Wang-Evers, M., Casper, M.~J., Glahn, J., Luo, T., Doyle, A.~E., Karasik, D., Kim, A.~C., Phothong, W., Nathan, N.~R., Heesakker, T., Kositratna, G., and Manstein, D. (2021).
\newblock Assessing the impact of aging and blood pressure on dermal microvasculature by reactive hyperemia optical coherence tomography angiography.
\newblock {\em Scientific Reports}, 11(1).

\bibitem[Wong et~al., 2006]{Wong2006}
Wong, B.~J., Williams, S.~J., and Minson, C.~T. (2006).
\newblock Minimal role for h1and h2histamine receptors in cutaneous thermal hyperemia to local heating in humans.
\newblock {\em Journal of Applied Physiology}, 100(2):535–540.

\bibitem[Wong et~al., 2015]{Wong2015}
Wong, R., Geyer, S., Weninger, W., Guimberteau, J.-C., and Wong, J.~K. (2015).
\newblock The dynamic anatomy and patterning of skin.
\newblock {\em Experimental Dermatology}, 25(2):92--98.

\bibitem[Yalcin et~al., 2015]{Yalcin2015}
Yalcin, O., Ortiz, D., Williams, A.~T., Johnson, P.~C., and Cabrales, P. (2015).
\newblock Perfusion pressure and blood flow determine microvascular apparent viscosity.
\newblock {\em Experimental Physiology}, 100(8):977–987.

\bibitem[Yazdi and Baqersad, 2022]{MostafaviYazdi2022}
Yazdi, S. J.~M. and Baqersad, J. (2022).
\newblock Mechanical modeling and characterization of human skin: A review.
\newblock {\em Journal of Biomechanics}, 130:110864.

\bibitem[Zhang, 2011]{ZHANG2011}
Zhang, J. (2011).
\newblock Effect of suspending viscosity on red blood cell dynamics and blood flows in microvessels.
\newblock {\em Microcirculation}, 18(7):562–573.

\end{thebibliography}

\end{document}